\RequirePackage{ifpdf}
\documentclass[hyper,12pt,letterpaper]{JHEP3}

\usepackage{graphicx}

\usepackage{latexsym,amssymb,amsmath,amsfonts}
\usepackage{mathrsfs}
\usepackage[makeroom]{cancel}

\usepackage{bbm}
\usepackage{bm}
\usepackage{subfigure}
\usepackage{paralist}
\usepackage{url}
\usepackage{youngtab}

\numberwithin{equation}{section}

\newcommand{\nn}{\nonumber}
\newcommand{\mat}[1]{\begin{pmatrix} #1 \end{pmatrix}}

\newcommand{\be}{\begin{equation}} 
\newcommand{\ee}{\end{equation}}
\newcommand{\bea}{\begin{equation} \begin{aligned}} \newcommand{\eea}{\end{aligned} \end{equation}}

\newcommand{\bit}{\begin{itemize}} 
\newcommand{\eit}{\end{itemize}}

\newcommand{\bC}{\mathbb{C}}

\newcommand{\bZ}{\mathbb{Z}}

\newcommand{\Z}{\mathbb{Z}}
\newcommand{\C}{\mathbb{C}}
\newcommand{\R}{\mathbb{R}}
\newcommand{\N}{\mathbb{N}}

\renewcommand{\t}{\widetilde }
\renewcommand{\d}{\partial }

\renewcommand{\b}{\bar }

\newcommand{\half}{{1\over 2}}

\newcommand{\bz}{{\b z}}

\newcommand{\by}{{\bf y}}
\newcommand{\ps}{{(s)}}
\newcommand{\psp}{{(s+1)}}

\newcommand{\CA}{\mathcal{A}}
\newcommand{\CB}{\mathcal{B}}
\newcommand{\CC}{\mathcal{C}}

\newcommand{\CF}{\mathcal{F}}

\newcommand{\CH}{\mathcal{H}}
\newcommand{\CI}{\mathcal{I}}

\newcommand{\CK}{\mathcal{K}}
\newcommand{\CL}{\mathcal{L}}
\newcommand{\CM}{\mathcal{M}}
\newcommand{\CN}{\mathcal{N}}

\newcommand{\CR}{\mathcal{R}}
\newcommand{\CS}{\mathcal{S}}

\newcommand{\CU}{\mathcal{U}}
\newcommand{\CV}{\mathcal{V}}
\newcommand{\CW}{\mathcal{W}}

\newcommand{\CZ}{\mathcal{Z}}

\newcommand{\FR}{\mathfrak{R}}
\newcommand{\Fg}{\mathfrak{g}}
\newcommand{\Fh}{\mathfrak{h}}

\newcommand{\m}{\mathfrak{m}}
\newcommand{\n}{\mathfrak{n}}

\newcommand{\bb}{{\bb b}}

\newcommand{\GG}{\mathbf{G}}

\newcommand{\GH}{\mathbf{H}}

\newcommand{\rk}{{{\rm rk}(\GG)}}

\newcommand{\h}{\hat}

\newcommand{\hs}{\hat{\sigma}}

\newcommand{\oneloop}{\text{1-loop}}

\DeclareMathOperator{\Tr}{Tr}
\DeclareMathOperator{\tr}{tr}

\DeclareMathOperator{\sign}{sign}

\newcommand{\SL}{{\mathscr L}}

\newcommand{\ov}{\over}

\newcommand{\fM}{\mathfrak{M}}
\newcommand{\tfM}{{\widetilde{\mathfrak{M}}}}

\newcommand{\setcond}[2]{\{\,#1\,:\,#2\,\}}

\newcommand{\cu}{{\bf u}}
\newcommand{\ch}{{\bf h}}

\title{Comments on twisted indices \\ in  3d  supersymmetric gauge theories}

\author{Cyril~Closset$^{\flat}$ and Heeyeon Kim$^{\sharp}$  \\

{}$^{\flat}$ Simons Center for Geometry and Physics\\ 
State University of New York, Stony Brook, NY 11794, USA\\

{}$^{\sharp}$ Perimeter Institute for Theoretical Physics\\
31 Caroline Street North, Waterloo, N2L 2Y5, Ontario, Canada

}

\preprint{}
\keywords{Supersymmetry, Topological Field Theory, Wilson, 't Hooft and Polyakov loops}

\abstract{
We study three-dimensional $\CN=2$ supersymmetric gauge theories on ${\Sigma_g \times S^1}$ with a topological twist along  $\Sigma_g$, a genus-$g$ Riemann surface. The twisted supersymmetric index at genus $g$ and the correlation functions of half-BPS loop operators on $S^1$ can be computed  exactly by supersymmetric localization. For $g=1$, this  gives a simple UV computation of the 3d Witten index.
Twisted indices provide us with a clean derivation of the quantum algebra of supersymmetric Wilson loops, for any Yang-Mills-Chern-Simons-matter theory, in terms of the associated Bethe equations for the theory on $\R^2 \times S^1$. This also provides a powerful  and simple tool to study  3d $\CN=2$ Seiberg dualities. 
Finally, we study $A$- and $B$-twisted indices for $\CN=4$ supersymmetric gauge theories, which turns out to be very useful for quantitative  studies of three-dimensional mirror symmetry. We also briefly comment on a relation between the $S^2 \times S^1$ twisted indices and the Hilbert series of $\CN=4$ moduli spaces.
}

\begin{document}

\tableofcontents


\section{Introduction}
Supersymmetric indices \cite{Witten:1982df} are simple yet powerful tools for studying supersymmetric field theories \cite{Witten:1999ds, Kinney:2005ej, Romelsberger:2005eg, Gadde:2013ftv, Benini:2013xpa, Hori:2014tda}. In this paper, we consider the twisted index of  three-dimensional  $\CN=2$ supersymmetric theories with an $R$-symmetry on a closed orientable Riemann surface $\Sigma_g$ of genus $g$:
\be\label{general index}
I_g\left(y_i\,;\, \n_i\right) = \Tr_{\left[\Sigma_g\,;\;  \n_i\right]}\left( (-1)^F \prod_i y_i^{Q_i}\right)~.
\ee
The theory is topologically twisted along $\Sigma_g$ by the $U(1)_R$ symmetry in order to preserve two supercharges, and one can introduce complexified fugacities $y_i$ and  quantized background fluxes $\n_i$  for any continuous global symmetries with conserved charges $Q_i$ commuting with supersymmetry. This index was recently computed by supersymmetric localization for any $\CN=2$ (ultraviolet-free) gauge theory in the $g=0$ case \cite{Benini:2015noa}. 
 In this paper, we  discuss the generalization to higher-genus Riemann surfaces. We also use the index, and similar  localization results for line operators, to study infrared dualities for theories with $\CN=2$ and $\CN=4$ supersymmetry.

\paragraph{Twisted index, localization and Bethe equations.} The quantity \eqref{general index} was first computed in \cite{Nekrasov:2014xaa}  in the context of the Bethe/gauge correspondence \cite{Nekrasov:2009uh, Nekrasov:2009ui}, using slightly different topological field theory methods. In this work, we recompute the twisted index for generic $\CN=2$ supersymmetric Yang-Mills-Chern-Simons (YM-CS) gauge theories with matter, using supersymmetric localization on the classical Coulomb branch \cite{Benini:2015noa, Closset:2015rna}. The index is equal to the supersymmetric partition function of the $\CN=2$ theory on $\Sigma_g\times S^1$, which can be computed as:
\be\label{Z formula intro}
Z_{\Sigma_g\times S^1}(y)= \sum_{\m}  \oint_{\rm JK} {dx\ov 2 \pi i x}Z_{\m}(x, y)~,
\ee
schematically.
Here the sum is over GNO-quantized fluxes $\m$ for the gauge group $\GG$, and the integral is a Jeffrey-Kirwan residue at the singularities of the classical Coulomb branch $\fM \cong (\C^\ast)^\rk/{{\rm Weyl}(\GG)}$, including singularities `at infinity' associated to semi-classical monopole operators. The integrand $Z_\m(x,y)$ contains classical and one-loop contributions. The derivation of \eqref{Z formula intro} closely follows previous  localization computations in related contexts  \cite{Doroud:2012xw, Benini:2012ui, Benini:2013nda, Benini:2013xpa, Hori:2014tda, Benini:2015noa, Closset:2015rna, Closset:2015ohf}. By summing over the fluxes $\m$ in \eqref{Z formula intro}, one recovers the result of \cite{Nekrasov:2014xaa}:
\be\label{Z formula BE intro}
Z_{\Sigma_g\times S^1}(y)=   \sum_{ \h x \,\in\, \CS_{\rm BE}}  \CH(\h x; y)^{g-1}~,
\ee
where $\CH$ is the so-called handle-gluing operator.%
~\footnote{This is a slight simplification valid for vanishing background fluxes. The  general case will be discussed in the main text.} 
The sum in \eqref{Z formula BE intro} is over solutions to the Bethe equations of the $\CN=2$ theory on $\R^2 \times S^1$, which are essentially the saddle equations for the two-dimensional twisted superpotential  $\CW(x; y)$ of the theory compactified on a finite-size circle. It is clear from \eqref{Z formula BE intro}  that much of the physics of  the twisted indices is encoded in  the twisted superpotential.~\footnote{For the same reason, the twisted superpotential $\CW$ plays an important role in the study of holographic black holes at large $N$ in 3d $\CN=2$ quiver theories with an holographic dual \cite{Benini:2015eyy, Hosseini:2016tor, Hosseini:2016ume}.}  $\CN=2$ theories on $\Sigma_g$ have also been studied recently in \cite{Gukov:2015sna, Gukov:2016gkn}.

\paragraph{The 3d Witten index.} In the special case $g=1$ and $\n_i=0$, the index \eqref{general index} specializes to the Witten index on the torus:
\be\label{intro W index}
I_{g=1}\left(y_i\,;\, 0\right) = \Tr_{T^2} \,(-1)^F~.
\ee
Note that no twisting is necessary in this case.
While the standard Witten index is generally not defined for the theories of interest, which have interesting vacuum moduli spaces in flat space, it turns out to be well-defined in the presence of general real masses~$m_i$~\cite{Intriligator:2013lca}, which enter the index  through the complexified fugacities $y_i$ with $|y_i| = e^{-2\pi \beta \,m_i}$. For any generic-enough choice of $m_i$ so that all the vacua are isolated, the index counts the total number of  massive and topological vacua, which does not change as we cross codimension-one walls in parameter space. 
We will compute the Witten index of a large class of abelian and non-abelian  theories, generalizing previous results \cite{Witten:1999ds, Ohta:1999iv, Intriligator:2013lca}. Note that the localization computation is an ultraviolet computation, complementary to the infrared  analysis of \cite{Intriligator:2013lca}. Whenever it is well-defined,  the Witten index of an $\CN=2$ YM-CS-matter theory  is the number of gauge-invariant solutions to the Bethe equations  \cite{Nekrasov:2009uh, Nekrasov:2014xaa}, as we can see from \eqref{Z formula BE intro}. The Witten index can also be computed from \eqref{Z formula intro} truncated to $\m=0$, because the terms with $\m\neq 0$ do not contribute to \eqref{intro W index}.

\paragraph{Dualities and Wilson loop algebras.}
The twisted index on $\Sigma_g \times S^1$ is a powerful tool to study infrared dualities, since the twisted indices of dual theories must agree.%
~\footnote{Up to a possible sign ambiguity that we will discuss below. 
} 
One of the most interesting such dualities is the Aharony duality between a $U(N_c)$ Yang-Mills theory with $N_f$ flavor and a dual $U(N_f-N_c)$ gauge theory \cite{Aharony:1997gp}. More generally, we will consider a general three-dimensional $U(N_c)_k$ YM-CS-matter theories with $N_f$ fundamental and $N_a$ antifundamental chiral multiplets, which we can call SQCD$[k, N_c, N_f, N_a]$. This three-dimensional $\CN=2$ SQCD enjoys an intricate pattern of Seiberg dualities \cite{Seiberg:1994pq} depending on $k$ and $k_c= \half(N_f-N_a)$ \cite{Giveon:2008zn, Cremonesi:2010ae, Benini:2011mf}, which can be precisely recovered by manipulating the twisted index. This provides a new powerful check of all of these dualities. 

We will also study half-BPS Wilson loop operators wrapped on the $S^1$ for any $\CN=2$ YM-CS-matter theory. The quantum algebra of Wilson loops is encoded in the twisted superpotential $\CW$ and corresponds to the $S^1$ uplift of the two-dimensional twisted chiral ring \cite{Witten:1993xi, Kapustin:2013hpk}. In particular, we will give an explicit description of the quantum algebra of supersymmetric Wilson loops in SQCD$[k, N_c, N_f, N_a]$, generalizing the results of \cite{Kapustin:2013hpk}.

\paragraph{$\CN=4$ mirror symmetry} Another useful application for the twisted index  is to three-dimensional $\CN=4$ gauge theories and mirror symmetry. 
We consider the twisted index with a topological twist by either factor of the $SU(2)_H\times SU(2)_C$ $R$-symmetry \cite{Rozansky:1996bq}. More precisely, we shall consider the $\CN=2$ subalgebra with either $U(1)_R= 2 U(1)_H$ or $U(1)_R= 2 U(1)_C$. The corresponding $\CN=2$ twists along $\Sigma_g$ are called the $A$- or $B$-twist, respectively. Let $H$ and $C$ denote the generators of $U(1)_H \subset SU(2)_H$ and $U(1)_C\subset SU(2)_C$, respectively. We define the $A$- and $B$-twist (integer-valued) $R$-charges:
\be\label{RA RB def intro}
R_A= 2 H~,\qquad \qquad R_B= 2 C~.
\ee
Either twist on $\Sigma_g$ preserves two supercharges commuting with $H-C$. We can introduce a
 fugacity $t$ for $U(1)_t\equiv  2 \left[U(1)_H -  U(1)_C\right]$, and consider the twisted index:
\be\label{Neq4 twisted index}
I_{g, \, A/B}\left(y_i,\, t\right) =  \Tr_{\Sigma_g}\left( (-1)^F \, t^{2(H-C)} \prod_i y_i^{Q_i}\right)~.
\ee
for either choice \eqref{RA RB def intro} of $U(1)_R$. Here all the background fluxes $\n_i$, $\n_t$  are left implicit. The fugacity $t\neq 1$ breaks $\CN=4$ supersymmetry to $\CN=2^\ast$, and is  necessary in order to apply the localization formula. 

Three-dimensional  $\CN=4$ mirror symmetry \cite{Intriligator:1996ex} is an infrared duality of 3d $\CN~=4$ theories, composed with an exchange of $SU(2)_H$ and $SU(2)_C$. The latter operation maps any supermultiplet of $\CN=4$ supersymmetry to the corresponding `twisted' supermultiplet. Consequently, the $A$-twisted index of a theory $T$ must equal the $B$-twisted index of its mirror $\check T$ according to:
\be\label{mirror index intro}
I^{[T]}_{g, \, A}(y,t) =I^{[\check T]}_{g, \, B}(\check y,\, t^{-1}) 
\ee
where $y$ and $\check y$ are the flavor fugacities and their mirror---for instance, real masses are exchanged with Fayet-Iliopoulos (FI) parameters. Similarly, we can study the mapping of half-BPS line operators wrapped on the $S^1$ under mirror symmetry. We will verify in a simple but non-trivial example that half-BPS Wilson loops in the $B$-twisted theory are mirror to half-BPS  vortex loops in the $A$-twisted theory, as  recently studied in \cite{Assel:2015oxa}.

Finally, we will  argue that the genus-zero $A$- and $B$-twisted indices---the $A$- and $B$-twisted $S^2\times S^1$ partition functions \cite{Benini:2015noa}--- with vanishing background fluxes are equal to the Coulomb and Higgs branch Hilbert series, respectively~\footnote{This was also observed by \cite{Noppadol:2016}.} \cite{Benvenuti:2006qr,Benvenuti:2010pq,Hanany:2011db,Cremonesi:2013lqa,Cremonesi:2014uva,Hanany:2016ezz}. 
It is relatively easy to show, for a large class of theories, that the $B$-twisted $S^2 \times S^1$ partition function only receives contribution from the $\m=0$ flux sector in \eqref{Z formula intro} and is indeed equal to the Higgs branch Hilbert series. 
Similarly, we conjecture that the $A$-twisted $S^2 \times S^1$ partition function, which generally receives contribution from an infinite number of flux sectors, is equal to the Coulomb branch Hilbert series  \cite{Cremonesi:2013lqa}. (Naturally, this would follow from mirror symmetry \eqref{mirror index intro} when a mirror theory exists.) We will show in some examples that the $A$-twisted index reproduces the Coulomb branch monopole formula of \cite{Cremonesi:2013lqa}. It would be very interesting to study this correspondence further.

\vskip0.5cm
\noindent {\it Note added:}  During the final stage of writing,  we became aware of another closely related work by F.~Benini and A.~Zaffaroni \cite{Benini:2016hjo}. We are grateful to them for giving us a few more days to finish writing our paper, and for coordinating the arXiv submission.

\vskip0.5cm
\noindent This paper is organized as follows. In section~\ref{sec: susy n all}, we study 3d $\CN=2$ theories on $\Sigma_g\times S^1$ preserving two supercharges and we present the $\CN=2$ localization formula \eqref{Z formula intro}
and explain some of its key properties. We also discuss the quantum algebra of Wilson loops. Much of the details of the derivation of \eqref{Z formula intro} are relegated to Appendix~\ref{app: derivation formula}. In sections~\ref{sec: expl1} and~\ref{sec: expl2} we consider  the twisted index of some of the simplest $U(1)$ and $U(N)$ theories, respectively. We also briefly discuss  how \eqref{Z formula BE intro} reproduces the $SU(N)$ Verlinde formula. In section~\ref{sec: sqcd and seiberg duals}
 we discuss 3d $\CN=2$ SQCD in great details, including an explicit description of its Wilson loop algebra. In section~\ref{sec: 3d Neq4}, we study $\CN=4$ theories and the index \eqref{Neq4 twisted index}. We also consider mirror symmetry for line operators, and the relation between the genus-zero twisted index and Hilbert series. Various appendices summarize our conventions and contain useful complementary material.

%
%
%
%

\section{Three-dimensional $\CN=2$ gauge theories on $\Sigma_g\times S^1$}\label{sec: susy n all}
In this section, we summarize some useful results about supersymmetric field theories on $\Sigma_g\times S^1$, and we present the explicit formula for the twisted index and for correlation functions of supersymmetric Wilson loops  wrapped on $S^1$ in the case of $\CN=2$ Yang-Mills-Chern-Simons-matter theories.

\subsection{Supersymmetry with the topological twist}
Consider any  three-dimensional $\CN=2$ supersymmetric gauge theories with an $R$-symmetry $U(1)_R$ on $\Sigma_g\times S^1$, with $\Sigma_g$ a closed orientable Riemann surface of genus $g$. 
Let us take the product metric:
\be\label{3d metric}
ds^2 = \beta^2 d t^2 + 2 g_{z\bz}(z, \bz) dz d\bz = (e^0)^2 + e^1 e^{\b1}~.
\ee
with $t\sim t+ 2\pi$ the circle coordinate, and $z, \bz$ the local complex coordinates on $\Sigma_g$ with Hermitian metric $g_{z\bz}$. We also choose a canonical frame $(e^0, e^1, e^{\b1})$.  (See Appendix \ref{Appendix: conv and susy} for our conventions.)
One can preserve two supercharges on $\CM_3=\Sigma_g\times S^1$, corresponding to the  uplift of the topological $A$-twist on $\Sigma_g$. In the formalism of  \cite{Closset:2012ru}, this corresponds to choosing a transversely holomomorphic foliation (THF) of  $\CM_3$ along the circle:
\be\label{def K and eta}
K = \eta^\mu \d_\mu  =  {1\over \beta} \d_t~.
\ee
The full supergravity background is given by: 
\be\label{3d sugra background}
H=0~, \qquad \qquad V_\mu=0~,\qquad\qquad \epsilon^{\mu\nu\rho} \d_\nu A_\rho^{(R)}  = -{1\over 4 } R\, \eta^\mu~.
\ee
The last equation  in \eqref{3d sugra background}  determines the $R$-symmetry gauge field $A_\mu^{(R)}$ up to flat connections, which must vanish to preserve supersymmetry. In other words,  $A_\mu^{(R)}$ is taken to vanish along $S^1$ and is equal to $\half \omega_\mu^{\rm (2d)}$ along $\Sigma_g$, with $\omega_\mu^{\rm (2d)}$ the two-dimensional spin connection. Due to the $A_\mu^{(R)} $ flux:
\be
{1\ov 2 \pi}\int_{\Sigma_g} d A^{(R)} = (g-1)~, 
\ee
the $R$-charges are quantized in units of ${1\ov g-1}$.
This background preserves two covariantly-constant Killing spinors $\zeta$ and $\t\zeta$ of $R$-charge $\pm 1$, respectively:
\be
 (\nabla_\mu - i A_\mu^{(R)}) \zeta=0~, \qquad \qquad (\nabla_\mu + i A_\mu^{(R)}) \t\zeta=0~.
\ee
In the canonical frame, the Killing spinors are given by:
\be\label{KS explicit}
\zeta=\mat{0\cr 1}~, \qquad\qquad \t\zeta=\mat{1\cr 0}~, 
\ee
The real Killing vector $K= \t\zeta\gamma^\mu\zeta \d_\mu$ constructed out of \eqref{KS explicit} is equal to \eqref{def K and eta}.

\subsubsection{Supersymmetry algebra and supersymmetry transformations}
Let us denote by $\delta$ and $\t\delta$ the action of the two supercharges on fields. We have the supersymmetry algebra:
\be\label{susy algebra 3d}
\delta^2= 0~, \qquad \t\delta^2=0~,\qquad \{\delta, \t\delta\} = -2i \left(Z+ \CL_K\right)~,
\ee
with $Z$ the real central charge of the $\CN=2$ superalgebra in flat space, and $\CL_K$ the Lie derivative along $K$. For a vector multiplet $\CV$ in Wess-Zumino (WZ) gauge, the real scalar component $\sigma$ also enters \eqref{susy algebra 3d} as  $Z= Z_0- \sigma$, where $Z_0$ is the actual central charge and  $\sigma$ is valued in the appropriate gauge representation.
All supersymmetry transformations and supersymmetric Lagrangians are easily obtained by specializing the results of \cite{Closset:2012ru}. We will use a convenient ``$A$-twisted'' notations for all the fields \cite{Closset:2015rna}.

Let $\GG$ and $\Fg={\rm Lie}(\GG)$ denote a compact Lie group and its Lie algebra, respectively.
In  WZ gauge, a $\Fg$-valued vector multiplet $\CV$ has components:
\be\label{V components}
\CV= \left(a_\mu~,\, \sigma~, \,\Lambda_\mu~, \, \t\Lambda_\mu~,\,  D\right)~.
\ee
The $A$-twisted fermions $\Lambda_\mu$ are holomorphic and anti-holomorphic one-forms with respect to the THF \eqref{def K and eta},~\footnote{See Appendix \ref{Appendix: conv and susy} and especially \cite{Closset:2013vra} for a general discussion.}
 which means that:
\bea\label{gauginiA} 
&\Lambda_\mu dx^\mu \;=\; \Lambda_t dt + \Lambda_z dz \;=\; \Lambda_0 e^0 + \Lambda_1 e^1~, \cr
&\t\Lambda_\mu dx^\mu \;=\; \t \Lambda_t dt + \t \Lambda_\bz d\bz \;=\; \t \Lambda_0 e^0 + \t \Lambda_{\b 1} e^{\b 1}~.
\eea
We mostly use the frame $e^0, e^1, e^{\b1}$ in the following.
Let us define the field strength
\be
f_{\mu\nu} = \d_\mu a_\nu - \d_\nu a_\mu -i [a_\mu, a_\nu]~.
\ee
We denote by $D_\mu$ the covariant and gauge-covariant derivative. 
The supersymmetry transformations of \eqref{V components} are
\bea\label{susyVector twisted}
&\delta a_\mu = i \t\Lambda_\mu~,   
\qquad &&  \t\delta a_\mu = - i \Lambda_\mu\cr 
&\delta\sigma =\t\Lambda_0~,
\qquad &&  \t\delta \sigma = -\Lambda_0~,\cr  
& \delta \Lambda_0 =i \left(D - 2 i f_{1\b 1}\right)+ i D_0 \sigma~,
\qquad  && \t\delta \Lambda_0 =0~,\cr
& \delta \Lambda_1 =2 f_{01} + 2 i D_1 \sigma~,
\qquad &&   \t\delta \Lambda_1 =0~,  \cr
& \delta \t\Lambda_0 =0~,
\qquad  && \t\delta \t\Lambda_0 =i \left(D - 2 i f_{1\b 1}\right)- i D_0 \sigma~,\cr
&\delta \t\Lambda_{\b1} =0~,
\qquad &&\t\delta \t\Lambda_{\b1}= -2 f_{0\b 1} - 2 i D_{\b1}\sigma~, \cr
&\delta D= -D_0\t\Lambda_0 - 2 D_1 \t\Lambda_{\b 1} + [\sigma, \t\Lambda_0] ~,\quad
\qquad  && \t\delta D= -D_0\Lambda_0- D_{\b 1} \Lambda_1  +[\sigma, \Lambda_0]~.
\eea
The explicit form of the super-Yang-Mills Lagrangian  $\SL_{YM}$ can be inferred from \cite{Closset:2012ru} and will not be needed in the following. The important fact for our purposes is that  the YM action is $Q$-exact,
\be
\SL_{YM}= \delta(\cdots)~.
\ee
like all $D$-terms.
The Chern-Simons (CS) term is given by:
\be\label{classical CS Lag}
\SL_{CS} ={k \over 4 \pi} \left(i \epsilon^{\mu\nu\rho} \left(a_\mu \d_\nu a_\rho- {2i\ov 3}a_\mu a_\nu a_\rho\right) - 2 D \sigma + 2 i \t \Lambda_0 \Lambda_0 + 2 i  \t\Lambda_{\b1}\Lambda_1 \right)~,
\ee
for any gauge group $\GG$.~\footnote{In general, we have a distinct CS level for each simple factor and for each $U(1)$ factor in $\GG$.}
 In the presence of an abelian sector, we can  also have mixed CS terms between $U(1)_I$ and $U(1)_J$, with $I\neq J$:
\be
\SL_{CS} = {k_{IJ} \over 2 \pi} \left(i \epsilon^{\mu\nu\rho} a^{(I)}_\mu \d_\nu a_\rho^{(J)} -  D^{(I)} \sigma^{(J)}- D^{(J)}\sigma^{(I)} +  i \t\lambda^{(I)}\lambda^{(J)}+ i \t\lambda^{(J)}\lambda^{(I)} \right)~,
\ee
with $\t\lambda^{(I)}\lambda^{(J)}=   \t \Lambda_0^{(I)} \Lambda_0^{(J)} +   \t\Lambda_{\b1}^{(I)}\Lambda_1^{(J)}$.  

Matter fields enter as chiral multiplets coupled to the vector multiplet $\CV$.
Consider a chiral multiplet $\Phi$ of $R$-charge $r$, transforming in a representation $\FR$ of $\Fg$. In $A$-twisted notation \cite{Closset:2015rna}, we denote the components of $\Phi$ by
\be\label{compo Phi}
\Phi= \left(\CA~,\, \CB~,\, \CC~,\, \CF\right)~.
\ee
The supersymmetry transformations are:
\bea\label{susytranfoPhitwistBis}
& \delta \CA = \CB~, \qquad\qquad  && \t\delta \CA=0~,\cr
& \delta \CB=0~, \qquad \qquad && \t\delta \CB= -2i\big(- \sigma +D_0\big)\CA~,\cr
& \delta \CC=\CF~, \qquad \qquad && \t\delta \CC= 2i D_{\b1}\CA~,\cr
& \delta \CF=0~, \qquad \qquad && \t\delta \CF=- 2i \big(-\sigma +D_0\big)\CC -2i D_{\b1}\CB -2i \t\Lambda_{\b1}\CA~,
\eea
where $D_\mu$ is appropriately gauge-covariant and $\sigma$ and $\t\Lambda_{\bz}$ act in the representation $\FR$.
Similarly, the charge-conjugate antichiral multiplet $\t\Phi$ of $R$-charge $-r$ in the representation  $\b\FR$ has components
\be\label{compo tPhi}
\t\Phi= \left(\t\CA~,\, \t\CB~,\, \t\CC~,\, \t\CF\right)~,
\ee
with
\bea\label{susytranfotPhitwistBis}
& \delta \t\CA = 0~, \qquad\qquad\qquad\qquad\qquad\qquad\qquad
& \t\delta \t\CA=\t\CB~,\cr
& \delta \t\CB= -2i \big(\sigma +D_0 \big)\t\CA~, \qquad\qquad\qquad \qquad\quad\quad &\t\delta \t\CB=0~,\cr
& \delta \t\CC= -2i D_1 \t\CA~,
\qquad\qquad\qquad \qquad\quad 
&\t\delta \t\CC=\t\CF~,\cr
& \delta \t\CF= -2i \big(\sigma +D_0 \big)\t\CC +2i D_1  \t\CB + 2i \Lambda_1\t\CA~,
& \t\delta \t\CF=0~.
\eea
Using the  vector multiplet transformation rules \eqref{susyVector twisted}, one can check that \eqref{susytranfoPhitwistBis}-\eqref{susytranfotPhitwistBis} realize the supersymmetry algebra 
\be\label{susy with gauge field i}
\delta^2=0~, \qquad \t\delta^2=0~, \qquad 
\{\delta,\t\delta\}=- 2i  \left(- \sigma+ \CL_K^{(a)} \right)~,
\ee
where $\CL_K^{(a)}$ is the gauge-covariant Lie derivative, and $\sigma$ acts in the appropriate representation of the gauge group. 
The standard kinetic term for the chiral multiplet reads:
\bea\label{kin chiral}
&\SL_{\t\Phi\Phi} &=&\; \t\CA \left(-D_0 D_0 - 4 D_1 D_{\b1} + \sigma^2 + D - 2 if_{1\b1}  \right)\CA - \t\CF \CF \cr
&&&  -{i\ov 2} \t\CB (\sigma+ D_0) \CB + 2 i \t\CC (\sigma-D_0)\CC + 2 i \t \CB D_1 \CC - 2 i \t\CC D_{\b1} \CB\cr
&&& - i \t\CB \t\Lambda_0 \CA + i \t\CA \Lambda_0 \CB - 2 i \t\CA \Lambda_1 \CC + 2 i \t\CC \t\Lambda_{\b1} \CA~.
\eea
The trace over gauge indices is implicit.
This Lagrangian is $\delta$-exact:
\be
\SL_{\t\Phi\Phi}  = \delta\t\delta \left({i\ov 2} \t\CA (\sigma+ D_0)\CA - \t\CC \CC\right)~.
\ee

\subsection{YM-CS-matter theories, twisted superpotential and localization}
Consider a generic $\CN=2$ YM-CS theory coupled to matter fields in chiral multiplets. The theory contains a vector multiplet $\CV$ for the gauge group $\GG$ with Lie algebra $\Fg$, and some matter multiplets in chiral multiplets $\Phi_i$  transforming in  representations $\FR_i$ of $\Fg$ and with $R$-charges $r_i$. We can also have a superpotential $W(\Phi)$ of $R$-charge $2$.

The UV description of  the theory includes Yang-Mills terms with dimensionful gauge couplings, as well as arbitrary Chern-Simons terms. For definiteness, consider a gauge group
\be\label{gauge group decomp}
\GG \cong  \prod_\gamma \GG_\gamma \times \prod_I U(1)_I
\ee
possibly up to discrete identifications, where $\GG_\gamma$ are simple Lie groups. For each $\GG_\gamma$, we have a Chern-Simons level $k_{\gamma}$, while we can have arbitrary mixed CS levels $k^{IJ}= k^{JI}$ in the abelian sector. 
In addition to these CS interactions for the gauge fields, we must also specify ``global'' CS levels for all the global symmetries of the theory, including the $R$-symmetry \cite{Closset:2012vp}. This might  include mixed CS terms between the abelian gauge and global symmetries. 
All the CS levels are either integer or half-integer, depending on parity anomalies.

For future reference, let us  introduce the Cartan subgroup
 $\prod_{a=1}^\rk U(1)$, and the corresponding symmetric matrix of CS levels $k ^{ab}$, which is given by
 \be
 k^{ab} \big|_\gamma =k_\gamma h^{ab} \big|_\gamma~, \qquad a, b \in \gamma~, 
 \ee
on each semi-simple factor, with $h^{ab}|_\gamma$ the Killing form of $\Fg_\gamma$, and by $k^{ab}=k^{IJ}$ ($a=I, b=J$) in the abelian sector. (Moreover, $k^{ab}=0$ for $a\in \gamma$ and $b=I$.)

It is natural to  couple the theory to an arbitrary supersymmetric background vector multiplet for any global symmetry $U(1)_F$. This includes a background flux $\n_F$ over $\Sigma_g$ as well as the real mass $\sigma_F=m_F$ paired together with a background $U(1)_F$ flat connections along $S^1$ into a complex parameter  $\nu_F$.  In particular, for any $U(1)_I$ gauge group there exists a  topological symmetry $U(1)_{T_I}$. The corresponding background real mass corresponds to a Fayet-Iliopoulos (FI) parameter  for the abelian gauge group $U(1)_I$, provided we turn on a unit mixed CS level between $U(1)_I$ and $U(1)_{T_I}$:
\be\label{FI as real mass}
\sigma_{T, I} =\xi_I~, \qquad \quad k_{I T_{I}} =1~,
\ee
using a convenient normalization for the FI parameters:
\be
\SL_{\rm FI} = -{\xi_I\over 2\pi} \tr_I(D)~.
\ee

\subsubsection{Classical Coulomb branch}\label{sec: coulomb branch}
The `classical Coulomb branch' of any YM-CS-matter theory on $\R^3$ is spanned by the constant expectation values of the real field $\sigma$, such that:
\be\label{sigma CB}
\sigma= {\rm diag}(\sigma_a)~, \qquad \qquad a= 1, \cdots, \rk~,
\ee
 and of the dual photons $\varphi_a$ of the effective $\prod_a U(1)_a$ abelian theory, modulo the Weyl group $W_\GG$. The fields $\sigma_a$ and $\varphi_a$ are paired into  chiral `bare' monopole operators, which take the form:
 \be\label{monopole operators}
 T_a^\pm= e^{\pm\phi_a}~, \qquad\qquad  \phi_a =-{2 \pi \ov e^2}\sigma_a +i\varphi_a~,
 \ee
 semi-classically, with $e^2$ the Yang-Mills coupling and $T^+_a T^-_a=1$, $\forall a$. Here $\phi_a$ is the lowest-component of a chiral multiplet $\Phi_a$ related to the field-strength linear multiplet by $\Sigma_a = -{e^2\ov 4 \pi}(\Phi_a + \t\Phi_a)$. In particular, the dual photon is defined by:
 \be\label{dual photon}
-{e^2\ov 2 \pi} \d_\mu \varphi = {i\over 2} {\epsilon_{\mu}}^{\nu\rho} f_{\nu\rho}+ i\eta_\mu D~.
 \ee
 Since the dual photons are periodic, the classical Coulomb branch has the topology of $(\C^*)^\rk/ W_\GG$,  a cylinder quotiented by the Weyl group.

Consider instead the same theory compactified on a circle $S^1$ of radius $\beta$. In this case, one can turn on flat connections $a_0$ for the gauge field along $S^1$, and the Coulomb branch coordinates \eqref{sigma CB} have a natural complexification:
\be\label{def ua}
u_a = i \beta (\sigma_a + i a_{0, a})~.
\ee
Due to the periodicity $a_{0, a}\sim  a_{0, a}+\beta^{-1}$ under large $U(1)_a$ gauge transformations around $S^1$, it is natural to define the complexified fugacities:
\be
x_a = e^{2\pi i u_a}~.
\ee
Similarly, for any global symmetry $U(1)_F$ we can turn on some background flat connections and  background real field $\sigma_F$, and we denote the corresponding fugacity by $y_F= e^{2\pi i \nu_F}$.
Under the supersymmetry \eqref{susyVector twisted}, we have:
\bea
&\delta u =0~,   
\qquad &&  \t\delta u=0\cr 
&\delta\b u =2 i \beta \t\Lambda_0~,
\qquad &&  \t\delta \b u=-2 i \beta \Lambda_0~.
\eea
Note that $u$  transforms as the lowest component of a twisted chiral multiplet of two-dimensional $\CN=(2,2)$ supersymmetry on $\Sigma_g$ with the $A$-twist. Let us denote by
\be
\t\fM\cong \{(u_a)\} \cong (\C^*)^\rk
\ee
the covering space of the complexified classical Coulomb branch $\fM \cong \t\fM/W_\GG$, spanned by the $u_a$'s. This classical moduli space has the same topology as the one spanned by  the chiral monopole operators \eqref{monopole operators}. This is no coincidence, as  the two descriptions are essentially related by a T-duality transformation \cite{Aganagic:2001uw, Intriligator:2013lca}, mapping chiral multiplets (of lowest component $\phi_a$) to twisted chiral multiplets (of lowest component $u_a$) in the  two-dimensional description.

The `holomorphic' properties of the low-energy theory on $\fM$ are determined by the effective twisted superpotential, which can be obtained by integrating out all the massive fields at generic values of $u_a$, including all the Kaluza-Klein (KK) modes on $S^1$ \cite{Nekrasov:2009uh, Dimofte:2011jd}. One finds:
\bea\label{CW general}
&\CW&=&\;\;\half k^{ab} u_a u_b+  k^{a F}_\text{g-f} u_a \nu_F\cr
&&&+\sum_i\sum_{\rho_i\in \FR_i} \left[ {1\ov (2 \pi i)^2}  {\rm Li}_2(x^{\rho_i} y_i)+ {1\ov 4} \left(\rho_i(u)+ \nu_i\right)^2 \right]+\sum_{\alpha>0} \half \alpha(u)~.
\eea
Here the last sum is over the positive roots of $\Fg$,
and we introduced fugacities for the flavor symmetries, with $\nu_i= \nu_F[\Phi_i]$ and $y_i= e^{2\pi i \nu_i}$. The mixed gauge-flavor CS levels are denoted schematically by $k^{a F}_\text{g-f}$, which includes  the FI terms according to \eqref{FI as real mass}.  We also introduced the convenient notation $x^{\rho_i} = \prod_a x_a^{\rho_i^a} = e^{2 \pi i \rho_i(u)}$.
The physically meaningful quantities are the first derivatives:
\bea\label{duW explicit}
&\d_{u_a} \CW &=&\; \; k^{ab} u_b+ k^{a F}_{\text{g-f}}  \nu_F \cr
&&& - {1\ov 2 \pi i}\sum_i\sum_{\rho_i} \rho_i^a\left[ \log(1- x^{\rho_i} y_i) - \pi i \left(\rho_i(u)+ \nu_i\right)\right]+\half \sum_{\alpha >0} \alpha^a~.
\eea
Note that this is invariant under large gauge transformations $u_a\sim u_a+1$ (and $\nu_F\sim \nu_F+1$ for background gauge fields) if and only if the CS levels are properly quantized (that is, integer or half-integer depending on the parity anomalies).
We shall also need the Hessian matrix of $\CW$:
\be
\d_{u_a} \d_{u_b} \CW= k^{ab}+\sum_i\sum_{\rho_i} \rho_i^a \rho_i^b \; \half \left({1+ x^{\rho_i}y_i\ov 1- x^{\rho_i}y_i}\right)~,
\ee
whose determinant we denote by:
\be\label{def Hu}
H(u) \equiv \det_{ab}\d_{u_a} \d_{u_b} \CW~.
\ee
Much of physics of the supersymmetric indices, and of correlation functions of supersymmetric Wilson loops, is encoded in this twisted superpotential.

\subsubsection{Localization, fugacities and classical actions}
The path integral of any $\CN=2$ YM-CS-matter theory on $\Sigma_g\times S^1$ can be localized onto the simplest supersymmetric configurations for the vector multiplet.
Since the YMs action is Q-exact, we can take the $e\rightarrow 0$ limit so that the vector multiplet localizes to \cite{Kapustin:2009kz, Benini:2012ui, Benini:2015noa}:
\be\label{susy eq vector}
\sigma = {\rm constant}~, \qquad \qquad D= 2 if_{1\b1}~, \qquad\qquad f_{01}=f_{0\b1}=0~. 
\ee 
We can diagonalize the background field $\sigma$ as in \eqref{sigma CB},  which Higgses the gauge group to the Cartan subgroup $\GH \cong \prod_a U(1)_a$ at generic values of $\sigma_a$. 
As discussed in \cite{Blau:1994rk, Blau:1995rs}, there is an obstruction to diagonalizing the vector multiplet globally on $\Sigma_g\times S^1$ due to the presence of non-trivial principal $\GH$-bundles (even for a trivial $\GG$ bundle, for instance if $\GG$ is simple), and we must therefore sum over all such non-trivial $\GH$-bundles. For $\GG$ Abelian, we just have a standard sum over topological sectors.
As a result,  the localization locus is divided into topological sectors indexed by GNO-quantized fluxes over $\Sigma_g$:
\be
\m ={1\ov 2 \pi} \int_{\Sigma_g} da= {1\ov 2 \pi} \int_{\Sigma_g} d^2 x\sqrt{g}(-2 i f_{1\b1})\; \in \Gamma_{\mathbf{G}^\vee}~.
\ee
The fluxes take value in the magnetic lattice  $\Gamma_{\mathbf{G}^\vee} \cong \bZ^\rk$, which  can be obtained from $\Gamma_\mathbf{G}$, the weight lattice of electric charges of $\mathbf{G}$ within  $i\mathfrak{h}^*$ by \cite{Englert:1976ng,Kapustin:2005py}
\be
\Gamma_{\mathbf{G}^\vee} = \setcond{k}{\rho(k) \in \bZ ~~ \forall \rho \in \Gamma_\mathbf{G}}~.
\ee
We denote by $(\m_a)$ the projection of $\m$ onto the magnetic flux lattice $\Z^\rk$ of the Cartan subgroup $\prod_a U(1)_a$.

Note that  \eqref{susy eq vector} implies that the dual photon appearing  in \eqref{dual photon} is constant. In other words, we are localizing onto the classical Coulomb branch using the `T-dual' variables \eqref{def ua}, in every topological sector.
The $U(1)_a$ flat connections along $S^1$,
\be
a_{0,a} = {1\over 2 \pi \beta} \int_{S^1} a_\mu dx^\mu~,
\ee
 are included into the complex variables \eqref{def ua}. One must also sum over arbitrary flat connections on $\Sigma_g$, but  we will see that they have little impact on the final answer. (Similarly, the final answer cannot depend on flat connections along  $\Sigma_g$ for background vector multiplets \cite{Closset:2013vra}, therefore we set these to zero from the start.)

 For future reference,  it is interesting to evaluate the classical action onto the supersymmetric locus \eqref{susy eq vector}. We  will also turn on general background fluxes, real masses and Wilson lines for flavor symmetries. For any $U(1)_F$ global symmetry (which might be part of the Cartan of a non-abelian group) with background vector multiplet $\CV_F$, we have:
 \be
 \n_F = {1\ov 2 \pi} \int_{\Sigma_g} da_F~, \qquad\qquad \nu_F= i \beta (\sigma_F + i a_{0, F})~, \qquad  y_F= e^{2\pi i \nu_F}~.
 \ee
 The only terms in the action that contributes on the supersymmetric locus are the Chern-Simons levels for gauge and global symmetries (including the FI terms), which  are not $Q$-exact. In a given topological sector,
 \be\label{full Z classical}
 Z^\text{classical}_\m(u)= \exp{\left(-S_{\rm CS}^{\rm gauge} - S_{\rm CS}^{\text{gauge-flavor} }- S_{\rm CS}^{\text{flavor} }- S_{\rm CS}^{\text{gauge-R} } -S_{\rm CS}^{\text{flavor-R} } \right) }~.
 \ee
 The gauge CS terms reads:
 \be\label{CS term clas}
 e^{-S_{\rm CS}^{\rm gauge} } = \prod_{a,b} (x_a)^{k^{ab} m_b}~.
 \ee
 Similarly, for the mixed flavor-gauged and flavor CS terms:
\be
e^{ -S_{\rm CS}^{\text{gauge-flavor} }}= \prod_{a, m}\left( y_m^{\m_a} x_a^{\n_m}\right)^{k_\text{g-f}^{a m}}~, \qquad\qquad
 e^{- S_{\rm CS}^{\text{flavor} }} =  \prod_{m,n} (y_m)^{k_{\text{f-f}}^{mn} \n_n}
\ee
where the indices $m, n$ run over the flavor group, including the topological symmetries. For each topological symmetry $U(1)_{T_I}$, we introduce the fluxes $\n_{T_I}$ and the fugacities:
\be
q_I = e^{2 \pi i \tau_I}~, \qquad \tau_I = {\theta_I\ov 2 \pi} + i \beta \xi_I~,
\ee
where $\theta_I$, the $U(1)_{T_I}$ Wilson line, is also a two-dimensional $\theta$-angle.
The last  two terms in \eqref{full Z classical} are mixed CS terms between abelian vector multiplets and  the $R$-symmetry gauge field in the new-minimal supergravity multiplet \cite{Closset:2012vp, Closset:2012ru}, which is given by
\be
\SL_\text{CS}^\text{R}= {k_R\over 2\pi } \left( i \epsilon^{\mu\nu\rho} a_\mu \d_\nu A_\rho^{(R)} - {1\ov 4} \sigma R\right)
\ee 
on the background \eqref{3d sugra background}.
This gives:
\be
e^{- S_{\rm CS}^{\text{gauge-R} }} = \prod_{I} x_I^{(g-1) k^I_R}~,\qquad \qquad
e^{- S_{\rm CS}^{\text{flavor-R} }} = \prod_{M} y_M^{(g-1) k^M_R}~,
\ee
with $g$ the genus of $\Sigma_g$, where $I$ runs over the abelian part of $\GG$ \eqref{gauge group decomp} and $M$ runs over the abelian part of the flavor group. Finally, we note that the purely gravitational CS terms of  \cite{Closset:2012vp} evaluate to zero on our  $\Sigma_g\times S^1$ background. This implies that the overall  phase of the twisted index is unambiguous (except possibly for a sign ambiguity to be discussed below), unlike for instance the phase of the $S^3$ partition function \cite{Closset:2012vg}.


\subsection{Induced charges of the monopole operators}
Consider the `bare' monopole operators $T_a^{\pm}$ in the abelianized  $\prod_a U(1)_a$ theory. Each operator $T_a^{\pm}$ carries charges under any abelian (gauge or global) symmetry  which can mix with the gauge symmetry $U(1)_a$, either classically through Chern-Simons interactions, or at one-loop in the presence of matter fields \cite{Aharony:1997bx, deBoer:1997kr, Borokhov:2002ib, Borokhov:2002cg}. 
These charges are:
\bea\label{induced charge T}
&Q^{b}[T_a^\pm] \equiv {Q_{a\pm}}^b &=&\; \pm k^{ab} - \half \sum_{i}\sum_{\rho_i\in \FR_i} |\rho_i^a| \,\rho_i^b~, \cr
&Q^{F}[T_a^\pm] \equiv {Q_{a\pm}}^F &=&\; \pm k^{aF}_{\text{g-f}} - \half \sum_{i}\sum_{\rho_i\in \FR_i}  |\rho_i^a| \, Q_i^F~,
\eea
under the gauge and flavor symmetries, where $Q_i^F$ is the charge of the chiral multiplet $\Phi_i$ under a flavor symmetry $U(1)_F$. 
The monopole operators also acquire  an induced $R$-charge (see {\it e.g.} \cite{Gaiotto:2009tk, Jafferis:2009th, Benini:2009qs, Benini:2011cma}) given by:
\be\label{induced Rcharge}
R[T_a^{\pm}] \equiv   r_{a\pm}=   \pm k_R^a - \half \sum_{i}\sum_{\rho_i\in \FR_i} |\rho_i^a| \, (r_i-1) - \half \sum_{\alpha\in \Fg} |\alpha^a|~,
\ee
with $r_i$ the $R$-charge of $\Phi_i$. The last term in \eqref{induced Rcharge} is the contribution from the gaugini (which carry $R$-charge $1$).

 For a generic value of $\sigma_a$ on the Coulomb branch, we can also compute the effective CS levels by integrating out the massive fields:
\bea
&k^{ab}_{\rm eff}(\sigma)&=& \;k^{ab} + \half\sum_i \sum_{\rho_i\in \FR_i} \sign(\rho_i(\sigma)+ m_i) \,\rho_i^a\rho_i^b~,\cr
&k^{aF}_{\text{g-f},  \rm eff}(\sigma)&=&\; k^{aF}_\text{g-f} +\half \sum_i \sum_{\rho_i\in \FR_i} \sign(\rho_i(\sigma)+ m_i) \, \rho_i^a Q_i^F~,
\eea
with $m_i = \sigma^F[\Phi_i]$, and
\be
k^{a}_{\text{R}, \rm eff}(\sigma)=k^a_R +\half \sum_i \sum_{\rho_i\in \FR_i} \sign(\rho_i(\sigma)+ m_i)\, \rho_i^a(r_i-1) +\half \sum_{\alpha\in \Fg} \sign(\alpha(\sigma))\, \alpha^a~.
\ee
We directly see that 
\be\label{Qa QF monopole}
{Q_{a\pm}}^{b} = \pm \lim_{\sigma_a \rightarrow \mp\infty}  k^{ab}_{\rm eff}(\sigma)~, \qquad \qquad
{Q_{a\pm}}^{F} = \pm \lim_{\sigma_a \rightarrow \mp\infty} k^{aF}_{\text{g-f},  \rm eff}(\sigma)~, 
\ee
and similarly for the induced $R$-charge \eqref{induced Rcharge}.
Equivalently, the charges \eqref{Qa QF monopole} can  be extracted from the twisted superpotential:
\be\label{Q from W}
{Q_{a\pm}}^{b} = \pm \lim_{\sigma_a \rightarrow \mp\infty}  \d_{u_a}\d_{u_b} \CW~,  \qquad
{Q_{a\pm}}^{F} = \pm \lim_{\sigma_a \rightarrow \mp\infty}  \d_{\nu_F}\d_{u_a} \CW~.
\ee
 It is therefore natural to associate the asymptotics of the Coulomb branch with the monopole operators $T^\pm_a$ \cite{ Aharony:1997bx, Intriligator:2013lca}.

\subsection{The algebra of Wilson loops}\label{subsec: Wilson loop alg}
In any YM-CS-matter theory with $\CN=2$ supersymmetry on $\R^2\times S^1$, one can define half-BPS Wilson loop operators wrapped over the circle. For a Wilson loop in the representation $\FR$ of $\GG$, we have~\footnote{Note that a Wilson loop is defined in terms of a representation $\FR$ of the gauge group $\GG$ instead of the algebra $\Fg$, although we will not discuss any of the interesting subtleties associated to this fact---see for instance \cite{Kapustin:2005py, Aharony:2013hda}. }
\be\label{Wilson loop R expl}
W_{\FR} = \Tr_{\FR} {\rm Pexp}{\left( -i \int_{S^1} dx^\mu \left(a_\mu - i \eta_\mu \sigma\right)\right)}~,
\ee
which preserves half of the supersymmetry. Such operators also preserve the $A$-twist supersymmetry on $\Sigma_g \times S^1$, as one can see using \eqref{susyVector twisted}. When evaluated  on the Coulomb branch covering space $\t \fM$, the Wilson loop \eqref{Wilson loop R expl} becomes a Laurent polynomial in $x$, corresponding to the character of the representation $\FR$:
\be\label{Wilson loop trace x}
W_{\FR} = \Tr_{\FR}\left(x\right)= \sum_{\rho\in \FR} x^\rho~.
\ee
More generally, we can consider any insertion of Wilson loops wrapping $S^1$ at distinct points on $\Sigma_g$. Any such insertion corresponds to a Weyl-invariant Laurent polynomial in $x$:
\be\label{def Wx}
W(x)  \in \C[x_1, x_1^{-1},  \cdots, x_\rk, x_\rk^{-1}]^{W_\GG}~.
\ee
While the classical algebra of Wilson loops is infinite dimensional, corresponding to the algebra of representations of $\GG$, the  quantum algebra of supersymmetric Wilson loops of an ${\CN=2}$ YM-CS-matter theory is generally finite dimensional, with relations encoded in the twisted superpotential \eqref{CW general}.
The quantum algebra relations are the relations satisfied by the solutions to:
\be\label{rel CW}
\exp{\left(2 \pi i \,\d_{u_a} \CW\right)}= 1~, \quad a=1, \cdots, \rk~,
\qquad \qquad x^\alpha \neq 1~,\quad \forall \alpha\in \Fg~,
\ee
with the second condition imposing that we stay away from the Weyl chambers walls in $\t\fM$. These equations are known as the Bethe equations of the theory compactified on $S^1$ \cite{Nekrasov:2009uh}. The quantum algebra takes the form:
\be\label{CAW general}
\CA_W=   \C[x_1, x_1^{-1},  \cdots, x_\rk, x_\rk^{-1}]^{W_\GG}/I_\CW~,
\ee
with the ideal $I_\CW$ generated by the relations determined from \eqref{rel CW}. 
We will derive these relations directly by localization on $\Sigma_g\times S^1$, and we will give an explicit presentation of \eqref{CAW general} in some interesting examples.
Note that the quantum algebra generally depends on all the fugacities for the global symmetries of the theory.
 Closely related discussions have appeared previously in  \cite{Witten:1993xi, Kapustin:2013hpk}.

Note that the Verlinde algebra \cite{Verlinde:1988sn, Witten:1988hf} of Wilson loops in pure  Chern-Simons theory with gauge group $\GG$ at level $\h k$ is a special case of \eqref{CAW general}. It  can be obtained by considering an $\CN=2$ supersymmetric Chern-Simons theory  with gauge group $\GG$ and CS level $k$, with  $\h k = k- h \sign(k)$ and  $h$ the dual Coxeter number of $\GG$.~\footnote{More generally, the matrix of CS levels $k^{ab}$  shifts to $\h k^{ab} = k ^{ab}-  \sign{(k^{ab})} \half\sum_{\alpha \in \Fg} \alpha^a\alpha^b$ after integrating out the gaugini. For $\GG$ semi-simple, we have $k^{ab}= h^{ab} k$  and $\half \sum_{\alpha\in \Fg} \alpha^a \alpha^b = h^{ab} h$,  with $h^{ab}$ the Killing form.} In the absence of matter fields, the ordinary Wilson loops are equivalent to  the supersymmetric Wilson loops \eqref{Wilson loop R expl} because $\sigma=0$ on-shell.


\subsection{The localization formula on $\Sigma_g\times S^1$}\label{subsec: loc formula}
One can use supersymmetric localization to compute the $\Sigma_g\times S^1$ partition function of a generic $\CN=2$ YM-CM-matter theory. More generally, we can consider a correlation function of Wilson loops along $S^1$, collectively denoted by $W$ as in \eqref{def Wx}. The localization formula reads:
\bea\label{main formula}
\langle W \rangle_g 
= {1 \ov |W_\GG|} \sum_{\m \in \Gamma_{\mathbf{G}^\vee}}
\sum_{u_* \in \tfM^{\m}_\text{sing}}
\underset{u=u_*}{ {\text{JK-Res}}} \left[ \mathbf{Q}(u_*),  \eta  \right]
I_\m (W)~,
\eea
in terms of a  Jeffrey-Kirwan (JK) residue on the differential form:
\bea\label{density}
&I_\m (W) =(-2\pi i)^\rk \; Z^\text{classical}_\m(u)\cr
&\qquad\qquad \times \left(\prod_i Z^{\Phi_i}_{\m}(u)\right)\; Z^\text{vector}_{\m} (u) \;H(u)^g\;   W(x)\; 
 d u_1 \wedge \cdots \wedge d u_\rk~,
\eea
on  $\tfM \cong (\bC^*)^\rk$, in each topological sector $\m$. The first factor is the  classical contribution $Z^\text{classical}(u)$ given by \eqref{full Z classical}. The second factor is the product of   the one-loop determinants
\be\label{oneloop matter}
Z^{\Phi_i}_{\m}(u) =  \prod_{\rho_i \in \mathfrak{R}_i } \left({ x^{\half \rho_i} y_i^\half\ov 1- x^{\rho_i} y_i}\right)^{\rho_i(\m) + \n_i +(g-1)(r_i-1)}~,
\ee
for chiral multiplets $\Phi_i$ in the representation $\FR_i$ of $\Fg$, of $R$-charge $r_i$, and with the appropriate fugacities $y_i$ and background fluxes $\n_i$ for the global symmetries. The third factor is the one-loop determinant for the $W$-bosons and their superpartners,
\be\label{oneloop vector}
Z^\text{vector}_{\m} (u) =(-1)^{\sum_{\alpha>0} \alpha(\m)}\prod_{\alpha \in \Fg} \left(1- x^\alpha\right)^{1-g}~,
\ee
with $\alpha$ the simple roots of $\Fg$. These one-loop determinants were computed in \cite{Benini:2015noa}.
Finally, the function $H(u)$ appearing in \eqref{density} is the Hessian of the effective twisted superpotential $\CW$ as defined in \eqref{def Hu}, that is:
\be
H(u) =\det_{ab}\left( k^{ab}+\sum_i\sum_{\rho_i} \rho_i^a \rho_i^b \; \half \left({1+ x^{\rho_i}y_i\ov 1- x^{\rho_i}y_i}\right)\right)~,
\ee
while $W(x)$ is a Laurent polynomial in $x$ corresponding to the Wilson loop insertion \eqref{def Wx}. The contribution $H(u)^g$ in \eqref{density}, which arises because of  additional gaugino zero-modes on $\Sigma_g$, is the main new ingredient with respect to the $S^2\times S^1$ computation of \cite{Benini:2015noa}.

\subsubsection{Singular hyperplanes and JK residue}
There are three types of singularities of the integrand \eqref{density} on the classical Coulomb branch covering space $\t\fM$: 

\paragraph{Matter field singularities.} Whenever $x^{\rho_i} y_i=1$, the one-loop determinant \eqref{oneloop matter} may develop a pole (depending on the flux sector $\m$). For any field component $\rho_i$ of a chiral multiplet $\Phi_i$, we define the hyperplanes:
\be\label{hyperplane 1}
H_{\rho_i, n} = \{ u \in \t\fM \; |\; \rho_i(u) + \nu_i = n~, \;\; n\in \Z~\}~.
\ee
 These singularities signal the presence of massless modes associated to vortices, which can appear at these loci. See for instance \cite{Intriligator:2013lca} for a detailed discussion of BPS vortices. 

\paragraph{Monopole operator singularities.} The singularities of the second type are located at $x_a = \infty$ and $x_a= 0$ (that is, at $\sigma_a= \mp \infty$) and correspond to the monopole operators $T_{a}^+$ and $T_{a}^-$, respectively, which can condense in those limits:
\be\label{hyperplane 2}
H_{a\pm} = \{ u \in \t\fM \; |\; u_a = \mp i \infty\}~.
\ee
It is useful to think of $\t\fM$ as a $(\C P^{1})^\rk$ by including these hyperplanes at infinity.
The integrand \eqref{density} has singularities of the form
\be\label{singularity monopole}
I_\m \sim x_a^{\pm\left(Q_{a\pm}(\m)+ Q_{a\pm} \n_F + (g-1) r_{a \pm}\right)}\; du_a \qquad {\rm } \quad {\rm as}\qquad \sigma_a \rightarrow \mp \infty~,
\ee
which are determined in terms of the induced charges \eqref{induced charge T}-\eqref{induced Rcharge}   of $T_{a}^\pm$.

\paragraph{W-boson singularities.} The singularities of the third type are the zeros of the vector multiplet one-loop determinant \eqref{oneloop vector} (if $g>1$). They are located at:
\be\label{hyperplane 3}
H_{\alpha, n} =   \{ u \in \t\fM \; |\; \alpha(u) = n~, \;\; n\in \Z~\}~,
\ee
for any simple root $\alpha$. These hyperplanes are the  walls of the Weyl chambers in the covering space $\t\fM$, where part of the non-abelian symmetry is restored. 
Poles including this hyperplane need a special treatment in the path integral. Indeed one
can easily check that singularities involving $H_\alpha$ are always non-projective (see below for a definition) so that the JK-residue operation
is ill-defined. We claim that we should simply exclude these poles from the residue integral.
We checked in many examples that this prescription gives the expected answer.
This is consistent with discussions in previous literature, in particular with the study of CS theory \cite{Blau:1993tv,Ohta:2012ev,Gukov:2015sna} and  two-dimensional theories \cite{Hanany:1997vm}.

\vskip0.5cm

Consider the Coulomb branch covering space compactified as $\t \fM \cong \left(\C P^1\right)^\rk$, with the hyperplanes at infinity included as the poles of each $(\C P^1)_a$. In each topological sector, we denote by  $\tfM^{\m}_\text{sing}$ the set of codimension-$\rk$ singularities coming from the intersection of $s\geq \rk$ hyperplanes \eqref{hyperplane 1} and/or \eqref{hyperplane 2}, and such that they are {\it not} located on the hyperplanes \eqref{hyperplane 3}.

The localization formula in \eqref{main formula} is given by a contour integral  on $\t\fM$ in each topological sector. The contour of integration is determined by the  Jeffrey-Kirwan residue prescription \cite{JK1995, 1999math......3178B,2004InMat.158..453S} around each singularity $u_\ast \in \tfM^{\m}_\text{sing}$. Consider any singular point $u_*$ at the intersection of  $s$ singular hyperplanes $H_{Q_1}, \cdots, H_{Q_s}$, whose directions and orientations are determined by the charge vectors 
\be\label{Qudef}
{\bf Q}(u_\ast)= \{Q_1, \cdots, Q_s\} \in \Gamma_\GG \subset i \Fh^*
\ee in the electric weight  lattice.  These charge vectors $Q_j$ are either weights $\rho_i$ from matter field singularities, or induced charges $Q_{a\pm}$ from monopole operator singularities. For the JK residue to exist, we assume that all the relevant singularities are {\it projective}. This means that, for any $u_\ast$,  the $s$ charges \eqref{Qudef} are contained within a {\it half-space} of $i \Fh^\ast$. A singularity with $s=\rk$ is said to be non-degenerate.

For completeness, we briefly review the definition of the JK residue.
(We refer to \cite{Benini:2013xpa, Hori:2014tda, Benini:2015noa, Closset:2015rna} for further discussions.)
We consider the case $u_\ast=0$, while the general case can be obtained by translation. Let us denote by $Q_S$ any subset of $\rk$ distinct charges in ${\bf Q}(\hs_*)$, and let us define:
\be
\omega_S=  \prod_{Q_j \in Q_S}{1\over Q_j(u)} \; du_1 \wedge \cdots\wedge du_\rk~, 
\ee  
the corresponding singular holomorphic $\rk$-form. 
The JK residue on $\omega_S$ is defined by
\be\label{def JK}
\underset{u=u_*}{ {\text{JK-Res}}}\left[{\bf Q} (u_\ast), \eta \right]\, \omega_S =  
 \left\{ 
\begin{array}{ll} {1\over |\det(Q_S)| }  &\qquad {\rm if}\quad \eta\in \text{Cone}(Q_S)~, \\
 0 &\qquad {\rm if}\quad \eta\notin \text{Cone}(Q_S)~,
\end{array} \right.
\ee
in terms of an auxilliary vector $\eta \in \Fh^\ast$, which we can choose at our convenience as long as it is not parallel to any of the charge vectors. 
For degenerate singularities where more than $\rk$ hyperplane meets, 
we refer to the prescription in \cite{1999math......3178B,2004InMat.158..453S,Benini:2013xpa} for an algorithmic determination of the JK contour.   The definition \eqref{def JK} is often sufficient to determine the JK contour in practice.

\subsection{Relation to the Bethe equations and to the Wilson loop algebra}\label{sec: bethe equ}
Note that all the factors in \eqref{density} that depend on the gauge flux $\m$ organize themselves into the twisted superpotential:
\be
e^{2\pi i \,\d\CW(\m)}\equiv \exp \left(2\pi i \sum_a  {\d \CW\ov \d u_a}\, \m_a\right)~, 
\ee
reproducing \eqref{duW explicit}.  
We can formally perform the sum over fluxes (see  \cite{Closset:2015rna} for a similar discussion) to obtain:
\bea\label{Wg inter}
&\langle W \rangle_g =\cr
&\quad\sum_{\h x  \,\in\, \CS_{\rm BE}} \oint_{x= \h x}  \prod_a \left[ {dx_a \ov 2 \pi i \, x_a}\, {1\ov e^{2\pi i \d_{u_a}\CW}-1} \right] \det_{ab}( \d_{u_a}\d_{u_b}\CW) \; \CU(x) \, \CH(x)^{g-1}\, W(x)~,
\eea
where we pick the Grothendieck residues at $x=\h x \in \CS_{\rm BE}$, with $\CS_{\rm BE}$ the set of distinct solutions (up to Weyl equivalences)  of the Bethe equations \eqref{rel CW}. Here we defined:
\be
\CU(x)=e^{-S_{\rm CS}^{\rm gauge} - S_{\rm CS}^{\text{gauge-flavor} }- S_{\rm CS}^{\text{flavor} }}\Big|_{\m=0}\,  \prod_{\rho_i \in \mathfrak{R}_i } \left({ x^{\half \rho_i} y_i^\half\ov 1- x^{\rho_i} y_i}\right)^{ \n_i }
\ee
and
\be
\CH(x) =e^{- S_{\rm CS}^{\text{gauge-R} } -S_{\rm CS}^{\text{flavor-R} }}\Big|_{g=2}  \, \prod_i \prod_{\rho_i \in \mathfrak{R}_i } \left({ x^{\half \rho_i} y_i^\half\ov 1- x^{\rho_i} y_i}\right)^{r_i-1} \; \prod_{\alpha \in \Fg} {1\ov 1- x^\alpha}\;\, H(u)~.
\ee
We assume that the two-dimensional theory is fully massive,  such that $H(\h x)\neq 0$, $\forall \h  x \in \CS_{\rm BE}$. 
This leads to:
\be\label{Index and BE res}
\langle W \rangle_g =  \sum_{ \h x \,\in\, \CS_{\rm BE}} \CU(\h x)\; \CH(\h x)^{g-1}\;  \; W(\h x)~.
\ee
This result was first obtained in \cite{Nekrasov:2014xaa} in the case $\CU=1$---that is, for vanishing background fluxes. The quantity $\CH(x)$ is the three-dimensional handle-gluing operator \cite{Nekrasov:2014xaa}, allowing us to write down  genus-$g$ correlation functions in terms of the genus-zero result:
\be
\langle W \rangle_g =  \langle W \CH^g \rangle_0~.  
\ee

According to \eqref{Index and BE res}, the Witten index \eqref{intro W index} is given by the number of distinct solutions to the Bethe equations:
\be\label{Windex general form}
 \Tr_{T^2} \,(-1)^F = \sum_{ \h x \,\in\, \CS_{\rm BE}}  1~.
\ee
As we will see in the examples, this directly reproduces the results of \cite{Witten:1999ds, Ohta:1999iv, Intriligator:2013lca}. This simply  reflects the one-to-one correspondence between three-dimensional vacua in the presence of generic real masses and  two-dimensional  vacua of the theory compactified on a circle of finite size \cite{Intriligator:2013lca}. 

The formula \eqref{Wg inter} directly implies the quantum algebra \eqref{CAW general} of Wilson loops. By definition of the ideal $I_\CW$ in  \eqref{CAW general}, any insertion of an element $Z$ of this ideal has vanishing correlation function with any other Wilson loop:
\be
\langle W Z \rangle_g  = 0~, \qquad \qquad {\rm if}\qquad Z(x) \in I_\CW~.
\ee
Conversely, if $\langle W Z \rangle_g=0$ for every possible insertion $W$, it implies that $Z(x)|_{x=\h x}=0$, $\forall \h x$, so that $Z(x)\in I_\CW$.

\subsection{Sign ambiguities of the twisted index and dualities}
We just explained how to compute the twisted index \eqref{general index} as a path integral on $\Sigma_g\times S^1$:
\be
I_g =\Tr_{\Sigma_g} \left((-1)^F \prod_i y_i^{Q_i}\right) = Z_{\Sigma_g \times S^1}~.
\ee
The overall sign of the 3d partition function seems ambiguous, although we have chosen it  such that the Witten index \eqref{Windex general form} is a non-negative integer. 
Whenever the gauge group $\GG$ contains abelian factors $U(1)_I$, the index suffers from a sign ambiguity in the sum over topological sectors, corresponding to shifting the fugacities $q_I$ for the topological symmetries $U(1)_{T_I}$ by arbitrary signs \cite{Gaiotto:2013bwa, Benini:2015noa}, $q_I \rightarrow (-1)^{n_I} q_I$ with $n_I \in \Z$. 
This can be thought of as a shift of the two-dimensional $\theta$-angles by multiples of $\pi$. These sign ambiguities lead to a possible ambiguity when checking dualities, and generally we will find that, for any pair of dual theories $T$ and $T_D$, we have~\footnote{Here the fugacities $q, y$ are mapped to $q_D, y_T$ in some way, which might involve some convenient choice of sign for $q$, $q_D$.}
\be\label{Z Zdual puzzle}
Z_{\Sigma_g \times S^1}^{[T]}(q, y) = (-1)^{(g-1)n_r +\sum_i \n_i n_i}\, Z_{\Sigma_g \times S^1}^{[T_D]}(q_D, y_D)
\ee
for some theory-dependent integers $n_r, n_i$.
 In principle,  any such ambiguity should be accounted for by an appropriate supersymmetric counterterm \cite{Closset:2012vg,Closset:2012vp} but the precise mechanism in this case is unclear to us at this point.~\footnote{Potentially related issues have been discussed in \protect\cite{Imbimbo:2014pla}.} An interesting special case of  \eqref{Z Zdual puzzle} is for a theory of two chiral multiplets $\Phi_1$, $\Phi_2$ with $R$-charges $r$ and $2-r$, gauge charge $Q$ and $-Q$ under a flavor $U(1)$ with fugacity $y$ and background flux $\n$, and a superpotential $W= \Phi_1 \Phi_2$. This theory is infrared ``dual'' to an empty theory, but the partition function reads:
 \be
 Z_{\Sigma_g \times S^1}^{[\Phi_1 \Phi_2]}(y) = (-1)^{Q\n + (g-1)(r-1)}~.
 \ee
  We leave a more precise understanding of these signs as an interesting question for future work.

\section{$\CN=2$  $U(1)$ theories and elementary dualities}\label{sec: expl1}
In this section, we study $\CN=2$ CS-matter theories with a gauge group $\GG=U(1)$. These theories were recently studied extensively in \cite{Intriligator:2013lca}.
This will serve as an interesting warm-up to the non-abelian theories of the next sections.

\subsection{$U(1)_k$ CS-matter theory}
Consider a $U(1)$ theory with CS level $k>0$ and charged chiral multiplets $Q_i$ and $\t Q_j$,  of gauge charge $n_i$ and $-\t n_j$, respectively, with $n_i>0$ and $\t n_j >0$.~\footnote{The gauge charges $n_i, \t n_j$ should not be confused with the background fluxes $\n_i, \t \n_j$. Moreover, here and in later sections we often use $\t\Phi$ to denote chiral multiplets of negative charges and {\it not} anti-chiral multiplets like in the last section. This should cause no confusion.} It is useful to define:
\be
k_c = \half\sum_i n_i^2 -\half \sum_j \t n_j^2~.
\ee
Without loss of generality, we assume that $k_c \geq 0$.
The theory has a large flavor symmetry, depending the choice of $n_i, \t n_j$, but we can focus on the axial symmetry $U(1)_A$ defined in Table \ref{tab:U1ktheory}.
\begin{table}[t]
\centering
\be\nn
\begin{array}{c|c|ccc}
    &  U(1)_\GG& U(1)_A &U(1)_T& U(1)_R  \\
\hline
Q_i        & n_i & n_i &0& r_i \\
\t Q_j    &- \t n_j & \t n_j&0& \t r_j
\end{array}
\ee
\caption{Gauge, axial, topological and $R$-charges of the matter fields in the $U(1)_k$ CS-matter theory.}
\label{tab:U1ktheory}
\end{table}
Let us denote by $y_i^{-1}$ and $\t y_j$ the flavor fugacities for $Q_i$ and $\t Q_j$, respectively. If we are only interested in $U(1)_A$, then $y_i= y_A^{-n_i}$ and $\t y_j = y_A^{\t n_j}$. 
To cancel a potential parity anomaly for $U(1)_A$, we also turn on the mixed flavor-CS term $
k_{gA}= - k_c$. We also redefine $q\rightarrow (-1)^{\sum_i n_i^2} q$ for convenience. 
The Bethe equation~\eqref{rel CW} for this theory reads:
\be
P(x) = \prod_i \left(x^{n_i} - y_i \right)^{n_i} - q y_A^{- \sum_i n_i^2} x^{k+k_c} \prod_j (x^{\t n_j} - \t y_j)^{\t n_j}=0~.
\ee
The twisted index is easily evaluated using the general results of the previous section. In particular, it follows from  \eqref{Windex general form} that  the Witten index  of this theory is equal to the degree of the polynomial $P(x)$:
\be
 \Tr_{T^2} \,(-1)^F =  \deg{(P)} =  \begin{cases}
k+\half \sum_i n_i^2+\half \sum_j \t n_j^2 &\qquad {\rm if} \qquad k\geq  k_c~,\\
\sum_i n_i^2 &\qquad {\rm if} \qquad k_c\geq  k~.
\end{cases}
\ee
This reproduces the Witten index computed in  \cite{Intriligator:2013lca} by a careful analysis of the vacuum structure of the theory.

\subsection{SQED/$XYZ$-model duality}\label{subsec:SQED XYZ}
As an interesting special case, consider three-dimensional SQED, a $U(1)$ gauge theory without CS interaction and with two charged scalar multiplets $Q, \t Q$ of charges $\pm 1$ and $R$-charge $r$. The theory has an axial symmetry $U(1)_A$ and a topological symmetry $U(1)_T$, with associated fugacities $y_A$ and $q$ (and background fluxes $\n_A$ and $\n_T$), respectively, and we have an FI parameter turned on according to \eqref{FI as real mass}. 

The twisted index \eqref{main formula} for SQED reads:
\bea
&Z_{\Sigma_g \times S^1}^{\rm SQED} = -\sum_{\m \in \Z} \oint_{\rm JK}   {dx \ov 2 \pi x}   (-q)^\m x^{\n_T} \left({x^\half y_A^\half\ov 1- x y_A}\right)^{\m+\n_A + (g-1)(r-1)}\cr
&\qquad\quad\qquad\times \left({x^\half y_A^\half\ov x-  y_A}\right)^{-\m+\n_A + (g-1)(r-1)} \left[\half \left({1+ x y_A\ov 1- x y_A}\right)+ \half \left({x+ y_A\ov x- y_A}\right)\right]^g~.
\eea
Note that we introduced a convenient sign in front of $q$. With $\eta >0$, the JK residue picks the pole at $x= y_A^{-1}$ for $\m \geq  -\n_A - r(g-1)$. There is no contribution from infinity on $\fM$ because the monopole operators $T^\pm$ are gauge invariant.  Following \cite{Benini:2015noa},
we can perform the sum over $\m$ first, which gives:
\bea\label{Zindex SQED inter}
&Z_{\Sigma_g \times S^1}^{\rm SQED} = \oint_{x= \h x} {dx \ov 2 \pi x} {P'(x)\ov P(x)} x^{\n_T}\left({x y_A \ov (1- x y_A)(x-  y_A)} \right)^{\n_A + (g-1)(r-1)}\cr
&\qquad\quad\qquad\qquad\times  \left[\half \left({1+ x y_A\ov 1- x y_A}\right)+ \half \left({x+ y_A\ov x- y_A}\right)\right]^{g-1}~,
\eea
where $\h x \equiv (1- q y_A)/(y_A-q)$ is the solution to the Bethe equation:
\be
P(x) = x- y_A^{-1} - q y_A^{-1}(x- y_A)=0~.
\ee
The expression \eqref{Zindex SQED inter} gives:
\bea
&Z_{\Sigma_g \times S^1}^{\rm SQED}= (-1)^{\n_T}
 \left(y_A \ov 1- y_A^2 \right)^{2 \n_A+(g-1)(2r-1)} \,\cr
&\qquad\times\left( q^{\half} y_A^{-\half}\ov 1- q y_A^{-1}\right)^{\n_{T} - \n_A+ (g-1)(r_T-1)} \left( q^{-\half} y_A^{-\half }\ov 1- q^{-1} y_A^{-1}\right)^{-\n_{T} -  \n_A+ (g-1)(r_T -1)} ~,
\eea
with $r_T=-r+1$.
Up to a sign $(-1)^{\n_T}$, this is simply the twisted index of three chiral multiplets $(X, Y, Z)= (M, T^+, T^-)$ with charges:
\be
\begin{array}{c|ccc}
    & U(1)_A & U(1)_T&  U(1)_R  \\
\hline
M        & 2& 0& 2 r \\
T^+    &- 1 & 1& -r+1\\
T^-    &- 1 & -1& -r+1
\end{array}
\ee
These charges are compatible with the cubic superpotential
$W= M T^+ T^-$.
This is expected since  3d $\CN=2$ SQED is dual to the $XYZ$ model \cite{Aharony:1997bx}.

\subsection{$U(1)_\half$ with a single chiral multiplet}\label{subsec: U1half}
Consider a $U(1)$ theory with a Chern-Simons level $k=\half$ and a single chiral multiplet $Q$ of  gauge charge $1$. We also choose $Q$ to have $R$-charge $r$, and we turn on a mixed gauge-$R$ CS level $k_{g R}= -\half (r-1)$, although the $R$-charge can be set to any value by mixing with the gauge symmetry. This theory has a flavor symmetry $U(1)_T$, the topological symmetry of the $U(1)$ gauge group. It is dual to a single free chiral multiplet $T^+$ of $U(1)_T$ charge $1$, corresponding to the lowest gauge-invariant  monopole operator for the `half' Coulomb branch of the gauge theory \cite{Dimofte:2011ju, Benini:2011mf}. Importantly, the dual free theory also contains the flavor CS terms:
\be\label{rel CS U1half dual}
\Delta k_{TT} = -\half~, \qquad \qquad \Delta k_{TR}= - {r\ov 2}~.
\ee
This is a special case of a more general Seiberg duality \cite{Benini:2011mf}, that we shall discuss in more details in section \ref{subsec: sqcd dual 4} below (and in Appendix \ref{appendix: Seiberg duals}).
The twisted index of the $U(1)_\half$ theory reads:
\bea
&Z_{\Sigma_g \times S^1}^{U(1)_\half, Q} = -  \sum_{\m \in \Z} \oint_{\rm JK}   {dx \ov 2 \pi x}   (-q)^\m x^{\half \m + \n_T- \half (g-1)(r-1)} \left({x^\half \ov 1-x}\right)^{\m+ (g-1)(r-1)}\cr
&\qquad\qquad\qquad\quad\qquad\qquad\times \left[ \half + \half\left({1+x\ov 1-x}\right)\right]^g~,
\eea
where we redefined $q\rightarrow -q$ for convenience.
Note that the monopole operators $T^\pm$ of this theory have gauge charges $0$ and $-1$, respectively. 
If we take $\eta>0$, the JK residue has contributions from $Q$ only, at $x=1$. If we take $\eta <0$ instead, we pick the poles at $x=0$. Either way, we can perform the sum over the fluxes as above, to obtain:
\be
Z_{\Sigma_g \times S^1}^{U(1)_\half, Q} = (-1)^{(g-1)r} \, q^{\n_T \Delta k_{TT}+ (g-1) \Delta k_{TR}}  \left({q^\half \ov 1-q}\right)^{\n_T + (g-1)(r-1)}~,
\ee
with the CS levels $\Delta k_{TT}, \Delta k_{TR}$ given in \eqref{rel CS U1half dual}, 
in perfect agreement with the duality. 

\section{Chern-Simons theories and the Verlinde formula}\label{sec: expl2}
In this section, we consider a supersymmetric Chern-Simons theory without matter. Consider the $\CN=2$ Chern-Simons theory with gauge group $\GG$ at level $k>0$. As we recalled at the end of section \ref{subsec: Wilson loop alg}, that theory is IR-equivalent to an ordinary  CS theory at level:
\be
\h k = k- h~,
\ee
with $h$ the dual Coxeter number of $\GG$.
The genus-$g$ supersymmetric index should give the dimension of the Hilbert space of a $\GG_{\h k}$ CS theory on $\Sigma_g$:
\be\label{Z as dimH CStheory}
Z_{\Sigma_g \times S^1}^{[\GG, k]} = { \rm dim}\, \CH\left(\Sigma_g; \GG_{k - h}\right)~,
\ee
which is famously given by the Verlinde formula \cite{Verlinde:1988sn, Blau:1993tv}. 
This provides an nice consistency check of our localization formula at higher genus.
Here we shall focus on $\GG= U(N)$ and $\GG= SU(N)$, for simplicity.  

\subsection{$U(N)$ $\CN=2$ supersymmetric CS theory}
Consider the $\CN=2$ $U(N)$ vector multiplet with Chern-Simons interaction at level $k>0$. Due to the $U(1)$ factor, the theory has a topological symmetry $U(1)_T$, and we can turn on the associated fugacity $q$ and background flux $\n_T$. The twisted index reads:~\footnote{Here we used the freedom to shift $q$ to $(-1)^{N-1} q$, for convenience.  This cancels the sign factor in front of \eqref{oneloop vector}.}
\be
Z_{\Sigma_g\times S^1}^{[N, k]}(q) = {(-1)^N\ov N!}   \sum_{\m \in \Z^N}q^\m  \oint_{\rm JK} \prod_{a=1}^N \left[{dx_a\ov 2 \pi i x_a}  x_a^{k\m_a + \n_T}\right] 
\prod_{\substack{a,b=1\\ a\neq b}}^{N} \left(x_a\ov x_b-x_a\right)^{g-1} \; k^{gN}~.
\ee
The factor of $k^{gN}$ is the contribution from $H=k^N$ for a $U(N)$ CS theory. 
The monopole operators $T_a^\pm$ have gauge charges ${Q_{a{\pm}}}^b = \pm \delta^a_b k$. If we take $\eta=(1, \cdots, 1)$ in the JK residue, we only have contributions from $x_a =\infty$. 
After performing the sum over the fluxes explicitly, the pole at $x_a=\infty$ are all relocated to the
solutions of the Bethe equations:
\be
P(x_a)=0~, \qquad a=1, \cdots, N~, \qquad x_a\neq x_b~,\; \; {\rm if} \; \;a\neq b~, \qquad
P(x)\equiv 1- q x^k~.
\ee
(One can check that the solutions to the Bethe equations go to $x_a\rightarrow \infty$ as $q\rightarrow 0$.)
Using the fact that:
\be
\sum_{\m_a= M}^\infty \left(q x_a^k \right)^\m = {(x_a^k q)^M\ov P(x)}~,
\ee
for any fixed integer $M$, we find:
\be
Z_{\Sigma_g\times S^1}^{[N, k]}(q) ={1\ov N!} \oint \prod_{a=1}^N \left[{dx_a\ov 2 \pi i}{P'(x_a)\ov P(x_a)} x_a^{ \n_T}\right] 
\prod_{\substack{a,b=1\\ a\neq b}}^{N} \left(x_a\ov x_b-x_a\right)^{g-1} \; k^{(g-1)N}~,
\ee
where the integral  becomes a sum of iterated residues at $x_a = \h x_\alpha$ with
\be
\h x_\alpha = q^{-{1\ov k}}\, \omega_\alpha~, \qquad  \alpha= 1, \cdots, k \qquad\qquad  \omega_\alpha \equiv e^{2\pi i \alpha \ov k}~,
\ee
the roots of $P(x)$.
The partition function thus reduces to a sum over choices of $N$ distinct integers among $\{\alpha\}=\{1, \cdots, k\}$.
 Let  $\CC_N^k$ denotes the set of all choices of $N$ distinct integers among $\{\alpha\}$, and let $I=\{\alpha_1, \cdots, \alpha_N\}$ be any element of $\CC_N^k$. We have:
 \be\label{UN Verlinde formula}
Z_{\Sigma_g\times S^1}^{[N, k]}(q) = k^{(g-1)N} \,  q^{-\n_T} \sum_{I\in \CC_N^k }  \left(\prod_{\alpha\in I} \omega_\alpha\right)^{\n_T} \prod_{\substack{\alpha, \beta \in I \\\alpha\neq \beta}}\left(1 - {\omega_\alpha\ov \omega_\beta}\right)^{1-g}~.
 \ee
In particular, $Z_{\Sigma_g\times S^1}^{[N, k]}(q)=0$ if   $N > k$. As a small consistency check, we note that \eqref{UN Verlinde formula} implies the Witten index:
\be
\Tr_{T^2}(-1)^F  =  \mat{k\cr N}~,
\ee
in agreement with  \cite{Ohta:1999iv}. In the case $\n_T=0$, the dependence on $q$ drops out from  \eqref{UN Verlinde formula} and it turns out  that the resulting numbers $Z_{\Sigma_g\times S^1}^{[N, k]}$ for $k \geq N$ are positive integers, consistent with the interpretation \eqref{Z as dimH CStheory}.

 The $U(N)_k$ supersymmetric CS theory enjoys level rank duality:
\be
U(N)_k\qquad    \longleftrightarrow \qquad   U(k-N)_{-k}~.
\ee
The duality also exchanges the sign of the topological current. We can show that:
\be\label{levelrank indices}
Z_{\Sigma_g\times S^1}^{[N, k]}(q)  = (-1)^{(k-1)\n_T+ (k-N)(g-1)}  q^{-\n_T} \; Z_{\Sigma_g\times S^1}^{[k-N, -k]}(q^{-1})~.
\ee
The factor $q^{-\n_T}$  is interpreted as a relative CS level $\Delta k_{TT} = -1$ for the $U(1)_T$ background gauge field. This duality is a special case of a more general three-dimensional Seiberg duality \cite{Giveon:2008zn, Benini:2011mf}, which we shall study thoroughly in section \ref{sec: sqcd and seiberg duals}.
To prove \eqref{levelrank indices}, we write down the twisted index as:
\be
Z_{\Sigma_g\times S^1}^{[N, k]}(q) = \sum_{\h x \in \CS_{\rm BE} } \CU(\h x) \,\CH(\h x)^{g-1}~,
\ee
with
\be\label{UH in pure CS}
 \CU(x)= \prod_{a=1}^N x_a^{\n_T}~, \qquad\qquad \CH(x) = k^N  \prod_{\substack{a,b=1\\ a\neq b}}^{N} \left(x_a\ov x_b-x_a\right)~,
\ee
following the notation of section \ref{sec: bethe equ}.
We can easily show that, for $\h x= \{\h x_a\}_{a=1}^N \subset \{\h x_\alpha \}_{\alpha=1}^k$ a set of $N$ distinct roots of $P(x)$, and $\h x_D= \{\h x_{\b a}\}_{\b a=1}^{k-N}$ its complement, we have:
\be
 \CU(\h x) = (-1)^{(k-1)\n_T}  q^{-\n_T}   \CU_D(\h x_D)~, \qquad\qquad
  \CH(\h x) = (-1)^{k-N}\CH_D(\h x_D)~,
\ee
where
\be
 \CU_D(x_D)= \prod_{\b a=1}^{k-N} x_{\b a}^{-\n_T}~, \qquad\qquad \CH_D(x_D) =(-k)^{k-N}  \prod_{\substack{\b a,\b b=1\\ \b a\neq \b b}}^{k-N} \left(x_{\b a} \ov x_{\b b}-x_{\b a}\right)~,
\ee
are the quantities \eqref{UH in pure CS} in the dual $U(k-N)_{-k}$ theory. The duality relation \eqref{levelrank indices} follows by exchanging any set $I \in \CC_N^k $ with its complement $I^c$ in $\{\alpha\}$.
One can similarly study Wilson loop correlation functions and verify that they satisfy the Verlinde algebra \cite{Kapustin:2013hpk, Benini:2015noa}. (See section \ref{subsec: wilson loops sqcd}  below for a general discussion in 3d $\CN=2$ SQCD.)

\subsection{The $SU(N)$ Verlinde formula}
It was noted in \cite{Benini:2015noa} that the $S^2\times S^1$ twisted index for an $U(N)$ theory with matter fields neutral under its center is equivalent to the $S^2\times S^1$ twisted index for the corresponding $SU(N)$ theory (if $\n_T=0$). On $\Sigma_g\times S^1$,  we can similarly show that:
\be
Z_{\Sigma_g\times S^1}^{SU(N)_k}= \left({N\ov k}\right)^g\; Z_{\Sigma_g\times S^1}^{[N, k]}(q)\Big|_{\n_T=0}~.
\ee
From \eqref{UN Verlinde formula}, we directly find:
\be\label{SUN verlinde}
Z_{\Sigma_g\times S^1}^{SU(N)_k} =  \left({N\ov k}\right)^g k^{(g-1)N} \,  \prod_{\substack{\alpha, \beta \in I \\\alpha\neq \beta}}\left(1 - e^{2\pi i (\alpha-\beta)\ov k}\right)^{1-g}~.
\ee
In particular, this reproduces the correct Witten index \cite{Witten:1999ds,Ohta:1999iv}:
\be
\Tr_{T^2}(-1)^F  =  \mat{k-1\cr N-1}~.
\ee
One can check that \eqref{SUN verlinde} agrees precisely with the Verlinde formula for $SU(N)_{k-N}$ pure CS theory on $\Sigma_g$. 
 In particular, it is easy to show that
\be
Z_{\Sigma_g\times S^1}^{SU(2)_k}=  V_{g, \h k}^{SU(2)} = \left({\h k +2\ov 2}\right)^{g-1} \sum_{j=0}^{\h k}\,\left( \sin{(j+1)\pi \ov \h k +2}\right)^{2-2 g}~.
\ee
in the special case $N=2$, with $\h k= k-2$. 
 One can also check level-rank duality like in~\cite{Zagier}.

\subsection{The equivariant Verlinde formula}
Another interesting theory is the $\CN=2$ Chern-Simons theory at level $k$ with an adjoint chiral multiplet $\Phi$ of real mass $m >0$ and $R$-charge $r$. For $r=2$, the $\Sigma_g\times S^1$ twisted index computes  the ``equivariant Verlinde formula'' introduced in \cite{Gukov:2015sna}. That formula was also computed in \cite{Gukov:2015sna} using the results of \cite{Nekrasov:2014xaa}, 
therefore it is obvious from  the general discussion in section \ref{sec: bethe equ} that we should reproduce this result as well. We briefly show this here. 

Consider $\GG=U(N)$ at CS level $k>0$. Let $U(1)_t$ be the symmetry that rotates the chiral multiplet $\Phi$ with charge $1$, and let us introduce the corresponding fugacity $t$ and background flux $\n_t$.  (We have $|t|= e^{-2 \pi \beta m}$ with $m$ the real mass.)  To make contact with \cite{Gukov:2015sna}, we  choose to turn on a mixed $U(1)_t$-$R$ CS level:
\be\label{tR mixed CS level exp GdP} 
k_{t R}= -\half N^2 (r-1)~.
\ee
We also allow for an arbitary gauge-$R$ CS level $k_{gR}$ for the $U(1)\subset U(N)$ gauge group.
The twisted index reads:
\bea
&Z_{\Sigma_g\times S^1}^{[U(N)_k, \Phi]}(q, t)={ t ^{(g-1)k_{tR}}(-1)^N\ov N!}   \sum_{\m \in \Z^N}q^\m  \oint_{\rm JK} \prod_{a=1}^N \left[{dx_a\ov 2 \pi i x_a}  x_a^{k\m_a + \n_T+ (g-1)k_{gR}}\right] 
 \cr
&  \;\times\prod_{\substack{a,b=1\\ a\neq b}}^{N} \left(x_a\ov x_b-x_a\right)^{g-1} \prod_ {\substack{a,b=1\\ a\neq b}}^{N} \left(x_a^{\half} x_b^{\half} t^\half \ov x_b-x_a t\right)^{\m_a-\m_b + \n_t +(g-1)(r-1)}\; \left(\det_{ab}\h H_{ab}(x)\right)^g~,
\eea
where we defined:
\be
\h H_{ab}(x) = k \delta_{ab} + \half \sum_{\substack{c,d=1\\ c\neq d}}^{N} \left(\delta_{ab}\delta_{ac} - \delta_{ac}\delta_{bd}\right) {x_c x_d (1-t^2)\ov (x_c-  x_d t)(x_d- x_c t)}~.
\ee
The Bethe equations of this theory are:
\be\label{BE equiv verlinde}
P_a(x) \equiv  \prod_{c=1}^N (x_c -x_a t) - q x_a^k \prod_{c=1}^N (x_a- x_c t)=0~, \qquad a=1, \cdots, N~, 
\ee
and $x_a\neq x_b$ if $a\neq b$. By resumming the fluxes and using the property:
\be
\d_{x_b} P_a\Big|_{x=\h x}= -{1\ov x_b}\prod_{c=1}^N (x_c- x_a t)\;  \h H_{ab}\Big|_{x=\h x}~,
\ee
satisfied by the solutions to the Bethe equations,
we indeed find:
\bea
&Z_{\Sigma_g\times S^1}^{[U(N)_k, \Phi]}(q, t)= \sum_{\h x \in \CS_{BE}} \oint_{x= \h x} 
\prod_{a=1}^N \left[{dx_a\ov 2 \pi i} { x_a^{ \n_T+ (g-1)k_{gR}}\ov P_a(x)}\right] \det_{ab}(\d_{x_b} P_a)~\,  \prod_{\substack{a,b=1\\ a\neq b}}^{N} \left(x_a\ov x_b-x_a\right)^{g-1} \cr
& \qquad\qquad \times  t ^{(g-1)k_{tR}} \prod_{\substack{a,b=1\\ a\neq b}}^{N}  \left(x_a^{\half} x_b^{\half} t^\half \ov x_b-x_a t\right)^{ \n_t +(g-1)(r-1)}\; \left(\det_{ab}\h H_{ab}(x)\right)^{g-1}~.
\eea
The sum is over the distinct solutions $\h x$ to the Bethe equations \eqref{BE equiv verlinde}, and each residue is taken at the isolated singularity $x=\h x$. (More precisely, at each $x=\h x$ we have  a local Grothendieck residue for the ideal $\{P_a\}_{a=1}^{N_c}$ in $\C[x_a]$.)
This gives:
\be
Z_{\Sigma_g\times S^1}^{[U(N)_k, \Phi]}(q, t)=  \sum_{ \h x \,\in\, \CS_{\rm BE}} \CU(\h x)\; \CH(\h x)^{g-1}~,
\ee
with:
\bea
&\CU(x) = \prod_{a=1}^N x_a^{\n_T}  \prod_{a, b=1}^{N} \left(x_a t^\half \ov x_b-x_a t\right)^{ \n_t }~,\cr
&\CH(x)= (1- t)^{-N(r-1)}   \prod_{a=1}^N x_a^{k_{gR}}  \prod_{\substack{a,b=1\\ a\neq b}}^{N} \left[{x_a\ov x_b-x_a}\left(x_a \ov x_b -x_a t\right)^{r-1}\right]~.
\eea
This formula precisely  agrees~\footnote{Note that, following \cite{Hori:2014tda, Benini:2015noa}, we have slightly different one-loop contributions from the ones in \cite{Nekrasov:2014xaa,Gukov:2015sna}. In the present case,  this difference was accounted for by turning on the mixed CS levels $k_{tR}$ and $k_{gR}$.} 
 with \cite{Gukov:2015sna} in the case $\n_T=\n_t=0$ and $q=1$, provided that we choose $k_{gR}= -(N-1)r$. 

\section{$\CN=2$ $U(N_c)_k$  YM-CS-matter theories and Seiberg dualities}\label{sec: sqcd and seiberg duals}

In this section, we study the three-dimensional $\CN=2$ supersymmetric version of SQCD on $\Sigma_g\times S^1$.
This theory consists of a  $U(N_c)$ vector multiplet with a Yang-Mills kinetic term and an overall Chern-Simons level $k$, coupled to $N_f$ chiral multiplets $Q_i$ ($i=1, \cdots, N_f$) in the fundamental representation of the gauge group and to  $N_a$ chiral multiplets $\t Q^j$ ($j=1, \cdots, N_a$) in the antifundamental  representation.  The global symmetry group is:  
\be\label{flavor sym SQCD}
SU(N_f) \times SU(N_a) \times U(1)_A \times U(1)_T\times U(1)_R~.
\ee
 Here $U(1)_A$ is the axial symmetry (which becomes trivial  if $N_f=0$ or $N_a=0$),  $U(1)_T$ is the topological symmetry of $U(N_c)$, and $U(1)_R$ is the $R$-symmetry.
Both $Q_i$ and $\t Q^j$ are taken to have $R$-charge $r \in \Z$, and the superpotential vanishes. 
\begin{table}[t]
\centering
\be\nn
\begin{array}{c|c|ccccc}
    &  U(N_c)& SU(N_f) & SU(N_a)  & U(1)_A &  U(1)_T & U(1)_R  \\
\hline
Q_i        & \bm{N_c}& \bm{\overline{N_f}} & \bm{1}& 1   & 0   &r \\
\tilde{Q}^j   & \bm{\overline{N_c}}  &  \bm{1}& \bm{N_a}  & 1   & 0   &r \\
\hline
T^\pm      &\; (\bm{N_c})^{\pm k-k_c} &  \bm{1} & \bm{1} & Q_\pm^A  & \pm 1   & r_\pm
\end{array}
\ee
\caption{Charges of the chiral multiplets of 3d $\CN=2$ SQCD. We also indicated the  charges  of the bare monopole operators $T^\pm$.}
\label{tab:SQCD charges}
\end{table}
To cancel the parity anomaly for the gauge symmetry, we must have:
\be
k+k_c\in \Z~, \qquad \qquad \quad k_c \equiv \half (N_f-N_a)~.
\ee
In order to cancel  potential parity anomalies for the flavor symmetry, we turn on  some mixed CS terms:
\be\label{kgA and kgR}
k_{g A}~, \qquad \qquad k_{gR}~,
\ee
between the $U(1) \subset U(N_c)$ factor of the gauge group and the $U(1)_A$  and $U(1)_R$ symmetry, respectively.  Note that the choice of mixed CS levels \eqref{kgA and kgR} is an  important part of the definition of the theory. In particular, it affects the quantum numbers of the monopole operators. We will make a convenient choice in the next subsection.
Finally, we also need to specify the global  CS levels for \eqref{flavor sym SQCD}.~\footnote{See Appendix \ref{appendix: Seiberg duals} and  especially \cite{Closset:2012vp} for a detailed discussion.}

As we will show momentarily, the Witten index  of three-dimensional $\CN=2$ SQCD  is given by:
\be\label{SQCD index}
\Tr_{T^2}(-1)^F = \mat{n \\ N_c}~, \qquad  {\rm with}\qquad 
n = \begin{cases}
|k|+ {N_f+ N_a\ov 2} &\qquad {\rm if} \qquad |k|\geq |k_c|~,\\
\max(N_f, N_a) &\qquad {\rm if} \qquad |k|\leq |k_c|~.
\end{cases}
\ee
For $N_c=1$, this was computed in \cite{Intriligator:2013lca}. 
When $n> N_c$, there exists a Seiberg-dual description of the theory \cite{Aharony:1997gp, Giveon:2008zn, Benini:2011mf} with dual gauge group 
$U(n-N_c)$  at  CS level $-k$, leaving the Witten index invariant.  The details of the Seiberg dual theory depend  on the relative values of $k$ and $k_c$ in an interesting way. 
  The matching of the twisted indices on $\Sigma_g$ between dual theories provides a powerful and intricate test of these dualities, including  the matching of  contact terms  for the global symmetries, which necessitates turning on certain background CS terms \cite{Willett:2011gp, Benini:2011mf, Closset:2012vp, Benini:2015noa}.

For future reference, let us comment on the monopole operators of the $U(N_c)$ theory. We denote by $T^\pm$  the bare monopole operators  of charge $\pm 1$ under $U(1)_T$. Their induced charges under $U(1)_A$ and $U(1)_R$ are:
\be\label{def QA rpm}
Q^A_\pm = \pm k_{gA}  - \half (N_f+N_a)~, \qquad\quad 
r_\pm = \pm k_{gR} - \half(N_f+N_a)(r-1) - N_c+1~.
\ee
The monopole operators also have induced gauge charges, as indicated in  Table \ref{tab:SQCD charges}. In particular,  $T^\pm$ is gauge invariant if and only if  $k=\pm k_c$. In that case, the Seiberg-dual theory contains one extra singlet (or two extra singlets if $k=0$) with the same quantum numbers as $T^\pm$, which couples to a monopole operator of the dual gauge group through the superpotential \cite{Aharony:1997gp, Benini:2011mf}.

 \subsection{The $\Sigma_g\times S^1$ index of 3d $\CN=2$ SQCD}
Consider $\CN=2$ SQCD as defined above.
 Let us  introduce generic fugacities $y_i$, $\t y_j$ (with $i=1,\cdots, N_f$ and $j=1,\cdots, N_a$)  for the  $SU(N_f)\times SU(N_f)\times U(1)_A$ flavor symmetry, such that:
\be
\prod_{i=1}^{N_f} y_i = y_A^{-N_f}~, \qquad \qquad
\prod_{j=1}^{N_a} \t y_j = y_A^{N_a}~,
\ee
with $y_A$ the $U(1)_A$ fugacity. We also introduce background fluxes $\n_i$, $\t \n_j$ subject to 
\be
\sum_i \n_i = -N_f \n_A~, \qquad \sum_j \t\n_j=  N_a \n_A~,
\ee 
with $\n_A$ the $U(1)_A$ flux.
We denote by  $q$ and $\n_T$ the fugacity and background flux for the topological symmetry $U(1)_T$.
 The twisted index of $\CN=2$ SQCD reads:
 \bea\label{Neq2SQCD index}
 &Z^{{\rm SQCD}\, [k, N_c, N_f, N_a] }_{\Sigma_g\times S^1}(q, y, \t y) = \cr
 &\qquad\qquad\qquad{(-1)^{N_c}\ov N_c!} \sum_\m \oint_{\rm JK} \prod_{a=1}^{N_c} {dx_a \ov 2 \pi i x_a} \;  Z^\text{cl}(x) Z^\oneloop_{\rm matter}(x) Z^\oneloop_{\rm gauge}(x) H(x)^g~,
 \eea
where the sum is over the fluxes $\m_a\in \Z$, $a=1, \cdots, N_c$. The integrand contains the classical piece:
\be\label{Zcl sqcd}
 Z^\text{cl}(x)  =\prod_{a=1}^{N_c}  \left[ (-1)^{(N_f+N_c-1) \m_a}  q^{\m_a} x_a^{\n_T}x_a^{k \m_a} x_a^{ (g-1)k_{gR}} x_a^{ k_{gA} \n_A } y_A^{k_{gA}\m_a} \right]~, 
\ee
which includes the mixed gauge-$U(1)_A$ and gauge-$R$ Chern-Simons terms \eqref{kgA and kgR}. Any other flavor CS term factorizes out of the index and can be ignored for our purposes. Note that we introduced a sign $(-1)^{(N_f+N_c-1)\sum_a \m_a}$ in \eqref{Zcl sqcd} for later convenience.  The other factors in the integrand are the one-loop determinants:
\bea
& Z^\oneloop_{\rm matter}(x)  = \prod_{a=1}^{N_c} \left[ \prod_{i=1}^{N_f} \left(x_a^\half y_i^\half \ov y_i-x_a\right)^{\m_a - \n_i+ (g-1)(r-1)} \, \prod_{j=1}^{N_a} \left(x_a^\half\t y_j^\half \ov x_a-\t y_j\right)^{- \m_a + \t\n_j+ (g-1)(r-1)}   \right]~,\cr
& Z^\oneloop_{\rm gauge}(x)   =(-1)^{(N_c-1)\sum_a \m_a}\, \prod_{\substack{a,b=1\\ a\neq b}}^{N_c} \left(x_a\ov x_b-x_a\right)^{g-1}~,
\eea
and the Hessian  of the twisted superpotential $\CW$:
\be\label{hessian sqcd}
 H(x)= \prod_{a=1}^{N_c} \h H(x_a)~, \qquad \h H(x)\equiv  k +\half \sum_{i=1}^{N_f} \left( {x +y_i\ov  y_i-x} \right)+\half \sum_{j=1}^{N_a} \left( {x+\t y_j\ov   x- \t y_j} \right)~.
\ee
Note that the index \eqref{Neq2SQCD index} depends on the choice of $R$-charge $r$ through the combination $\n_A + (g-1)(r-1)$ only, therefore we could set $r=1$ without loss of generality. Nonetheless, we find it instructive to present the final formulas for an arbitrary $r$.

Since the gauge charges of the monopole operators $T_a^\pm$ are given by:
\be
Q_{a\pm}^b = \delta_a^b \left(\pm k- k_c \right)~,
\ee
different singularities contribute to the JK residue \eqref{Neq2SQCD index} depending on the relative values of $k$ and $k_c$.  
 Without loss of generality, we can consider $k\geq 0$, $k_c \geq 0$. There are four distinct cases:
\begin{itemize}
\item If $k=k_c=0$, we have a $U(N_c)$ gauge theory with $N_f=N_c$ and no Chern-Simons term. The  theory has a quantum Coulomb branch spanned by the gauge-invariant monopole operators $T^\pm$.  Aharony duality  \cite{Aharony:1997gp} provides a dual description with a $U(N_f-N_c)$ gauge group. 

\item If $k>k_c \geq 0$, the CS interactions lifts the Coulomb branch. The dual theory with gauge group $U(k+ N_f- k_c - N_c)$ is known as Giveon-Kutasov duality when $k_c=0$ \cite{Giveon:2008zn}.

\item If $k_c > k \geq 0$, there is no quantum Coulomb branch and the dual theory has a $U(N_f-N_c)$ gauge group \cite{Benini:2011mf}. 

\item If $k=k_c >0$, the theory has ``half'' a quantum Coulomb branch, spanned by $T^+$.

\end{itemize}
The dualities with $k_c\neq 0$ were introduced in \cite{Benini:2011mf}.
All the  dualities   of  \cite{Giveon:2008zn, Benini:2011mf} for YM-CS-matter theories with unitary gauge groups can be derived from the Aharony duality  through real mass deformations. Nonetheless,  it will be instructive to compute the twisted index in every case, especially because it is rather subtle to take the necessary decoupling limits between different values of $[k, N_c, N_f, N_a]$ at the level of the index.  For completeness, we consider those real mass deformations---in flat space---in Appendix \ref{appendix: Seiberg duals}, where we also re-derive the relative global CS levels that are crucial for precise checks of these dualities \cite{Benini:2011mf,Closset:2012vp}.

For definiteness, we  choose the mixed CS levels \eqref{kgA and kgR} to be:
\be\label{kgA explicit}
k_{gA}=  \begin{cases}
-k_c &\quad {\rm if} \quad k \geq k_c~,\\
-k &\quad {\rm if} \quad k \leq k_c~,
\end{cases}\qquad\quad
k_{gR}=  \begin{cases}
-k_c \,(r-1) &\quad {\rm if} \quad k \geq k_c~,\\
-k\, (r-1) &\quad {\rm if} \quad k \leq k_c~.
\end{cases}
\ee
There are the levels obtained by real mass deformations  from SQCD$[0, N_c, n_f, n_f]$ at $k_{gA}=k_{gR}=0$ (see Appendix \ref{appendix: Seiberg duals}).

\subsection{The Bethe equations of 3d $\CN=2$ SQCD and Seiberg duality}\label{subsec: bethe and duals}
Assuming $k_c \geq 0$, $k \geq 0$,  let us define the `characteristic polynomial':
\be\label{Px full}
P(x) = \prod_{i=1}^{N_f} (x-y_i ) - q\, y_A^{Q_+^A} x^{k+ k_c} \prod_{j=1}^{N_a} (x-\t y_j)~,
\ee
of degree:
\be
n  \equiv \deg(P)= \begin{cases}
k+ {N_f+ N_a\ov 2} &\qquad {\rm if} \qquad k \geq k_c~,\\
N_f &\qquad {\rm if} \qquad k \leq k_c~.
\end{cases}
\ee
It is easy to verify that the Bethe equations of  SQCD$[k, N_c, N_f, N_a]$ are given by:
\be\label{bethe eq sqcd}
P(x_a)=0~, \; \quad a=1, \cdots, N_c~, \qquad \qquad x_a \neq x_b \quad {\rm if} \quad a\neq b~.
\ee
Let $\{\h x_\alpha\}_{\alpha=1}^n$ be the set of roots of \eqref{Px full}, which are distinct for generic values of the parameters.
The set $ \CS_{\rm BE}$ of distinct solutions to the Bethe equations is the set of all  unordered subsets  $\{\h x_a\}_{a=1}^{N_c} \subset \{\h x_\alpha\}$ of $N_c$ elements.

We can easily perform the sum over gauge fluxes in  \eqref{Neq2SQCD index}. For definiteness, let us choose $\eta= (1, \cdots, 1)$ in the JK residue. The contributing poles are at $x_a= y_i$ and $x_a= \infty$, where the latter singularities contribute only if $k>k_c$. We first perform the sum over the fluxes $\m_a \geq M$, with $M\in \Z$ some fixed integer depending on the background fluxes, which cancels out of the computation. The geometric series for each $\m_a$ reproduces the characteristic polynomial \eqref{Px full}:
\be
\sum_{\m_a=M}^\infty e^{2 \pi i \d_{u_a} \CW\, \m_a} = (e^{2 \pi i \d_{u_a} \CW})^M {\prod_{i=1}^{N_f} (x_a-y_i)\ov P(x_a)}~,
\ee
and the   resulting contour integral has contributions from the poles at the roots of $P(x)$.  One can check that these roots go to $\h x_\alpha \rightarrow y_i$ and $\h x_\alpha \rightarrow \infty$ in the limit $q\rightarrow 0$.
Using the identity:
\be\label{dP H identity}
\d_x P(\h x_\alpha) = - {\h x_\alpha}^{-1} \prod_{i=1}^{N_f} (\h x_\alpha- y_i) \h H(\h x_\alpha)~,
\ee
for any root $\h x_\alpha$ of $P(x)$, we can rewrite the twisted index as:
\be\label{Zsqcd as sum}
Z^{{\rm SQCD}\, [k, N_c, N_f, N_a] }_{\Sigma_g\times S^1}(q, y, \t y) = \sum_{\h x \,\in\, \CS_{\rm BE}}  \CU(\h x)\,  \CH(\h x)^{g-1}~,
\ee
as anticipated in section \ref{sec: bethe equ}. This directly implies the formula \eqref{SQCD index} for the Witten index.
Here we have:
\be\label{CU sqcd}
\CU(x) = \prod_{a=1}^{N_c}\left[{ x_a^{n_T- Q_-^A \n_A}   \prod_{i=1}^{N_f}(y_i-x_a)^{\n_i} \prod_{i=1}^{N_f} y_i^{-\half\n_i} \prod_{j=1}^{N_a} \t y_j^{\half \t \n_j}  \ov \prod_{j=1}^{N_a} (x_a-\t y_j)^{\t \n_j} }    \right]~,
\ee
\be\label{CH sqcd}
\CH(x) =  \prod_{a=1}^{N_c}\left[{x_a^{- (r_- -1)} y_A^{-k_c (r-1)}\,  (-1)^{N_f-1} \,  \d_x P(x_a) \ov   \prod_{i=1}^{N_f}(y_i-x_a)^{r}  \prod_{j=1}^{N_a} (x_a-\t y_j)^{r-1} }    \right]
 \prod_{\substack{a,b=1\\ a\neq b}}^{N_c} {1\ov x_a-x_b}~,
\ee
with $Q_-^A$ and $r_-$ defined in \eqref{def QA rpm}. Note that  we used \eqref{dP H identity} to massage $\CH(x)$ in \eqref{CH sqcd}.

\begin{table}[t]
\centering
\be\nn
\begin{array}{c|c|ccccc}
    &  U(n-N_c)& SU(N_f) & SU(N_a)  & U(1)_A &  U(1)_T & U(1)_R  \\
\hline
q_j   & \bm{{n-N_c}}  &  \bm{1}& \bm{\overline{N_a}}  & -1   & 0   &1-r \\
{\t q}^i        & \bm{\overline{n-N_c}}& \bm{N_f} & \bm{1}& -1   & 0   &1-r \\
{M^j}_i  & \bm{1} &\bm{\overline{N_f}}& \bm{N_a} &  2   & 0   &2 r \\
\end{array}
\ee
\caption{Charges of the chiral multiplets for the Seiberg dual of 3d $\CN=2$ SQCD. There is also one extra singlet $T^\pm$ if $k= \pm k_c$ (or both,  if $k=k_c=0$), corresponding to the Coulomb branch operator of the `electric' theory.}
\label{tab:SQCD dual charges}
\end{table}
The expression  \eqref{Zsqcd as sum}  is the most convenient to study Seiberg dualities. The dual theory has a gauge group $U(n-N_c)$ with dual matter fields as indicated in Table \ref{tab:SQCD dual charges}. Let $x_{\b a}$ ($\b a= 1, \cdots, n-N_c$) denote the gauge fugacities for the dual gauge group. The corresponding Bethe equations take the form:
\be\label{dual bethe eqs sqcd}
P_D(x_{\b a})=0~, \quad \b a=1, \cdots, n-N_c~,  \qquad \qquad x_{\b a} \neq x_{\b b} \quad {\rm if} \quad \b a\neq \b b~.
\ee
with%
~\footnote{Note that we used the freedom to multiply $q$ and $q_D$ by a sign. We chose $q\rightarrow (-1)^{N_f+N_c-1} q$ in \eqref{Zcl sqcd}, and similarly $q_D \rightarrow (-1)^{N_a + n- N_c-1} q_D$ in the dual theory.} 
\be
 P_D(x) =  \prod_{j=1}^{N_a} (x-\t y_i ) - q_D\, y_A^{-Q_+^A} x^{-(k+ k_c)} \prod_{i=1}^{N_f} (x-y_i)~.
\ee
We directly see that $P(x)$ and $P_D(x)$ have the same roots $\{\h x_\alpha\}_{\alpha=1}^n$ if $q_D= q^{-1}$.  Indeed,  the duality identifies the topological currents of the $U(N_c)$ and $U(N_f-N_c)$ gauge groups, with a relative sign. If we denote by $U(1)_{T_D}$ the topological current of $U(N_f-N_c)$, we have $T_D= - T$, and therefore:
\be\label{rel q qD aharony}
q_D= q^{-1}~, \qquad \quad \n_{T_D}= - \n_T~,
\ee
for the fugacities and background fluxes, respectively.  We denote by $ \CS_{\rm BE}^D$ the set of distinct solutions to the dual Bethe equations \eqref{dual bethe eqs sqcd}, which is the set of all 
unordered subsets  $\{\h x_{\b a}\}_{\b a=1}^{n-N_c} \subset \{\h x_\alpha\}$ of $n-N_c$ elements.

The twisted index of the dual theory takes the form:
\be\label{Z dual}
Z^{\rm dual}_{\Sigma_g \times S^1}(q, y, \t y) = Z^{\text{CS}}_{\Sigma_g \times S^1} Z^{\rm singlets}_{\Sigma_g \times S^1} \, Z^{{\rm SQCD}\, [-k, n-N_c, N_a, N_f] }_{\Sigma_g\times S^1}(q^{-1},\t y, y) ~.
\ee
The first factor is the contribution from the relative flavor CS terms, which are discussed in more details below and in Appendix \ref{appendix: Seiberg duals}. The second factor $Z^{\rm singlets}_{\Sigma_g \times S^1}$ is the contribution from the gauge-singlet fields that are part of the dual theory. This includes  the contribution from the  `mesonic' gauge-singlet fields ${M^j}_i$, which reads:
\be\label{def Zm}
Z_{\Sigma_g \times S^1}^{M} = \cu_M (\ch_M)^{g-1}~, 
\ee
where we defined:
\be\label{def uM hM}
\cu_M \equiv  \prod_{i=1}^{N_f}\prod_{j=1}^{N_a} \left({y_i^\half \t y_j^\half \ov y_i-\t y_j} \right)^{-\n_i +\t\n_j}~, \qquad\quad
\ch_M \equiv  \prod_{i=1}^{N_f}\prod_{j=1}^{N_a} \left({1 \ov y_i-\t y_j} \right)^{2r-1}~.
\ee
We have $Z^{\rm singlets}_{\Sigma_g \times S^1}=Z_{\Sigma_g \times S^1}^{M}$ if $k\neq k_c$, while in the limiting case $k=k_c >0$ (or $k=k_c=0$) we must also include the contribution from an extra singlet   $T^+$  (or two extra singlets $T^\pm$, respectively).
The last factor in \eqref{Z dual} is the contribution from the dual  gauge group $U(n-N_c)$ with its charged matter fields. Note the exchange of the fugacities $y$ and $\t y$  in \eqref{Z dual}.

 By a similar reasoning as above, we can show that the gauge contribution in \eqref{Z dual} can be expressed as a sum over the solutions to the dual Bethe equations:
\be
Z^{{\rm SQCD}\, [-k, n-N_c, N_a, N_f] }_{\Sigma_g\times S^1}(q^{-1},\t y, y) = \sum_{\h x_D \,\in\, \CS_{\rm BE}^D}  \CU_D(\h x_D)\,  \CH_D(\h x_D)^{g-1}~,
\ee
with
\be\label{CU sqcd dual}
\CU_D(x_D) = \prod_{\b a=1}^{n-N_c}\left[{ x_{\b a}^{-n_T+ Q_-^A \n_A}   \prod_{j=1}^{N_a}(\t y_j-x_{\b a})^{\t \n_i} \prod_{i=1}^{N_f} y_i^{\half\n_i} \prod_{j=1}^{N_a} \t y_j^{-\half \t \n_j}  \ov \prod_{i=1}^{N_f} (x_{\b a}- y_i)^{\n_i} }    \right]~,
\ee
\be\label{CH sqcd dual}
\CH_D(x_D) =  \prod_{{\b a}=1}^{n-N_c}\left[{x_{\b a}^{ (r_- -1)} y_A^{k_c r - Q_+^A}\, q^{-1}\,  (-1)^{N_a} \,  \d_x P(x_{\b a}) \ov   \prod_{i=1}^{N_f}(x_{\b a}-y_i)^{-r}  \prod_{j=1}^{N_a} (\t y_j-x_{\b a})^{-r+1} }    \right]
 \prod_{\substack{{\b a},{\b b}=1\\ {\b a}\neq {\b b}}}^{n-N_c} {1\ov x_{\b a}-x_{\b b}}~.
\ee
This can be obtained from  \eqref{CU sqcd}-\eqref{CH sqcd} by exchanging $i$ and $j$ indices together with the substitutions:
\be
k\rightarrow -k~, \quad N_c\rightarrow n-N_c~, \quad  N_a  \leftrightarrow N_f~, \quad y_i \leftrightarrow \t y_j~,\quad r\rightarrow 1-r~, \quad q\rightarrow q^{-1}~,
\ee
and similarly for the background fluxes. 
The identity of twisted indices across Seiberg duality can be shown  by replacing the set of $N_c$ roots  $\h x= \{\h x_a\}_{a=1}^{N_c}$ of $P(x)$  by its complement $\h x_D =\{\h x_{\b a}\}_{\b a=1}^{n-N_c}\subset \{\h x_\alpha\}$.  In Appendix  \ref{appendix: proof equality indices}, we prove that:
\be\label{U H U H dual}
\CU(\h x)= \cu\; \CU_D(\h x_D)~,\qquad \qquad \CH(\h x)=\ch\; \CH_D(\h x_D)~, 
\ee
 for any partition $\{\h x_\alpha\}= \h x \cup \h x_D$ of the roots of $P(x)$.
The quantities $\cu$ and $\ch$ only depend on the fugacities (and fluxes) for the global symmetries, and are such that:
\be\label{ch as ZcsZsin}
\cu \, \ch^{g-1}= Z^{\text{CS}}_{\Sigma_g \times S^1} Z^{\rm singlets}_{\Sigma_g \times S^1}~,
\ee
with an extra (ambiguous) sign included in the definition of $Z^{\text{CS}}_{\Sigma_g \times S^1}$.   The exact expression for $\cu$ and $\ch$ are given in Appendix \ref{appendix: proof equality indices}. The relations \eqref{U H U H dual}-\eqref{ch as ZcsZsin} directly imply the equality of  twisted indices for the dual theories: 
\be\label{dual rel general}
Z^{{\rm SQCD}\, [k, N_c, N_f, N_a] }_{\Sigma_g\times S^1}(q, y, \t y) \, = \, Z^{\rm dual}_{\Sigma_g \times S^1}(q^{-1}, y, \t y)~.
\ee
These results also imply the equality of Wilson loop correlators. For any Wilson loop $W$ of the $U(N_c)$ theory, there exists a dual Wilson loop $W_D$ such that:
\be\label{W WD sqcd}
W(\h x) = W_D(\h x_D)~, 
\ee
where $\h x$ and $\h x_D$ of $P(x)$ are complementary sets of roots defined as above.
Using the expression \eqref{Index and BE res}, we easily see that  the dual correlation functions on $\Sigma_g\times S^1$ must coincide:
\be
\langle  W \rangle_g = \langle  W_D \rangle_g^{\rm dual}~.
\ee
We will discuss this duality map in more details in subsection \ref{subsec: wilson loops sqcd} below.

\subsection{Aharony duality ($k=k_c=0$)}
Consider SQCD with $k=k_c=0$. This is a  $U(N_c)$ YM theory  with $N_f$ pairs of fundamental and antifundamental chiral multiplets $Q_i, \t Q^j$ and a vanishing superpotential. We choose the mixed gauge-flavor CS terms $k_{gA}=k_{gR}=0$ according to \eqref{kgA explicit}. We also set all the global (flavor and $U(1)_R$) CS levels to zero.

The dual theory is a $U(N_f-N_c)$ YM theory with $N_f$ fundamental and antifundamental chiral multiplets $\t q_j, q^i$, $N_f^2$  singlets ${M^j}_i$ transforming under $SU(N_f)\times SU(N_a)$, and two extra singlets $T^\pm$ charged under the topological symmetry $U(1)_T$. These fields interact through the superpotential \eqref{W aharony} given in Appendix \ref{app subsec: aharony}. 
All the gauge and global CS levels vanish as well.
The gauge and global charges of all the dual matter fields are summarized in Table \ref{tab: Aharony duality charges}.
\begin{table}[t]
\centering
\be\nn
\begin{array}{c|c|ccccc}
    & U(N_f-N_c)& SU(N_f) & SU(N_f)  & U(1)_A &  U(1)_T & U(1)_R  \\
\hline
q^i   &\bm{N_f-N}&    \bm{1} & \bm{\overline{N_f}} & -1   & 0   &1-r \\
\t q_j       & \bm{\overline{N_f-N_c}}& \bm{N_f}    & \bm{1}    & -1   & 0   &1-r \\
{M ^j}_i      & \bm{1} &\bm{\overline{N_f}}& \bm{N_f} &  2   & 0   &2 r \\
T^+       & \bm{1} & \bm{1} & \bm{1} & -N_f   & 1   & -N_f(r-1) -N_c +1\\
T^-     & \bm{1} & \bm{1} & \bm{1} & -N_f   & -1   & -N_f(r-1) -N_c +1
\end{array}
\ee
\caption{Charges of the matter fields in the Aharony dual theory. }
\label{tab: Aharony duality charges}
\end{table}
The singlets ${M^j}_i$  and $T^\pm$ are identified with the gauge-invariant mesons $\t Q^j Q_i$  and with the lowest gauge invariant monopole operators of $U(N_c)$, respectively.

The $\Sigma_g\times S^1$ partition function of the  electric theory is given by:
\be
 Z^{{\rm SQCD}\, [0, N_c, N_f, N_f] }_{\Sigma_g\times S^1}(q, y, \t y)~, \qquad\qquad (k_{gA}=k_{gR}=0)~,
\ee
a special case of the  SQCD index   \eqref{Neq2SQCD index}. The partition function of the magnetic theory is given by:
\be\label{magn Aharony def}
Z^{\rm dual}_{\Sigma_g \times S^1}(q, y, \t y) = (-1)^{\n_T+ (g-1)(N_f-N_c)}\; Z^{\rm singlets}_{\Sigma_g \times S^1} \, Z^{{\rm SQCD}\, [0, N_f-N_c, N_f, N_f] }_{\Sigma_g\times S^1}(q^{-1},\t y, y) ~,
\ee
with the singlet contribution:
\bea\label{Zsin aha}
&Z^{\rm singlets}_{\Sigma_g \times S^1} = \prod_{i=1}^{N_f}\prod_{j=1}^{N_f} \left(y_i^\half \t y_j^\half \ov y_i-\t y_j \right)^{-\n_i+\t n_j+(g-1)(2r-1)}  \times\,\cr
&\left( q^{\half} y_A^{-\half N_f}\ov 1- q y_A^{-N_f}\right)^{\n_{T} - N_f \n_A+ (g-1)(r_+-1)} \left( q^{-\half} y_A^{-\half N_f}\ov 1- q^{-1} y_A^{-N_f}\right)^{-\n_{T} - N_f \n_A+ (g-1)(r_- -1)}  ~,
\eea
with $r_+= r_- = -N_f(r-1)- N_c+1$ the $R$-charge  of the  gauge-singlet chiral multiplets $T^\pm$. The first line in \eqref{Zsin aha} is the meson contribution \eqref{def Zm}-\eqref{def uM hM} and the second line is the contribution from $T^+$ and $T^-$, respectively.
To complete the proof of the equality \eqref{dual rel general} for the twisted indices, we need to show that:
\be
\cu\, \ch^{g-1} = (-1)^{\n_T+ (g-1)(N_f-N_c)}\; Z^{\rm singlets}_{\Sigma_g \times S^1}~.
\ee
One can check that this follows from the formula \eqref{u h explicit} in Appendix \ref{appendix: proof equality indices} when $k=0$, $N_f=N_a= n$.

\subsection{Duality for $k>k_c \geq 0$}\label{subsec: sqcd dual 2}
Consider SQCD$[k,N_c,N_f,N_a]$ with CS level $k>k_c\geq 0$. We choose the mixed CS levels $k_{gA}$, $k_{gR}$ according to \eqref{kgA explicit}. The dual gauge theory has a gauge group
${U(k + \half(N_f+N_a) -N_c)}$
 at CS level $-k$,  with mixed gauge-$U(1)_A$ and gauge-$R$ CS levels:
 \be
 k^D_{gA}= k_c~, \qquad \qquad
 k^D_{gR}= k_c r~.
 \ee
The dual matter sector consists of the dual charged chiral multiplets $q_j$, $\t q^i$ and the $N_f N_a$ gauge-singlet mesons ${M^j}_i$, with the standard Seiberg dual superpotential $W= \t q M q$. The gauge and global charges are summarized in Table \ref{tab:SQCD dual charges} above.

To fully state the duality, we need to specify the relative CS levels for the global symmetry group \eqref{flavor sym SQCD}. In Appendix \ref{appendix: Seiberg duals}, we show that:
\be\label{rel CS 02}
\Delta k_{SU(N_f)} = \half(k+k_c)~, \qquad 
\qquad   \Delta k_{SU(N_a)} = \half(k-k_c)~,
\ee
and
\bea\label{rel CS 2}
&\Delta k_{AA}= {N_f+ N_a\ov 2}n- 2 N_f N_a~,\qquad
&&\Delta k_{TT}=-1  \cr
&\Delta k_{AR}=  {N_f+ N_a\ov 2}(n-N_c)-  N_f N_a + (r-1)\Delta k_{AA}~,  \;
&& \Delta k_{AT}= \Delta k_{TR}=0~,
\eea
with $n= k + \half (N_f+N_a)$.
Here we omited $\Delta k_{RR}$ because it does not enter the $\Sigma_g\times S^1$ partition function.
Assembling all the pieces, the twisted index  of the Seiberg dual theory is given by:
\be\label{Z dual 2}
Z^{\rm dual}_{\Sigma_g \times S^1}(q, y, \t y) = Z^{\text{CS}}_{\Sigma_g \times S^1} Z^{M}_{\Sigma_g \times S^1} \, Z^{{\rm SQCD}\, [-k, n-N_c, N_a, N_f] }_{\Sigma_g\times S^1}(q^{-1},\t y, y)~,
\ee
with $Z_{\Sigma_g \times S^1}^{M}$ defined in \eqref{def Zm}, and:
\be
Z^{\text{CS}}_{\Sigma_g \times S^1}= (-1)^{n_\ast}  \,  \prod_{i=1}^{N_f} y_i^{{s_i}{\Delta k_{SU(N_f)}}} \,\prod_{j=1}^{N_f} y_i^{{\t s_j}{\Delta k_{SU(N_a)}}} \;  q^{\Delta k_{TT} n_T} y_A^{\Delta k_{AA} n_A + \Delta k_{AR} (g-1)}~,
\ee
with the relative CS levels \eqref{rel CS 02}-\eqref{rel CS 2}.
Here  $(-1)^{n_\ast}$ is an unimportant sign, and we defined 
the $SU(N_f)\times SU(N_a)$ fluxes  $s_i = \n_i +\n_A$ and $\t s_j= \n_j- \n_A$. Using the results of Appendix \ref{appendix: proof equality indices}, one can check that \eqref{ch as ZcsZsin} holds, which completes the proof of the equality of twisted indices in this case.

\subsection{Duality for $k_c>k\geq 0$}\label{subsec: sqcd dual 3}
Consider SQCD$[k,N_c,N_f,N_a]$ with non-negative CS level $ k < k_c$, with  the mixed CS levels $k_{gA}$, $k_{gR}$ given in \eqref{kgA explicit}. The dual gauge theory has gauge group:
$U(N -N_c)$   at CS level $-k$, and with mixed gauge-$U(1)_A$ and gauge-$R$ CS levels:
 \be\label{kmixed dual 3}
 k^D_{gA}= k~, \qquad \qquad
 k^D_{gR}= k r~.
 \ee
The dual matter sector is like in the last subsection, as  summarized in Table \ref{tab:SQCD dual charges} above. The relative CS levels for this duality are:
\bea\label{rel CS 3}
&\Delta k_{SU(N_f)} =  k~, \qquad  \qquad&&  \Delta  k_{SU(N_a)} =0~,\cr
&\Delta  k_{AA}= 3 k N_f~, \qquad &&\Delta k_{TT}=0~, \cr
&\Delta k_{AR}= k (N_c+ N_f) +(r-1) \Delta k_{AA}~, \qquad \quad
&&\Delta k_{AT}= - N_f~, \cr
& \Delta k_{TR} = - N_c + (r-1)  \Delta k_{AT}~,
\eea
as we explain in Appendix \ref{appendix: Seiberg duals}. 
The dual twisted index reads:
\be\label{Z dual 3}
Z^{\rm dual}_{\Sigma_g \times S^1}(q, y, \t y) = Z^{\text{CS}}_{\Sigma_g \times S^1} Z^{M}_{\Sigma_g \times S^1} \, Z^{{\rm SQCD}\, [-k, N_f -N_c, N_a, N_f] }_{\Sigma_g\times S^1}(q^{-1},\t y, y)~,
\ee
with $Z_{\Sigma_g \times S^1}^{M}$ defined in \eqref{def Zm}, and:
\bea\label{ZCS max chiral}
&Z^{\text{CS}}_{\Sigma_g \times S^1}&=& (-1)^{n_\ast}  \,  \prod_{i=1}^{N_f} y_i^{{s_i}{\Delta k_{SU(N_f)}}}\;  q^{\Delta k_{TT} n_T + \Delta k_{AT} \n_A + \Delta k_{TR}(g-1)} \cr
&&&\quad\times y_A^{\Delta k_{AA} n_A +\Delta k_{AT} n_T+ \Delta k_{AR} (g-1)}~,
\eea
with the relative CS levels \eqref{rel CS 3}, while  $(-1)^{n_\ast}$ is another unimportant sign. One can check that \eqref{ch as ZcsZsin} holds in this case as well.

\subsection{Duality for $k=k_c>0$}\label{subsec: sqcd dual 4}
The final case  to consider is SQCD$[k_c, N_c, N_f, N_a]$ with CS level $k=k_c >0$, the limiting case between subsections \ref{subsec: sqcd dual 2} and \ref{subsec: sqcd dual 3}. The dual gauge group is a $U(N_f-N_c)$ gauge group at CS level $-k$ and  mixed CS levels \eqref{kmixed dual 3}. In addition to the dual charged multiplets and mesons ${M^j}_i$, there is an extra singlet $T^+$ and a superpotential \eqref{W aharony 2}.
The relative CS levels for this duality  are:
\bea\label{rel CS 4}
&\Delta k_{SU(N_f)} =  k~, \qquad  \qquad&&  \Delta  k_{SU(N_a)} =0~,\cr
&\Delta  k_{AA}= 3 k N_f- \half N_f^2~, \qquad &&\Delta k_{TT}=-\half~, \cr
&\Delta k_{AR}= k (N_c+ N_f)- \half N_c N_f +(r-1) \Delta k_{AA}~, \qquad \quad
&&\Delta k_{AT}= - \half N_f~, \cr
& \Delta k_{TR} = - \half N_c + (r-1)  \Delta k_{AT}~.
\eea
The dual twisted index reads:
\be\label{Z dual 4}
Z^{\rm dual}_{\Sigma_g \times S^1}(q, y, \t y) = Z^{\text{CS}}_{\Sigma_g \times S^1} Z^{\rm singlets}_{\Sigma_g \times S^1} \, Z^{{\rm SQCD}\, [-k, n-N_c, N_a, N_f] }_{\Sigma_g\times S^1}(q^{-1},\t y, y)~.
\ee
The singlet contribution includes the contribution from $T^+$:
\be
 Z^{\rm singlets}_{\Sigma_g \times S^1} =  Z^M_{\Sigma_g \times S^1}\, \left( q^{\half} y_A^{-\half N_f}\ov 1- q y_A^{-N_f}\right)^{\n_{T} - N_f \n_A+ (g-1)(r_+-1)}~, 
\ee
with $r_+ = -\left(k +\half (N_f+N_a)\right) - N_c+1$ the $R$-charge of $T^+$ in this case. The factor $ Z^{\text{CS}}_{\Sigma_g \times S^1}$ in \eqref{Z dual 4} is given by \eqref{ZCS max chiral} with relative CS levels \eqref{rel CS 4}. 
 One can check that \eqref{ch as ZcsZsin} holds here as well, which completes the proof of \eqref{dual rel general}.

\subsection{Wilson loop algebra and the duality map}\label{subsec: wilson loops sqcd} 
As an illustration of the general discussion of section \ref{subsec: Wilson loop alg}, let us consider the quantum algebra of Wilson loops in SQCD$[k,N_c, N_f, N_a]$.  
A particularly interesting case is for $k=k_c=0$, which we consider in some more details below.
The Wilson loop algebra for SQCD  was studied previously in \cite{Kapustin:2013hpk} by considering the theory on $S^3$, and we  follow a similar logic on $\Sigma_g \times S^1$. As we emphasized in section \ref{subsec: Wilson loop alg}, the Wilson loop algebra is always   encoded in the Bethe equations of the theory on $\R^2\times S^1$.

\subsubsection{Wilson loops and Seiberg duality}
Wilson loops in a $U(N_c)$ theory are in one-to-one correspondence with symmetric Laurent polynomials in the coordinates $x_a$:
\be
W(x)\in \C[x_1, x_1^{-1},  \cdots, x_\rk, x_\rk^{-1}]^{S_{N_c}}~,
\ee
which are in one-to-one correspondence with Young tableaux graded by the $U(1)\subset U(N_c)$ charge ${\bf q} \in \Z$. For instance, Wilson loops in the fundamental and antifundamental representations correspond to:
\be
W_{{\tiny\yng(1)}_{+1}}(x) = \sum_{a=1}^{N_c} { x_a}~,\qquad\qquad
W_{\overline{\tiny\yng(1)}_{-1}}(x) = \sum_{a=1}^{N_c}  {1\ov x_a}~.
\ee
Products of Wilson loops are given by the corresponding tensor products of $U(N_c)$ representations. 
Consider first the representations $\FR$ with ${\bf q}\geq 0$, corresponding to all the symmetric polynomials $W(x) \in \C[x_1,  \cdots, x_{N_c}]^{S_{N_c}}$, which form a subalgebra. They are generated by the elementary symmetric polynomials:
\be
s_l^{(N_c)}(x) = \sum_{1\leq a_1 < \cdots< a_l \leq N_c} x_{a_1} x_{a_2}\cdots x_{a_l}~, \quad\qquad l=0, \cdots, N_c~,
\ee
which correspond to the Young tableaux with $l$ vertical boxes:
\be\label{s def sqcd}
s_0^{(N_c)}(x)  =  1~, \qquad  s_1^{(N_c)}(x)  = {\tiny\yng(1)}~, \qquad  s_2^{(N_c)}(x)  = {\tiny\yng(1,1)}~, \qquad  \cdots~.
\ee
Let us define the generating function:
\bea
&Q(z) = \prod_{a=1}^{N_c} (z-x_a) &=& \;\sum_{l=0}^{N_c} (-1)^{l} z^{N_c-l} \; s_l^{(N_c)}(x)\cr
&&=&\; z^{N_c} - z^{N_c-1} \, {\tiny\yng(1)} +  z^{N_c-2}\,  {\tiny\yng(1,1)} -\cdots + (-1)^{N_c} x_1\cdots x_{N_c}~,
\eea
where we identify any irreducible Wilson loop $W(x)$  with its corresponding Young tableau.

The quantum Wilson loop algebra is governed by the Bethe equations \eqref{bethe eq sqcd}, which are given  in terms of the polynomial $P(x)$ of degree $n$ \eqref{Px full}. The quantum algebra relations $f=0$ are the relations satisfied by any solution to the Bethe equations---that is, we have $f(\h x)=0$ for any set $\h x=\{\h x_a\}_{a=1}^{N_c}$ of $N_c$ distinct roots of $P(x)$. These relations can be conveniently written in a  gauge-invariant form \cite{Gaiotto:2013sma,Benini:2014mia}  as:
\be\label{full rels sqcd}
P(z) - C(q) \, Q(z) \,Q^D(z)=0~,
\ee
where we defined:
\be
C(q) = \begin{cases}
1 -q\, y_A^{-N_f} &\qquad {\rm if}\qquad k=k_c\geq 0 \cr
-q\, y_A^{-N_f} &\qquad {\rm if}\qquad k>k_c\geq 0 \cr
1 &\qquad {\rm if}\qquad k_c>k \geq 0 
\end{cases}~,
\ee
so that $P(z)/C(q)$ is monic in $z$.
Here $Q^D(z)$ is an auxilliary monic polynomial of degree $n-N_c$ in $z$. Recalling that the Bethe equations of the Seiberg dual theory with $U(n-N_c)$ gauge group are given in terms of the same polynomial $P(x)$:
\be\label{dual bethe eqs sqcd 2}
P(x_{\b a})=0~, \quad \b a=1, \cdots, n-N_c~,  \qquad \qquad x_{\b a} \neq x_{\b b} \quad {\rm if} \quad \b a\neq \b b~,
\ee
we are led to identify $Q^D(z)$ as the generating function for the  dual Wilson loops $W^D(x_D)$ with non-negative $U(1)$ charge:
\be
Q_D(z) = \prod_{\b a=1}^{n-N_c} (z-x_{\b a}) = \;\sum_{p=0}^{n-N_c} (-1)^{p} z^{n-N_c-p} \;s_p^{(n-N_c)}(x_D)~.
\ee
We also use the notation:
\be\label{sD def sqcd}
s_0^{(n-N_c)}(x_D)  =  1~, \qquad  s_1^{(n-N_c)}(x_D)  = {\tiny\yng(1)}^D~, \qquad  s_2^{(n-N_c)}(x_D)  = {\tiny\yng(1,1)}^D~, \qquad  \cdots~.
\ee
Expanding  both sides of \eqref{full rels sqcd} in $z$, one finds $n$ relations between the quantities \eqref{s def sqcd} and \eqref{sD def sqcd}.  Solving for $s_p^{(n-N_c)}(x_D)$ in terms of $s_l^{(N_c)}(x)$, we are left with the relations satisfied by the Wilson loops with ${\bf q}\geq 0$. To obtain the full quantum algebra of Wilson loops (corresponding to Laurent polynomials instead of polynomials), we just need to adjoin the elements $x_a^{-1}$. Following \cite{Kapustin:2013hpk}, we can write $P(x)$ as 
\be
P(x) = C(q) x^n + c_{n-1} x^n + \cdots + c_1 x + c_0~, 
\ee
and we have
\be
{1\ov \h x_a} = -{1\ov c_0}\left(C(q)\h x_a^{n-1} + c_{n-1} \h x_a^{n-2} +\cdots + c_1\right)
\ee
for $\{\h x_a\}$ any solution to the Bethe equations. Therefore these elements $x_a^{-1}$ are not independent in the quotient ring, and the quantum algebra \eqref{CAW general}  is the ring of $U(N_c)$ representations with ${\bf q}\geq 0$---labelled by Young tableaux of maximum $N_c$ rows---quotiented by the relations encoded in \eqref{full rels sqcd}.  The quotient ring is finite-dimensional, consisting of Young tableaux with a maximum of $N_c$ rows and $n-N_c$ columns. 

The relations \eqref{full rels sqcd} also encode the duality map \eqref{W WD sqcd} between the Wilson loops $W$ of $U(N_c)$ and the Wilson loops $W_D$ of the dual theory. Seiberg duality then acts as an isomorphism of the quantum Wilson loop algebra \cite{Kapustin:2013hpk}, which is rendered manifest in \eqref{full rels sqcd}.

\subsubsection{Wilson loops in Aharony duality}\label{subsec: Aharony Wilson}
To illustrate the above considerations, let us consider $U(N_c)$ with $k=0$ and $N_f=N_a$ in more details. The characteristic polynomial in this case reads:
\be
P(z) =  \prod_{i=1}^{N_f} (z-y_i ) - q\, y_A^{-N_f} \prod_{j=1}^{N_f} (z-\t y_j)~.
\ee
We have the quantum relations \eqref{full rels sqcd} with $C(q)=1-  q y_A^{-N_f}$. 
Note that we have:
\be
P(z)= \sum_{m=0}^{N_f} (-1)^m z^{N_f-m} \left(s_m^F - q\, y_A^{-N_f} {\t s}_m^F \right)~, 
\ee
where we defined:
\be
s_m^F = s_m^{(N_f)}(y)~, \qquad \qquad {\t s}_m^F = s_m^{(N_f)}(\t y)~,  \qquad \qquad m=0, \cdots, N_f~.
\ee
the elementary symmetric polynomials in the fugacities $y_i$ and $\t y_j$ for the $SU(N_f) \times SU(N_f)\times U(1)_A$ flavor group. We can think of these quantities as `flavor Wilson loops' for the background gauge fields. It follows that the quantum ring relations are given explicitly by:
\be\label{relations W Aharony duality}
\sum_{l=0}^m s_l^{(N_c)}(x) \; s_{m-l}^{(N_f-N_c)}(x_D) = {1\ov 1-  q y_A^{-N_f}} \left(s_m^F - q\, y_A^{-N_f} {\t s}_m^F \right)~, \quad  m=1, \cdots, N_f~.
\ee
Here it is understood that $s_l^{N_c}(x)= 0$ for $l> N_c$ and $s_p^{(N_f-N_c)}(x_D)= 0$ for $p > N_f-N_c$. 
For instance, the first relation reads:
\be
 {\tiny\yng(1)} +  {\tiny\yng(1)}^D = {1\ov 1  - q\, y_A^{-N_f} } \left(\sum_{i=1}^{N_f} y_i   - q\, y_A^{-N_f}   \sum_{j=1}^{N_f} \t y_j\right)~.
\ee
This is the relation between the Wilson loop $W_ {\tiny\yng(1)}$ in the fundamental  representation of $U(N_c)$ and the dual Wilson loop in the fundamental representation of $U(N_f-N_c)$.

The relations \eqref{relations W Aharony duality} have an interesting property in the limit $y_i= \t y_j$ ($i=j$), when we have :
\be\label{relations W Aharony duality bis}
\sum_{l=0}^m s_l^{(N_c)}(x) \; s_{m-l}^{(N_f-N_c)}(x_D) =s_m^F~, \qquad \quad m=1, \cdots, N_f~.
\ee
The number of summands in $x$, $x_D$ or $y$ is equal on either side; if we set $x_a= x_{\b a}=y_i=1$, we have a relation between dimensions of gauge and flavor representations.

\paragraph{Example: $U(3)$ with $N_f=5$.}  To illustrate the above, let us work out the case $N_c=3$ and $N_f=5$. We take $y_i=y_j=1$ for simplicity. In that case, the equations \eqref{relations W Aharony duality} read:
\bea\label{example U3}
 {\tiny\yng(1)}^D\; \;+\; \; {\tiny\yng(1)} &=& 5~, \cr
 {\tiny\yng(1,1)}^D \;\;+\;  \;{\tiny\yng(1)}^D \otimes  {\tiny\yng(1)}\;\;+\;  \; {\tiny\yng(1,1)} &=&10~,\cr
  {\tiny\yng(1,1)}^D \otimes  {\tiny\yng(1)} \;\;+\;  \; {\tiny\yng(1)}^D \otimes  {\tiny\yng(1,1)}  \;\;+\;  \; {\tiny\yng(1,1,1)}&=& 10~,\cr
   {\tiny\yng(1,1)}^D \otimes  {\tiny\yng(1,1)} \;\;+\; \;  {\tiny\yng(1)}^D \otimes  {\tiny\yng(1,1,1)}&=& 5~, \cr
    {\tiny\yng(1,1)}^D \otimes {\tiny\yng(1,1,1)}  &=& 1~.
\eea
The Aharony dual gauge theory has gauge group $U(2)$. From  the two first lines of \eqref{example U3} we find the duality relations:
\be
 {\tiny\yng(1)}^D= 5-  {\tiny\yng(1)}~, \qquad \qquad  {\tiny\yng(1,1)}^D= 10\;-\; 5\;  {\tiny\yng(1)}\;+\; {\tiny\yng(2)}~,
\ee
between Wilson loops in the dual theories.
We also find the quantum Wilson loop algebra relations:
\bea\label{rel U3 explicit}
{\tiny\yng(3)} &=& 10\;-\; 10 \;{\tiny\yng(1)} \;+\; 5\; {\tiny\yng(2)}~, \cr
{\tiny\yng(3,1)} &=& 5\;-\; 10 \;{\tiny\yng(1,1)} \;+\; 5\; {\tiny\yng(2,1)}~,\cr
{\tiny\yng(3,1,1)} &=&  1\;-\; 10\; {\tiny\yng(1,1,1)}\;+\; 5 \; {\tiny\yng(2,1,1)}~.
\eea
Using these relations repeatedly, any $U(3)$ Young tableaux with more than two columns can be written as a linear combinations of Wilson loops of one or two columns. As a further consistency check, we can verify that the total dimensions of the $U(3)$ representations on both sides of the relations \eqref{rel U3 explicit} agree, as expected from \eqref{relations W Aharony duality bis}.

\subsubsection{Wilson loops in Giveon-Kutasov duality}
As another example, consider the case $k>0$ and $k_c=0$, corresponding to Giveon-Kutasov duality \cite{Giveon:2008zn}. The characteristic polynomial is given by:
\be
P(z) =  \prod_{i=1}^{N_f} (z-y_i ) - q\, y_A^{-N_f} z^k \prod_{j=1}^{N_f} (z-\t y_j)~.
\ee
From \eqref{full rels sqcd}, we easily derive the $k+N_f$ quantum algebra relations:
\be\label{relations W GK duality}
\sum_{l=0}^m s_l^{(N_c)}(x) \; s_{m-l}^{(k+N_f-N_c)}(x_D) = (-1)^{k+1} q^{-1} y_A^{N_f} \, s_{m-k}^F +{\t s}_m^F~,
\ee
for $m=1, \cdots, k+N_f$,  similarly to subsection \ref{subsec: Aharony Wilson}. Here it is understood that $s^F_m=0$ if $m<0$. This case was studied previously in \cite{Kapustin:2013hpk}, where the Bethe equations $P(x_a)=0$ appeared as relations satisfied by BPS Wilson loops on $S^3$.

\paragraph{Example: $U(3)$ with $k=2$ and $N_f=2$.}  The dual theory is a  $U(1)$ theory with CS level $-2$. If we consider $y_i=\t y_j=1$ for simplicity, the relations \eqref{relations W GK duality} give:
\be\label{example U3 GK}
 {\tiny\yng(1)}^D = 2- {\tiny\yng(1)}~, \qquad
 \qquad {\tiny\yng(2,1)} = q^{-1}-1+2 \, {\tiny\yng(1,1)}~, \qquad\qquad
  {\tiny\yng(2,1,1)} = -2 q^{-1}~.
\ee


\section{$\CN=4$ gauge theories and mirror symmetry}  \label{sec: 3d Neq4}
Three-dimensional $\CN=4$ supersymmetric gauge theories are particularly interesting because they admit different choices of topological twisting  \cite{Rozansky:1996bq, Kapustin:2010ag, Gukov:2016gkn}, which are often related to each other by three-dimensional mirror symmetry \cite{Intriligator:1996ex}.  
In this section, we define the $A$- and $B$-twists of $\CN=4$ theories on $S^1\times \Sigma_g$---and a certain $\CN=2^\ast$ deformation thereof. 
We study the corresponding twisted indices and their behavior under mirror symmetry. We also briefly discuss the mirror map between Wilson loop and vortex loop operators following \cite{Assel:2015oxa}.

\subsection{The $A$- and $B$-twist of 3d $\CN=4$ gauge theories}
The 3d $\CN=4$ supersymmetry algebra in flat Euclidean space-time reads:
\be
\{ Q_\alpha^{A\b A}, Q_\beta^{B \b B}\} = 2 \epsilon^{AB}\epsilon^{\b A \b B}\, P_{\alpha\beta}~.
\ee
The eight supercharges $Q_\alpha^{A\b A}$ transform as $({\bf 2}, {\bf 2})$ under the R-symmetry group $SU(2)_H\times SU(2)_C$, and we introduced the indices $A, B= 1, 2$ for $SU(2)_H$ and $\b A, \b B=\b 1, \b 2$ for $SU(2)_C$.
We can preserve half of the supercharges on any three-manifold by twisting the $SU(2)_L$ Lorentz group with either $SU(2)_H$ or $SU(2)_C$ \cite{Rozansky:1996bq}. Let us denote by $U(1)_H \times U(1)_C$ the Cartan subgroup of $SU(2)_H\times SU(2)_C$, and by $H$ and $C$ the corresponding charges. We define the integer-valued $R$-charges:
\be\label{RA RB def}
R_A= 2 H~,\qquad \qquad R_B= 2 C~.
\ee
For a theory on $\Sigma_g \times S^1$, we can identify either $R_A$ or $R_B$ as the $U(1)_R$ symmetry of an $\CN=2$ subalgebra, and proceed as in section \ref{sec: susy n all}.

The $SU(2)_C$ twist is known as the Rozansky-Witten twist \cite{Rozansky:1996bq}.
It preserves four scalar supercharges on any three-manifold:
\be\label{Btwist Qs}
Q_+^{1\b 1}~, \quad Q_+^{2\b 1}~, \quad Q_-^{1\b 2}~, \quad Q_-^{2\b 2}~.
\ee
On $\Sigma_g \times S^1$, we preserve the supersymmetry algebra:
\be
\{Q_+^{A\b 1}, Q_-^{B \b 2}\} = 2 \epsilon^{AB} E~,
\ee
where $E$ is the generator of translation along $S^1$.
This is the algebra of an $\CN=4$ supersymmetric quantum mechanics (QM) with 
$U(1)_C\times SU(2)_H$  R-symmetry \cite{Hori:2014tda}. We call this $\Sigma_g \times S^1$ background the $B$-twist. It  corresponds to a topological twist along $\Sigma_g$ by the $R$-charge $R_B$ in \eqref{RA RB def}.
Similarly, the $SU(2)_H$ twist preserves the four scalar supercharges:
\be\label{Atwist Qs}
Q_+^{1\b 1}~, \quad Q_-^{2\b 2}~, \quad Q_+^{1\b 2}~, \quad Q_-^{2\b 1}~,
\ee
and preserves the  algebra:
\be\label{22algebra QM}
\{Q_+^{1\b A}, Q_-^{2\b B}\} = 2\epsilon^{\b A\b B}\, E~, 
\ee
on $\Sigma_g\times S^1$, which is the algebra of an $\CN=4$ supersymmetric quantum mechanics with $U(1)_H\times SU(2)_C$ R-symmetry. 
 We call this $\Sigma_g \times S^1$ background the $A$-twist, corresponding to a topological twist along $\Sigma_g$ by $R_A$ in \eqref{RA RB def}.

Both twists preserve the two supercharges $Q_+^{1\b 1}$ and $Q_-^{2\b 2}$, which satisfy the $\CN=2$ supersymmetric quantum mechanics algebra:
\be
\{Q_+^{1\b 1}, Q_-^{2\b 2}\} = 2 E~.
\ee
These are the two supercharges that we  use for supersymmetric localization. Importantly, they commute with the flavor symmetry $U(1)_t \equiv 2\left[U(1)_H- U(1)_C\right]$, 
with conserved charge:
\be\label{Qt def}
Q_t \equiv R_A- R_B~.
\ee
 We can therefore turn on a fugacity $t$ for $U(1)_t$, which breaks $\CN=4$ supersymmetry to $\CN=2^\ast$. Let us define the $A$-twisted index:
\be\label{Neq4 A twisted index def}
I_{g, \, A}\left(y_i,\, t\right) =  \Tr_{\Sigma_g^A}\left( (-1)^F \, t^{Q_t} \prod_i y_i^{Q_i}\right)~
\ee
with the $U(1)_R$ charge $R=R_A$, and the $B$-twisted index:
\be\label{Neq4 B twisted index def}
I_{g, \, B}\left(y_i,\, t\right) =  \Tr_{\Sigma_g^B}\left( (-1)^F \, t^{Q_t} \prod_i y_i^{Q_i}\right)~,
\ee
with $R=R_B$.
The fugacity $t$ will play a crucial role in our computation, since we generally need $t\neq 1$ for the localization formula of section \ref{sec: susy n all} to be well-defined.~\footnote{Technically, this is so that all the singularities entering the JK residue be projective. The fugacity $t$ regulates non-projective singularities by splitting  $\CN=4$ multiplet masses.}

\begin{table}[t]
\centering
\begin{equation}\nonumber
\begin{array}{c|ccc|cc|c}
&U(1)_L&U(1)_C&U(1)_H& S_A=L+H\;\;&S_B=L+C&Q_t\\
\hline
A_\mu&1&0&0&1&1&0\\
\sigma&0&0&0&0&0&0\\
D_0 & 0&0&0&0&0&0\\
 \lambda_\alpha^{1\b 1},  \lambda_\alpha^{2\b 2} &\mp\frac12 &\pm\frac12 &\pm\frac12 & 0,1,-1,0&0,1,-1,0 & 0\\
\hline
\phi,\bar\phi&0&\pm1&0& 0 & \pm 1&\mp 2 \\
D^\mp & 0&0&\pm 1&\mp 1&0 &\mp 2\\
\lambda_{\alpha}^{1\b 2},\lambda_{\alpha}^{2\b 1} &\mp\frac12 &\pm\frac12 &\mp\frac12 & 0,1,-1,0 &-1,0,0,1 & \mp 2\\
\end{array}
\end{equation}
\caption{Charges of the components fields of an $\CN=4$ vector multiplet. Here $U(1)_L$ is the spin along  $\Sigma_g$,  and the combinations $S_A= L+H$  and $S_B= L+C$ are the $A$-twisted and $B$-twisted spins, respectively. Here we used the notation $\lambda_\alpha^{A\b A}$ for the gaugini, while the auxiliary fields $(D_0, D^\mp)$ are in the $\bf 3$ of  $SU(2)_H$.}
\label{tab: charge Neq4 vector}
\end{table}

\subsubsection{3d $\CN=4$ supermultiplets and mirror symmetry}
We consider $\CN=4$ gauge theories built out of $\CN=4$ vector multiplets and hypermultiplets. 
The $\CN=4$ vector multiplet for a gauge group $\GG$ with Lie algebra $\Fg$ consists of an $\CN=2$ vector multiplet $\CV$ and a chiral multiplet  $\Phi$, valued in the adjoint representation of $\Fg$. An  $\CN=4$ hypermultiplet charged under $\GG$ consists of
two $\CN=2$ chiral multiplet $(Q_1, \tilde Q_2)$ in a representation $(\FR, \b \FR)$ of $\Fg$, together with the charge conjugate anti-chiral multiplets.
In $\CN=2$ language, the coupling of the hypermultiplet to the vector multiplet includes the superpotential:
\be
W=Q_1\Phi \tilde Q_2~.
\ee
The non-abelian R-charges are assigned in the UV and do not change under RG flow.
We summarized the field content and the charges of a $\Fg$-valued vector multiplet  in Table \ref{tab: charge Neq4 vector}, while the hypermultiplet field content is given in Table \ref{tab: charge Neq4 hyper}.
\begin{table}[t]
\centering
\begin{equation}\nonumber
\begin{array}{c|c|ccc|cc|c}
 & \quad\Fg \quad&U(1)_L& U(1)_C&U(1)_H& S_A=L+H\;\;&S_B=L+C&Q_t\\
\hline
q_1,\bar q_1& \FR  &0&0&\pm \frac12 &\pm \frac12&0&\pm 1 \\
\psi_{\alpha}^{1},\bar\psi_{\alpha}^{1} & \FR&\mp\frac12 &\mp\frac12 &0 & \mp \frac12 &-1,0,0,1 & \pm 1 \\
\hline
q_2,\bar q_2&\b\FR  &0&0&\pm \frac12&\pm\frac12&0&\pm 1 \\
\psi_{\alpha}^2,\bar\psi_{\alpha}^2 &\b\FR  &\mp\frac12 &\mp\frac12 &0& \mp\frac12&-1,0,0,1 & \pm 1
\end{array}
\end{equation}
\caption{Charges of the components fields of an hypermultiplet. Here $q_1$ and $q_2$ are the lowest components of the 3d $\CN=2$ chiral multiplets $Q_1$ and $\t Q_2$, respectively, and $\b q_1$, $\b q_2$ are their charge conjugates.}
\label{tab: charge Neq4 hyper}
\end{table}
Under the $B$-twist, the fields
\be
(A_0, \sigma, D_0, D^{\mp})
\ee
from the vector multiplet become scalars on $\Sigma_g$, which implies that the
resulting one-dimensional gauged quantum mechanics on $S^1$ enjoys $\CN=(0,4)$ supersymmetry, with $(D_0, D^\mp)$
transforming as a triplet under $SU(2)_H$.
On the other hand, under the $A$-twist the fields:
\be
(A_0, \sigma, \phi,\bar\phi, D_0)
\ee
become scalars, with $(\sigma, \phi,\bar\phi)$ transforming as a triplet under $SU(2)_C$.
The resulting one-dimensional gauge theory is an $\CN=(2,2)$ supersymmetric quantum mechanics.%
~\footnote{ The supersymmetry multiplets of $\CN=(0,4)$ and $\CN=(2,2)$
quantum mechanics can be obtained by dimensional reduction of the two-dimensional $\CN=(0,4)$  and $\CN=(2,2)$ multiplets, respectively.}

Some other useful representations of $\CN=4$ supersymmetry are the twisted vector multiplet and the twisted hypermultiplet.~\footnote{The use of the term `twisted' for these representations of $\CN=4$ supersymmetry is standard, and should not be confused with the $A$- and $B$-twist terminology.}  For any `ordinary' $\CN=4$ supermultiplet one can construct a `twisted' representation of supersymmetry by exchanging $SU(2)_H$ and $SU(2)_C$. This `mirror automorphism' of the supersymmetry algebra is a trivial statement, in the sense that a gauge theory containing only vector multiplets and hypermultiplets is isomorphic to the same theory with twisted vector multiplets and twisted hypermultiplets, by a simple relabelling of the $R$-symmetry representations. The mirror automorphism naturally exchanges the $A$- and $B$-twists.

On the other hand, $\CN=4$ {\it mirror symmetry} is a non-trivial infrared duality of two distinct gauge theories (of vector and hypermultiplets) \cite{Intriligator:1996ex} composed with the mirror automorphism of $\CN=4$ representations. Mirror symmetry therefore implies that the $A$-twisted index \eqref{Neq4 A twisted index def} of a theory $T$ must agree with the $B$-twisted index \eqref{Neq4 B twisted index def} of its mirror $\check T$:
\be\label{Mirror indices formula}
I_{g, \, A}^{[T]}\left(y,\, t\right) = I_{g, \, B}^{[\check T]}\left(\check y,\, t^{-1}\right)~,
\ee
and similarly with $A$- and $B$-twists exchanged. Here, $y_i$ are the fugacities for the  flavor symmetries of $T$ and $\check y_i$ are the mirror fugacities of $\check T$---as we will review in the examples below, mirror symmetry exchanges Coulomb branch parameters (FI parameters) with Higgs branch parameters (real masses).

\subsubsection{The Wilson  loop and vortex loop operators}\label{lineop}

Three-dimensional $\CN=4$ gauge theories contain very interesting half-BPS loop operators. The half-BPS Wilson loop on a closed loop $\gamma$  can be thought of as a 1d $\CN=(0,4)$ quantum mechanics living on $\gamma$
\cite{Assel:2015oxa}.
On $\Sigma_g \times S^1$, such Wilson loops can be studied by wrapping them over $S^1$. We have the Wilson loop $W_{\FR}$ given by \eqref{Wilson loop R expl} 
for any representation $\FR$ of $\GG$. 
This amounts to inserting a factor 
\be
W(x)= \Tr_{\FR}\left(x\right)
\ee
in the path integral localized on the classical Coulomb branch, as discussed in details in section \ref{subsec: Wilson loop alg}. 
Such  Wilson loops preserve the four supercharges \eqref{Btwist Qs} of the $B$-twist on $\Sigma_g^B \times S^1$, while they only preserve two supercharges in the $A$-twisted theory. Consequently, we can study half-BPS Wilson loops in the $B$-twisted theory, or more generally quarter-BPS Wilson loops in the $A$-twisted theory.  In this work, we will focus on the half-BPS loop Wilson loop operators in the $B$-twisted theory.

The half-BPS loop operator which preserves the full $\CN=(2,2)$ one-dimensional algebra \eqref{22algebra QM} of the A-twisted theory is the vortex loop $V$ along $S^1$. 
This loop operator can be realized  in the UV as a 1d $\CN=(2,2)$ supersymmetric
quantum mechanics living on the loop,  coupled non-trivially to the bulk three-dimensional theory by gauging a 1d global symmetry with 3d gauge fields \cite{Assel:2015oxa}. The insertion of such a vortex loop amounts to inserting an $\CN=(2,2)$ QM index  inside the localized path integral on $\Sigma_g^A \times S^1$. For any one-dimensional GLSM coupled to the 3d gauge field, we insert:
\be\label{ZQM index}
V(x) \equiv  Z_{S^1}^{\rm QM}(x, t, y)= \oint_{\text{JK}(\xi_{\rm 1d})} \prod_{u_{\rm 1d}^i} \frac{du^i_{\rm 1d}}{2\pi i u^i_{\rm 1d}} Z^{\rm 1d}_{\text{1-loop}}(u_{\rm 1d}, x, t, y)~,
\ee
into the $\Sigma_g^A\times S^1$ localization formula. 
The quantum mechanical index \eqref{ZQM index} is written in terms of a JK residue integral over $u_{\rm 1d}$ according to the results of \cite{Hori:2014tda}.
Here the $u_{\rm 1d}$'s are the complexified flat connections of the 1d gauge theory, $x$ stands for the 3d gauge fugacities,  and  $y$ stands for the other flavor fugacities. The fugacity $t$ is a fugacity for the $R_A- R_B$ fugacity of the one-dimensional $\CN=(2,2)$ algebra.  Since the vortex operator preserves the full supersymmetry algebra of the $A$-twisted theory, this can be identified with the $Q_t$ flavor symmetry \eqref{Qt def} of the three-dimensional theory.

It is clear from symmetry considerations that half-BPS Wilson loops $W$ should be mapped to half-BPS vortex loops under mirror symmetry:
\be\label{rel W to V}
 \langle   W  \rangle^{T}_{g, B}= \langle   V  \rangle^{\check T}_{g, A}~.
\ee
The precise mirror symmetry map between a Wilson loop $W$ and a  vortex loop $V$  has been thoroughly studied in \cite{Assel:2015oxa}, and we summarize some of these results in Appendix \ref{Appendix:loop}.
In section \ref{lineoperators} below, we will verify the relation \eqref{rel W to V} for loop operators on $\Sigma_g \times S^1$ in an interesting example. We leave a more systematic study of \eqref{rel W to V} using twisted indices for future work.

\subsection{The $\CN=4$ localization formula on $\Sigma_g \times S^1$}\label{localizationN=4}
We can easily compute the twisted indices \eqref{Neq4 A twisted index def} and \eqref{Neq4 B twisted index def}, and the corresponding expectation values of half-BPS loop operators, as a special case of the $\CN=2$ localization formula  of section \ref{subsec: loc formula}.
Consider an $\CN=4$ gauge theory with gauge group $\GG$ and charged hypermultiplets $(Q_{1, i}, \t Q_{2, i})$ in representations $\FR_i$ of $\Fg$, with fugacities and background fluxes $y_i, \n_i$. 

\subsubsection{The $A$-twisted index}
The $A$-twisted index takes the form:
\be\label{Atwisted index full}
Z_{\Sigma_g^A \times S^1} = {(-1)^{\rk}\ov |W_\GG|}  \sum_{\m \in \Gamma_{\mathbf{G}^\vee}} q^\m\oint_{\rm JK} \prod_{a=1}^\rk {dx_a\ov 2 \pi i x_a}\; Z_{\m, A}^{\rm hyper}(x)\, Z_{\m, A}^{\rm vector}(x)  \, H(x)^g~.
\ee
The factor $q^\m$ in \eqref{Atwisted index full} denotes the FI term contributions from the free subgroup $\prod_I U(1)_I$ of $\GG$:
\be
q^\m \equiv \prod_I q_I^{\m_I}~.
\ee
We could also turn on a background flux $\n_{T_I}$ for the topological symmetry $U(1)_{T_I}$, which would contribute an extra classical factor to \eqref{Atwisted index full} like in previous sections, but we will mostly set $\n_{T_I}=0$ in the following.~\footnote{These terms preserve $\CN=4$ supersymmetry. The correct mixed-CS term (also called BF term) involves the $\CN=4$ vector multiplet $(\CV, \Phi)$ and a background twisted vector multiplet  $(\CV_{\rm t}, \Phi_{\rm t})$ coupling to the topological conserved current of $\CV$ \cite{Brooks:1994nn, Kapustin:1999ha}. In $\CN=2$ language, this includes the superpotential $W= \Phi \Phi_t$.}
The one-loop determinants are given by:
\bea\label{oneloop Atwist}
&Z_{\m, A}^{\rm hyper}= \prod_i \prod_{{\rho_i} \in \FR_i} \left({x^{\rho_i} y_i - t\ov 1- x^{\rho_i} y_i t}\right)^{\rho_i(\m) + \n_i}\left[ {x^{\rho_i} y_i t\ov  (1- x^{\rho_i} y_i t)(x^{\rho_i} y_i - t)}\right]^{\n_t}~, \cr
&Z_{\m, A}^{\rm vector} = (t-t^{-1})^{(2 \n_t + (g-1))\rk} \prod_{\alpha\in \Fg}  \left({1- x^\alpha \ov t- x^\alpha t^{-1}}\right)^{\alpha(\m)-g+1} \left(t- x^\alpha t^{-1}\right)^{ 2 \n_t}~,
\eea
and  the Hessian determinant $H(x)$ reads:
\be\label{H Neq4}
H(x) = \det_{ab}\left[H^{\text{vector}}_{ab}+H^{\text{hyper}}_{ab}\right]~,
\ee
with
\bea
&H^{\text{vector}}_{ab} = \half\sum_{\alpha\in G} \alpha^{a}\alpha^{b}\left({t+ x^\alpha t^{-1}\ov t- x^\alpha t^{-1}}\right)~,\cr
&H^{\text{hyper}}_{ab}=\half\sum_{i}\sum_{\rho_i\in \FR_i} \rho_i^a\rho_i^b \left({1+ x^{\rho_i} y_i t\ov 1-x^{\rho_i} y_i t}+ {x^{\rho_i} y_i +t \ov x^{\rho_i} y_i - t}\right)~.
\eea

For $g=0$, an infinite number of flux sectors contribute to \eqref{Atwisted index full} in general. On the other hand, in the case $g>0$ and $\n_t=0$, we can argue that only a finite number of flux sectors contribute non-trivially. (A similar observation was first made in \cite{Gadde:2015wta} for the $T^2\times S^2$ partition function of 4d $\CN=1$ theories.) 
This follows from the fact that
\be
\lim_{x\rightarrow 0}H(x) = \lim_{x\rightarrow \infty}H(x)=0~,
\ee
while the one-loop determinants \eqref{oneloop Atwist} stay finite in that limit.
It implies that the contributions from the residue integral at $x=0$ and $x=\infty$ must vanish, meaning that there is no wall-crossing \cite{Hori:2014tda} as we vary the parameter $\eta$ of the JK residue integral. This allows us to choose a convenient $\eta$ for each $\m$. 
Consider the case $\GG= U(1)$ for simplicity.
 For non-zero flux $\m$, we choose $\eta=-\m$ such that for $\m >0$, so we have to pick the contributions from negatively charged fields.  These fields contribute poles only when $0<\m<g$, therefore there is no contribution for $\m\geq g$.
Similarly, if $\m<0$ we have a contribution from the positively-charged fields, which contribute  only when $-g<\m$.
To summarize, for $\GG=U(1)$ the $A$-twisted index with $g>0$ and $\n_t=0$ receives contributions from a finite  number of flux sectors $-g<\m<g$. 
Similar considerations apply for any $\GG$.
In particular, the Witten index ($g=1$) only receives contributions from the vanishing flux sector on $T^2$.

\subsubsection{The $B$-twisted index}
The $B$-twisted index reads:
\be\label{Btwisted index full}
Z_{\Sigma_g^B \times S^1} = {(-1)^{\rk}\ov |W_\GG|}  \sum_{\m \in \Gamma_{\mathbf{G}^\vee}} q^\m\oint_{\rm JK} \prod_{a=1}^\rk {dx_a\ov 2 \pi i x_a}\; Z_{\m, B}^{\rm hyper}(x)\, Z_{\m, B}^{\rm vector}(x)  \, H(x)^g~,
\ee
where $H(x)$ is the same as in \eqref{H Neq4}, and the one-loop determinants are:
\bea\label{oneloop Btwist}
&Z_{\m, B}^{\rm hyper}= \prod_i \prod_{{\rho_i} \in \FR_i} \left({x^{\rho_i} y_i - t\ov 1- x^{\rho_i} y_i t}\right)^{\rho_i(\m) + \n_i}\left[ {x^{\rho_i} y_i t\ov  (1- x^{\rho_i} y_i t)(x^{\rho_i} y_i - t)}\right]^{\n_t- g+1}~, \cr
&Z_{\m, B}^{\rm vector} = (t-t^{-1})^{(2 \n_t - (g-1))\rk} \cr
&\qquad\qquad\times\prod_{\alpha\in \Fg} \left({ 1-x^\alpha \ov t- x^\alpha t^{-1}}\right)^{\alpha(\m)} \left[{ 1\ov (1- x^\alpha)(t- x^\alpha t^{-1})}\right]^{g-1}  \left({ t- x^\alpha t^{-1}}\right)^{2 \n_t}~.
\eea
In contrast to the $A$-twisted index, the $B$-twisted theory with $\n_t=0$ at $g=0$ or $g=1$ gets contribution from the $\m=0$ sector only, because the residue at infinity vanishes. (See section \ref{Higgs branch HS}.)
This implies that those indices are independent of the fugacities $q_I$ associated to the  FI parameters.
On the other hand, when  $g>1$ the one-loop determinants \eqref{oneloop Btwist} in general have poles with non-vanishing residue at infinity on the classical Coulomb branch, and an infinite number of flux sectors  generally contribute.

\subsection{The simplest abelian mirror symmetry}\label{subsec: basic mirror}
The simplest 3d mirror symmetry is between $\CN=4$ SQED with one flavor and a free hypermultiplet.
Consider first a free hypermultiplet with fugacities $y$, $t$ and background fluxes $\n$, $\n_t$ for the $U(1)\times U(1)_t$ flavor symmetry. 
Its $A$-twisted index is given by:
\be\label{ZgA free hyper}
Z_{g, A}^{\rm hyper}(y, t)\equiv  \left(y-t\ov 1- y t \right)^{\n}\left( y t \ov (1- y t )(y-t)\right)^{\n_t}~,
\ee
and its $B$-twisted index reads:
\be\label{ZgB free hyper}
Z_{g,B}^{\rm hyper}(y, t)\equiv \left(y-t\ov 1- y t \right)^{\n}\left( y t \ov (1- yt )(y-t)\right)^{\n_t-(g-1)}~.
\ee
Consider next $\CN=4$ SQED$1$, a $U(1)$ theory with a single hypermultiplet. 
In $\CN=2$ notation, the field content can be summarized by:
\be\label{table SQED Neq4}
\begin{array}{c|c|cc|cc}
    & U(1)_{\rm gauge} &  U(1)_H & U(1)_C & U(1)_t &U(1)_T \\
\hline
Q       &1 &\half& 0 & 1 & 0\\
\t Q      &-1  &\half & 0 & 1 & 0\\
\Phi      &0  &0 & 1 & -2 & 0\\
\hline
T^+ &0 & 0 &\half &-1 &1\\
T^- &0 &0 &\half &-1 &-1\\
\end{array}
\ee
Here $U(1)_T$ is the topological symmetry.
The two last lines in \eqref{table SQED Neq4} stand for the two gauge-invariant monopoles operators of the theory. We see that $(T^+, T^-)$ sits in the {\it twisted hypermultiplet} representation of $\CN=4$ supersymmetry. In fact, $\CN=4$ SQED$1$ is infrared dual to this free twisted hypermultiplet, or equivalently, it is mirror to a free hypermultiplet \cite{Intriligator:1996ex}. 

The twisted index provides a nice check of this duality. 
Let us introduce the quantities:
\be
Z^\Phi_{g, A} (t) =  \left(t -t^{-1}\right)^{2 \n_t+(g-1)}~, \qquad\qquad
Z^\Phi_{g,B}(t) =  \left(t -t^{-1}\right)^{2 \n_t-(g-1)}~, 
\ee
and
\be
H(x)=  {x t (t-t^{-1}) \ov (1-x t)(t-x)}~.
\ee
We also introduce the fugacity $q$ and background flux $\n_T$ for $U(1)_T$.
The $A$-twisted index of SQED$1$ reads:
\be
Z^{{\rm SQED}1}_{g, A}(q, t) =- \sum_{\m\in \Z} \oint_{\rm JK} {dx\ov 2 \pi i  x} \,(-q)^\m x^{\n_T}\, Z^\Phi_{g, A} (t) \, Z_{g, A}^{\rm hyper}(x, t)\, H(x)^g~,
\ee
with $Z_{g, A}^{\rm hyper}$ defined in \eqref{ZgA free hyper}, 
and similarly for the $B$-twist.  We also introduced a convenient sign for $ q\rightarrow -q$. Using the same methods as in previous sections, it is easy to show that:
\bea\label{ZQQED1mirr}
&Z_{{\rm SQED}1}^A(q, t)  = (-1)^{g-1+ \n_T} \; Z_{\rm hyper}^B(q, t^{-1})~,\cr
& Z_{{\rm SQED}1}^B(q, t)  = (-1)^{g-1+ \n_T} \; Z_{\rm hyper}^A(q, t^{-1})~.
\eea
It was shown in \cite{Intriligator:1996ex} that this mirror symmetry is 
formally a Fourier transform of the free hypermultiplet path integral \cite{Intriligator:1996ex}. The relation \eqref{ZQQED1mirr} is the concrete realization of this fact on $\Sigma_g\times S^1$. A similar computation was done on $S^3$ in \cite{Kapustin:2010xq}.

\subsection{Other examples}\label{examples}
In this subsection, we evaluate the $A$- and $B$-twisted indices of several interesting examples. For simplicity, we will set all background fluxes to zero, $\n_i=\n_T=\n_t=0$, in the remainder of this section.

\subsubsection{The free hypermultiplet}
Consider the free hypermultiplet. We see from \eqref{ZgA free hyper} that 
\be
Z_{g, A}^{\rm hyper}(y,t)=1~,
\ee
 in the absence of background fluxes. On the other hand, the hypermultiplet $B$-twisted index reads:
\be\label{ZgB hyper no flux}
Z_{g, B}^{\rm hyper}(y,t)=  \left(t+t^{-1}-y-y^{-1}\right)^{g-1}~.
\ee

\subsubsection{$\GG=U(1)$ with $N_f$ flavors}\label{subsec: SQEDNf}
Let us consider $\CN=4$ SQED---a $U(1)$ vector multiplet coupled to $N_f$ hypermultiplets $(Q_i, \t Q_i)$ ($i=1, \cdots, N_f$) of charge $1$. We introduce the fugacities $y_i^{-1}$ such that $\prod_i y_i=1$ for the $SU(N_f)$ flavor group.

\paragraph{$A$-twisted $\CN=4$ SQED.} 
The $A$-twisted index reads:
\be\label{ZA SQED}
Z_{g, A}^{{\rm SQED}[N_f]} = - (t-t^{-1})^{g-1} \sum_{\m \in \Z}   \left((-1)^{N_f} q\right)^\m \oint_{\rm JK} {dx\ov 2 \pi i x } \, \prod_{i=1}^{N_f} \left({x - t y_i\ov y_i - x t}\right)^\m \; H(x)^g~,
\ee
with
\be\label{H SQED Neq4}
H(x) = \sum_{i=1}^{N_f} \half \left(\frac{x t + y_i}{y_i- x t}+\frac{ x + t y_i}{x- ty_i}\right)~.
\ee
We also introduced a sign $q \rightarrow (-1)^{N_f} q$ for  convenience, similarly to the $\CN=2$ case in section \ref{sec: sqcd and seiberg duals}.
For $\eta>0$, the JK residue picks the poles at $x= y_i t^{-1}$. The sum over fluxes $\m$ can be performed like in previous examples. The Bethe equation for this theory is given by:
\be
P(x) = \prod_{i=1}^{N_f} (xt - y_i)- q\prod_{i=1}^{N_f} (x - t y_i) = 0~.
\ee
We can then rewrite the index \eqref{ZA SQED} as:
\be
Z_{g, A}^{{\rm SQED}[N_f]} =  \sum_{\h x \in \CS_{\rm BE}} \CH_A(\h x)^{g-1}~,\qquad \qquad 
\CH_A(x) = (t- t^{-1})\, H(x)~,
\ee
where $\CS_{\rm BE}$ is the set of $N_f$ roots of $P(x)$, and $\CH_A$ is the $A$-twist handle-gluing operator. Let us evaluate $Z_{g, A}^{{\rm SQED}}$ explicitly in a few examples. For $N_f=1$, we have:
\be
Z_{g, A}^{{\rm SQED}[1]}(q,t) = (-1)^{g-1} \left(t+ t^{-1} - q - q^{-1}\right)^{g-1}~,
\ee
which is identified with the $B$-twisted hypermultiplet \eqref{ZgB hyper no flux} according to \eqref{ZQQED1mirr}.  At genus zero, we can evaluate \eqref{ZA SQED} for any $N_f$ as we shall explain in subsection \ref{subsec: Atwist HS} below. We find:
\be\label{Zg0ASQED exp1}
Z_{g=0, A}^{{\rm SQED}[N_f]}(t, y, q)  =  -{t^{-1} (1- t^{-2 N_f})\ov (1-t^{-2})(1- qt^{-N_f})(1- q^{-1} t^{-N_f})}~,
\ee
which is independent of $y_i$.
This happens to coincide with the Coulomb branch Hilbert series (HS) of $\CN=4$ SQCD \cite{Cremonesi:2013lqa}.%
~\footnote{More precisely, we have that $Z_{g=0, A}^{{\rm SQED}[N_f]}(t, q) =  -t^{\half}_{\rm HS} \, {\rm HS}(t_{\rm HS}, z_{\rm HS})$ with $t=t^{-\half}_{\rm HS}$ and $q= z_{\rm HS}$ in the notation of \cite{Cremonesi:2013lqa}---see equation (3.2) of that paper. The factor $t_{\rm HS}^{{1\ov 2}}$ could be cancelled by turning on an $\CN=2$ mixed CS level between $U(1)_R$ and the $U(1)_t$ flavor symmetry. \label{footnoteHS}}

At genus one, we have the Witten index:
\be
Z_{g=1, A}^{{\rm SQED}[N_f]}(t, y, q)  = \Tr_{T^2} (-1)^F = N_f~.
\ee

The $\CN=4$ SQED with $N_f=2$ case is particularly interesting, since it realizes the self-mirror $T[SU(2)]$ theory of Gaiotto-Witten \cite{Gaiotto:2008ak}. 
For $g=2$ we can write down an explicit formula:
\be
Z_{g=2, A}^{T[SU(2)]}(q, a ,t) =- \frac{(1+t^2)[t^2(a+a^{-1}-2)(q+q^{-1}-2)+ 4(1-t^2)^2]}{t(t^2-a)(t^2-a^{-1})}~,
\ee
where we defined $a = {y_1\ov y_2}$.
In the limit $t\rightarrow 1$, we find a simple result at any genus:
\be
\lim_{t\rightarrow 1} Z_{g, A}^{T[SU(2)]}(q, a ,t) = 2\left(q^{\half}-q^{-\half}\right)^{2g -2}~.
\ee

\vskip0.3cm
\paragraph{$B$-twisted $\CN=4$ SQED.} 
The $B$-twisted index reads:
\bea\label{ZB SQED}
&Z_{g, B}^{{\rm SQED}[N_f]} = - (t-t^{-1})^{-g+1}\sum_{\m \in \Z}   \left((-1)^{N_f} q\right)^\m \cr
&\qquad\times\oint_{\rm JK} {dx\ov 2 \pi i x } \, \prod_{i=1}^{N_f} \left({x - t y_i\ov y_i - x t}\right)^\m  \left[{(y_i-x t)(x-t y_i)\ov x y_i t}\right]^{g-1}\; H(x)^g~,
\eea
with $H(x)$ given in \eqref{H SQED Neq4}.
By the same reasoning as above, this can be massaged into:
\be
Z_{g, B}^{{\rm SQED}[N_f]} =  \sum_{\h x \in \CS_{\rm BE}} \CH_B(\h x)^{g-1}~,
\ee
with
\be
\CH_B(x) =\left( {1\ov t- t^{-1}} \prod_{i=1}^{N_f} {(y_i-x t)(x-t y_i)\ov x y_i t}\right) \, H(x)~.
\ee
For $N_f=1$, this gives:
\be
Z_{g, B}^{{\rm SQED}[1]} = (-1)^{g-1}~,
\ee
as expected from the mirror symmetry relation \eqref{ZQQED1mirr}.
For $N_f=2$ and $g=0$, we find:
\be
Z_{g=0, B}^{T[SU(2)]}=  -\frac{t^{-1}(1-t^{-4})}{(1-t^{-2})(1 -at^{-2})(1 - a^{-1}t^{-2})}~,
\ee
which can be identified with the Higgs branch HS of $T[SU(2)]$ up to a factor of $-t^{-1}$. This is a special case of a general relation that we discuss in section \ref{Higgs branch HS} below. 
For $g=2$, we have:
\be
Z_{g=2, B}^{T[SU(2)]}(q, a ,t)=  - \frac{(1+t^2)[t^2(a+a^{-1}-2)(q+q^{-1}-2)+ 4(1-t^2)^2]}{t(t^2-q)(t^2-q^{-1})}~,
\ee
and in the limit $t\rightarrow 1$:
\be
\lim_{t\rightarrow 1} Z_{g, B}^{T[SU(2)]}(q,a, t) = 2\left(a^{\half}-a^{-\half}\right)^{2g -2}~.
\ee
These expressions provide nice checks of the self-mirror property of $T[SU(2)]$. Mirror symmetry exchanges $q$ and $a$, and sends $t$ to $t^{-1}$, so that:
\be
Z_{g, B}^{T[SU(2)]}(q,a, t)= Z_{g, A}^{T[SU(2)]}(a, q, t^{-1})~.
\ee
This is indeed satisfied by the formulas above, and can be checked for any $t$ at higher genus as well.

\subsubsection{Linear quiver gauge theory}\label{subsec: ALquiver}
\begin{figure}[t]
\begin{center}
\includegraphics[width=7cm]{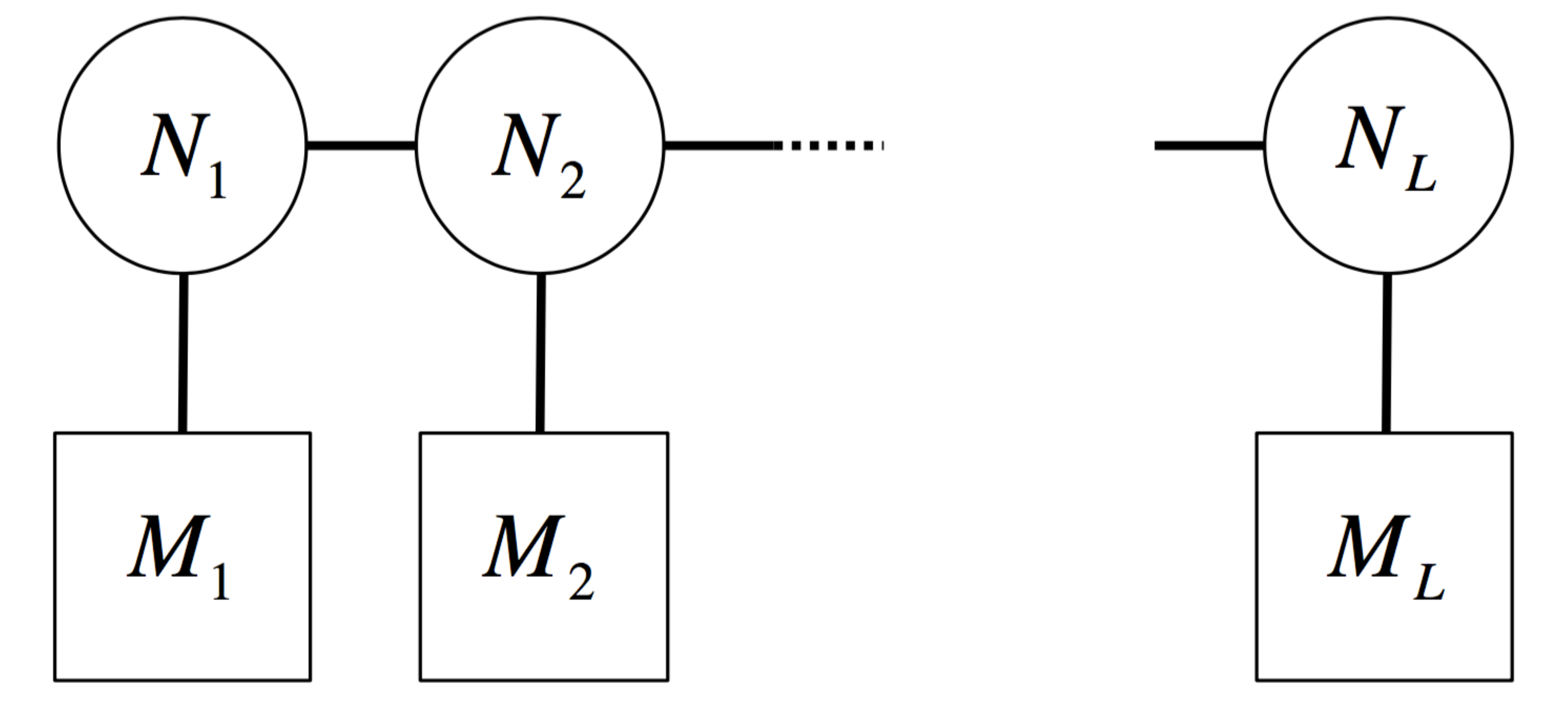}
\caption{A generic $A_L$-type linear quiver with $\CN=4$ supersymmetry. The circles and squares stand for $U(N_s)$ gauge groups and  $SU(M_s)$ flavor groups ($s=1, \cdots, L$), respectively.}
\label{LinearQuiver}
\end{center}
\end{figure}
We can generalize the computation of the last subsection to  the more general linear quiver theory in Figure \ref{LinearQuiver}, with gauge group
\be
\GG= \prod_{s=1}^L U(N_s)~.
\ee
The mirror properties of this class of theories are well understood from D-brane constructions \cite{Hanany:1996ie,Gaiotto:2008ak}.

\paragraph{$A$-twisted  $A_L$ quiver.} 
Following \eqref{Atwisted index full}, the integral expression of the $A$-twisted index reads:
\bea\label{ZA full AL}
Z^{[A_L]}_{\Sigma_g^A\times S^1} = \prod_{s=1}^L{(-1)^{N_s}\ov N_s!} \sum_{\m_a^\ps}  q_s^{\m^\ps}\oint_{\rm JK} \prod_{s=1}^L \prod_{a=1}^{N_s} {dx_a^\ps\ov 2 \pi i x_a^\ps} \, Z_{\m, A}^{\rm hyper}(x)\, Z_{\m, A}^{\rm vector}(x)  \, H(x)^g~,
\eea
with:
\bea\label{oneloopALquiv}
&Z_{\m, A}^{\rm hyper} = \prod_{s=1}^L \prod_{i=1}^{M_s} \prod_{a=1}^{N_s} \left[x_a^\ps - y_i^\ps t\ov y_i^\ps - x_a^\ps t\right]^{\m_a^\ps}\; \prod_{s=1}^{L-1} \prod_{a=1}^{N_s} \prod_{b=1}^{N_{s+1}} \left[x_a^\ps - x_b^\psp t\ov x_b^\psp - x_a^\ps t\right]^{\m_a^\ps- \m_b^\psp}~,\cr
&Z_{\m, A}^{\rm vector} = (t-t^{-1})^{(g-1)\sum_s N_s} \prod_{s=1}^L  \prod_{\substack{a,b=1\\ a\neq b}}^{N_s} \left[ x_b^\ps - x_a^\ps\ov x_b^\ps t - x_a^\ps t^{-1}\right]^{\m_a^\ps- \m_b^\ps- g+1}~,
\eea
and 
\bea\nn
&H(x)= \det_{RS}{H_{RS}(x)}~,\cr
&H_{RS} =    \half  \left(\delta_{r, s+1} + \delta_{r,s-1}\right) \left[{x_a^\ps t + x_b^{(r)}\ov  x_b^{(r)}-x_a^\ps t }+{x_a^\ps + t x_b^{(r)}\ov x_a^\ps  - t x_b^{(r)}}\right]\cr
&+ {\delta_{rs}\ov 2}\left(\delta_{ab} \sum_{i=1}^{M_s} \left[{x_a^\ps t+ y_i^\ps\ov y_i^\ps - x_a^\ps t}+{x_a^\ps +t y_i^\ps \ov x_a^\ps -t y_i^\ps}\right]+\sum_{\substack{c,d=1\\ c\neq d}}^{N_s} { \delta_{ad}(\delta_{ab}-\delta_{bc}) x_c x_d (t^2- t^{-2})\ov (x_c t- x_d t^{-1})(x_d t- x_c t^{-1})}\right)~,
\eea
where $R=(r, a)$ with $a=1, \cdots, N_r$ and $S=(s,b)$ with $b=1, \cdots, N_s$ (and $s,r=1, \cdots, L$).

Let us first consider the abelian $A_L$ quiver theory with $(N_1,\cdots, N_L)=(1,\cdots, 1)$
and $(M_1,\cdots, M_L)= (1,0,\cdots,0,1)$, for which $\rk= L$. This theory is mirror to $\CN=4$ SQED with $N_f= L+1$ flavors. In this case, the one-loop determinants \eqref{oneloopALquiv} simplify to:
\be
Z_{\m, A}^{\rm hyper} = \prod_{s=0}^L \left[x^\ps- x^\psp t \ov x^\psp - x^\ps t\right]^{\m^\ps- \m^\psp}~, \qquad\qquad
Z_{\m, A}^{\rm vector}  = (t-t^{-1})^{(g-1)L}~,
\ee
with the understanding that $x^{(0)}=x^{(L+1)}=y_1$ and $\m^{(0)}= \m^{(L+1)}=0$.
As we will explain momentarily, we can choose $\eta=(1,\cdots, 1)$ and 
sum over the flux sectors $\m^{(s)}>M$ for all $s$, for some integer $M$. This gives
\be\label{ZA abel}
Z^{[A_L]\,{\rm abel}}_{\Sigma_g^A\times S^1} =  (t-t^{-1})^{(g-1)L} \oint \prod_{s=1}^L {dx^\ps\ov 2 \pi i}\;
\frac{\underset{rs}{\det}\left(\partial_{x^{(r)} }P_{(s)}\right)}{\prod_{s=1}^L P_{(s)}(x)}\; H(x)^{g-1}~,
\ee
with
\be
P_{(s)}= (x^\psp - x^\ps t)(x^{(s-1)}- x^\ps t)-q_s (x^{(s)}-x^{(s+1)}t)(x^{(s)}-x^{(s-1)}t)~.
\ee
The Bethe equations of the abelian quiver are:
\be\label{BE abel quiver}
P_\ps(x)= 0~, \qquad {s=1, \cdots L}~.
\ee
Since the original JK residue selects only a subset of poles of the integrand, we need to show that all
the selected poles are mapped to the solutions of \eqref{BE abel quiver}.\footnote{Unlike the quiver with single $U(1)$ node, there can exist a rank $L$ singularity in \eqref{ZA abel} such that only a subset of the equations $P_{(s)}=0$ are satisfied. These singularities correspond to the poles of the original integrand (before summation over $\m$) that does not satisfy the JK condition.}
In order to show this, we  note that, in the large FI parameter limit ($q_s \rightarrow 0$), the solution to  the equations $P_{(s)}=0$ is continuously mapped to a particular pole of the original integrand before the flux summation, which enables us to track the displacement of the poles. (The trivial solutions which involve $x^{(s)}=x^{(s+1)}=0$ should be excluded since they are always located outside of the contour.)
Taking this limit, one can see that non-trivial
solutions of the equations $\lim_{q_s\rightarrow 0}P_{(s)}=0$ for all $s=1,\cdots, L$ are simply
classified by the $L$-tuple of charge sets such that, for every component $s$, there exists at least one
charge vector whose $s$-th component is positive. This is nothing but the charge sets selected by the
JK residue prescription.
Hence we can write the $A$-twisted index in terms of
the sum over the Bethe roots:
\be
Z^{[A_L]\,{\rm abel}}_{\Sigma_g^A\times S^1} =\sum_{\h x \in \CS_{\rm BE}} \CH_A(\h x)^{g-1}~, \qquad \qquad
\CH_A(x)= (t-t^{-1})^L  \; H(x)~.
\ee
These considerations can be straightforwardly generalized to the non-abelian $A_L$ quiver.
We obtain:
\bea
&Z^{[A_L]}_{\Sigma_g^A\times S^1}= \sum_{\h x \in \CS_{\rm BE}} \CH_A(\h x)^{g-1}~,\cr
&\CH_A(x) =  (t-t^{-1})^{\sum_s N_s} \prod_{s=1}^L  \prod_{\substack{a,b=1\\ a\neq b}}^{N_s} \left[ x_b^\ps t - x_a^\ps t^{-1}\ov  x_b^\ps - x_a^\ps\right]\; H(x)~,
\eea
with the Bethe equations:
\bea\nn
&P_{(s),a}(x)=0~,\qquad s=1, \cdots, L, \quad a=1, \cdots, N_s~, \cr
&P_{(s),a}(x)\equiv \prod_{i=1}^{M_s} (y_i^\ps - x_a^\ps t)\prod_{b=1}^{N_{s+1}} (x_b^\psp- x_a^\ps t) \prod_{c=1}^{N_{s-1}} (x_c^{(s-1)}- x_a^\ps t)\prod_{d\neq a}^{N_{s}}(x_d^\ps t - x_a^\ps t^{-1})\cr
&\qquad - q \prod_{i=1}^{M_s} (x_a^\ps -y_i^\ps t)\prod_{b=1}^{N_{s+1}} (x_a^\ps-x_b^\psp t) \prod_{c=1}^{N_{s-1}} (x_a^\ps- x_c^{(s-1)} t)\prod_{d\neq a}^{N_{s}}(x_a^\ps t-x_d^\ps t^{-1})~.
\eea
Note that  we should exclude the solutions with $x_a^{(s)}=x_b^{(s)}$
for $a\neq b$, as well as the trivial solutions of $P_{(s),a}(x)=0$. 
These equations are the Bethe equations of the XXZ $SU(L)$ spin chain.
The correspondence between quantum integrable models and  3d $\CN=4$ gauge theories has been  studied  extensively in the literature \cite{Nekrasov:2014xaa,Gaiotto:2013bwa}.

For this theory, the Witten index is most easily computed by considering the flux zero sector of \eqref{ZA full AL}, which also gives the number of gauge-inequivalent solutions to the Bethe equations.~\footnote{Due to the presence of solutions that trivially solve the equations (for instance $x_a^{(s)}= x_b^{(s')}=0$),
it is not straightforward to read off the number of non-trivial solutions from the order of the polynomials, unlike in the $U(1)$ example of the previous section.}
We have:
\be
Z^{[A_L]}_{T^3} = \prod_{s=1}^L{(-1)^{N_s}\ov N_s!} \sum_{\m_a^\ps}  q_s^{\m^\ps}\oint_{\rm JK} \prod_{s=1}^L \prod_{a=1}^{N_s} {dx_a^\ps\ov 2 \pi i} \; \det_{RS} \left({1\ov x_a^\ps} H_{RS}(x)\right)~.
\ee
with $R=(r,a)$ and $S=(s,b)$.
Since $H_{RS}/x_a^\ps$ is a sum over simple poles with residue $\pm 1$ (for `negatively' and `positively' charged field components, respectively), this quantity
counts the number of poles that passe the JK condition (including the exclusion of poles on the Weyl chamber walls), and the final answer is independent of the fugacities.
For instance, for $U(N_c)$ gauge theory with $N_f$ hypermultiplets (that is, $L=1$, $N_1=N_c$ and $M_1=N_f$),
one can explicitly check that only the charge sets  consisting of the positively charged part of the hypermultiplets only (for $\eta>0$) contribute non-trivially to the JK residue. Hence we have
\be
I_{g=1}^{U(N_c), N_f} = \left(\begin{array}{c}N_f\\N_c\end{array}\right)\ ,
\ee
which is the number of massive vacua of that theory. 

\paragraph{$B$-twisted  $A_L$ quiver.} 
The $B$-twisted index can be described similarly to \eqref{ZA full AL} using the general expression \eqref{Btwisted index full}. By the same reasoning as above, we find:
\be
Z^{[A_L]}_{\Sigma_g^B\times S^1} =\sum_{\h x \in \CS_{\rm BE}} \CH_B(\h x)^{g-1}~, \qquad \qquad
\CH_B(x)=\left(Z_{(0,4)}^{[A_L]} \right)^{-1} \; H(x)~,
\ee
where $Z_{(0,4)}^{[A_L]}$ can be written as:
\bea
 Z_{(0,4)}^{[A_L]}&=(t-t^{-1})^{\sum_s N_s} 
 \prod_{s=1}^L  \prod_{\substack{a,b=1\\ a\neq b}}^{N_s} \left(x_b^\ps - x_a^\ps\right)\left(x_b^\ps t - x_a^\ps t^{-1}\right)\cr
 &\qquad\times  \prod_{s=1}^L \prod_{i=1}^{M_s} \prod_{a=1}^{N_s} \left({(y_i^\ps x_a^\ps t)^\half\ov x_a^\ps - y_i^\ps t}\right)\left( {(y_i^\ps x_a^\ps t)^\half\ov y_i^\ps - x_a^\ps t}\right) \;\cr
 &\qquad\times   \prod_{s=1}^{L-1} \prod_{a=1}^{N_s} \prod_{b=1}^{N_{s+1}} \left({ (x_a^\ps x_b^\psp t)^\half \ov x_a^\ps - x_b^\psp t}\right) \left({(x_a^\ps x_b^\psp t)^\half\ov  x_b^\psp - x_a^\ps t}\right)~.
 \eea
This quantity coincides with  the one-loop determinant of a one-dimensional $\CN=(0,4)$ supersymmetric theory for  the same quiver \cite{Hori:2014tda}.

The mirror symmetry  relation \eqref{Mirror indices formula}  for twisted indices implies:
\be\label{mirror AL}
\sum_{\h x_T \in\CS_{\rm BE}^{[T]}} \CH_A^{[T]}(\h x_T)^{g-1} = \sum_{\h x_{\check T}\in  \CS_{\rm BE}^{[\check T]}} \CH_B^{[\check T]}(\h x_{\check T})^{g-1}~,
\ee 
possibly up to a sign, and similarly for $A$- and $B$-twists exchanged.
From the perspective of the Bethe-gauge correspondence \cite{Nekrasov:2014xaa},  mirror symmetry
between a pair of 3d ${\CN=2^*}$ theories is equivalent to a so-called bispectral duality between the corresponding
integrable models \cite{Gaiotto:2013bwa}.  It was argued in \cite{Gaiotto:2013bwa} that the solutions of the Bethe equations $P_{(s),a}(x)=0$ of two mirror quivers are in one-to-one correspondence. The relation  \eqref{mirror AL} further implies that the handle-gluing operators $\CH_A^{[T]}$ and 
$ \CH_B^{[\check T]}$ coincide when evaluated on pairs of mirror solutions $(\h x_T, \h x_{\check T})$ to the Bethe equations. This can be checked explicitly for $T[SU(2)]$, that we consider in the next subsection.

%

\subsection{Half-BPS line operators for ${T[SU(2)]}$}\label{lineoperators}
In this subsection we briefly discuss the matching between half-BPS Wilson loops and vortex loops in the case of the $T[SU(2)]$ self-mirror theory. As in other cases, much of this theory is governed by the Bethe equation. In the description in terms of $\CN=4$ SQED$[2]$ of section \ref{subsec: SQEDNf}, we have:
\be\label{Bethe TSU2}
P(x) \equiv  (xt-a^\half)(x t -a^{-\half})-q(x-t a^{\half})(x-t a^{-\half}) =0~.
\ee
The mirror Bethe equation $\check P(x)=0$ is obtained from \eqref{Bethe TSU2} by the substitution $a \leftrightarrow q$ and $t\rightarrow t^{-1}$.

\subsubsection{Wilson loops on $\Sigma_g^B\times S^1$}
As discussed in section \ref{lineop}, the B-twisted theory admits half-BPS Wilson line operators. The expectation value of the Wilson line operator $W(x)$ can be written as:
\be\label{wilsonN=4}
\langle W\rangle_{g, B}^{T[SU(2)]} = \sum_{\h x |P(\h x)=0} \CH_B(\h x)^{g-1}\, W(\h x)~,
\ee
with
\be
\CH_B(x)=(a^\half+ a^{-\half})(x+x^{-1}) - 2 (t+ t^{-1})~.
\ee
The quantum algebra of Wilson loops is therefore given by 
\be
\CA_{T[SU(2)]}  = \mathbb{C}[x_, x^{-1}]/\{P(x)=0\}~.
\ee 
In particular, $W_1(x)= x$ is the only independent Wilson loop. All other operators $W_k(x)= x^k$, $k \neq 0, 1$, can be written in terms $W_1$ using the relation $P(x)=0$.
For instance, we find:
\be
\left\langle x \right\rangle_{g=0, B}^{T[SU(2)]} = - \frac{t^2 (a^{1/2}+a^{-1/2})}{(t^2 - a)(t^2-a^{-1})}
\ee
at genus zero.

%

\subsubsection{Vortex loops on $\Sigma_g^A\times S^1$}
\begin{figure}[h]
\begin{center}
\includegraphics[width=6cm]{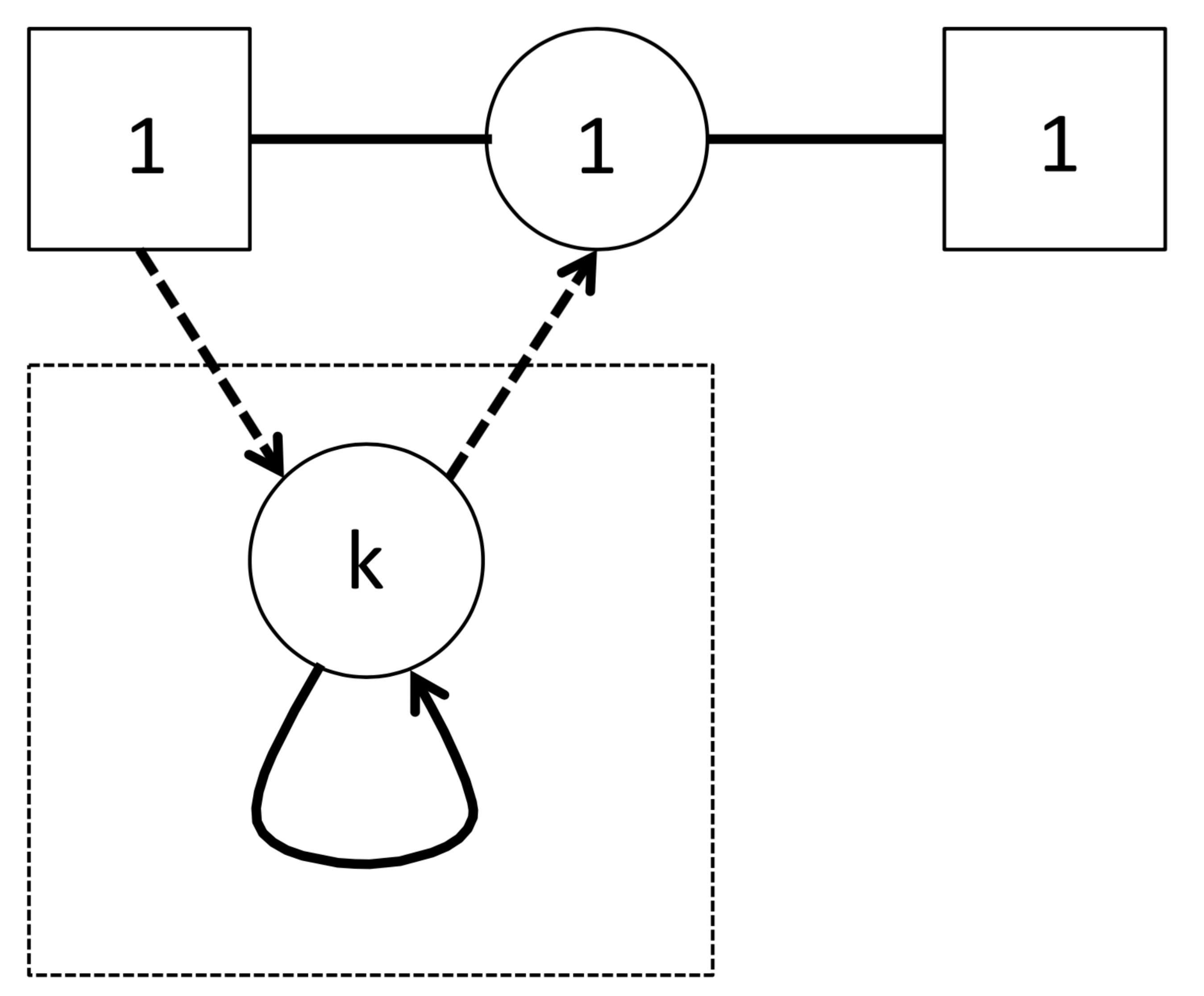}
\caption{The vortex operator $V_k$ dual to the Wilson loop $W_k$ in $T[SU(2)]$ theory. The quiver in the dotted box is a one-dimensional $\CN=(2,2)$ GLSM consisting of a gauge group $G=U(k)$ with
one fundamental, one anti-fundamental and one adjoint multiplet.}
\label{k_vortex}
\end{center}
\end{figure}

The expectation value of an half-BPS vortex loop in the $A$-twisted $T[SU(2)]$  is given by:
\be\label{vortexN=4}
\langle V\rangle_{g, A}^{T[SU(2)]} = \sum_{\h x |P(\h x)=0} \CH_A(\h x)^{g-1}\, V(\h x)~,
\ee
with
\be
\CH_A(x)= \left(t-t^{-1}\right)^2 \left({x t \ov a^\half t - (1+t^2) x + a^{-\half} x^2}+{x t \ov a^{-\half} t - (1+t^2) x + a^{\half} x^2} \right)~.
\ee
The vortex loop  $V(x)$ mirror to the $B$-twisted Wilson loop of charge $k$, $W_k(x)= x^k$,  can be
realized by coupling a  certain one-dimensional $\CN=(2,2)$ supersymmetric QM to the $A$-twisted theory \cite{Assel:2015oxa}. In the UV, the coupling of a 1d GLSM defines a singularity for the 3d gauge field as
\be
F_{z\bar z} = e^2\mu_{\rm 1d}~ \delta^2(x)~,
\ee
where $e$ is $3d$ gauge coupling and $\mu_{\rm 1d}$ is a moment map for 1d flavor symmetry. The precise field contents of the 1d GLSM dual to a given Wilson loop  was studied in \cite{Assel:2015oxa} by realizing the mirror symmetry as an $S$-duality on a system of $D$-branes.
We briefly review the relevant results in  Appendix \ref{Appendix:loop}.
The one-dimensional quiver theory mirror to the $W_k$ Wilson loop is summarized in Figure \ref{k_vortex}.
It consists of a 1d $U(k)$ theory with $\CN=(2,2)$ supersymmetry coupled to one fundamental, one anti-fundamental and one adjoint chiral multiplet. 
Due to the presence of a cubic superpotential coupling the 1d fundamental, anti-fundamental and the 3d fundamental multiplets,  we assign the $Q_t\equiv 2(H-C)$ charge to be $Q_t=0$ for the 1d fundamental and $Q_t=1$ to the anti-fundamental. Since the 1d adjoint field is not charged under
any of the global symmetry in this case, is has  $Q_t^{\text{adj}}=0$.\footnote{
However, we will keep $Q_t^{\text{adj}}$ turned on in the integrand and take the limit $Q_t^{\text{adj}}\rightarrow 0$ at the very end of the calculation. This is to avoid a non-projective singularity
in the JK residue.} 
The vortex loop of Figure \ref{k_vortex} contributes:
\bea\label{Vexpl}
&V_k (x,a,t)= \frac{1}{k!} \frac{q^{-{k\ov 2}}}{(t-t^{-1})^k}  \cr
&\quad\quad\times\int_{\text{JK}(\xi_{\rm 1d})} \prod_{i=1}^k\frac{du_i}{2 \pi i \,u_i}
\prod_{i\neq j}^k \frac{u_i-u_j}{u_i t^{-1} -u_j t}
\prod_{i\neq j}^k\frac{u_i t^{\half Q_t^{\text{adj}}-1}-u_j t^{-\half Q_t^{\text{adj}}+1}}{u_i t^{\half Q_t^{\text{adj}}}-u_j t^{-\half Q_t^{\text{adj}}}}\cr
&\qquad\qquad\qquad\qquad\times \prod_{i=1}^k\left(\frac{-u_i t^{-1}+ x t}{ u_i -x }\right)\prod_{i=1}^k\left(\frac{-a^{\half} t^{-\half} + u_it^{\half}}{a^{\half} t^{\half} - u_i t^{-\half}}\right)\ .
\eea
Note that we added a factor $q^{-{k\ov 2}}$ in front of the integral, which takes into account the flavor Wilson line associated to the `left NS5 branes'  of  \cite{Assel:2015oxa}.
Among the poles selected by the JK residue for $\xi_{\rm 1d}>0$, only one of the rank-$k$ singularities  gives a non-vanishing residue (up to the Weyl symmetry $S_k$).
The residue integral yields
\be
V_k (x,a,t) = q^{-{k\ov 2}} \left( \frac{x t- a^{\half}}{x  -a^{\half}t}\right)^k~,
\ee
which can be inserted in the formula \eqref{vortexN=4} for the vortex loop expectation value.
One can check by direct computation that $V_1(\h x)$ gives a solution of the dual Bethe equation. In other words, if $\h x$ is a root of the polynomial $P(x)$ defined in \eqref{Bethe TSU2}, then 
\be
\h x_M = V_1(\h x)= q^{-\half}\frac{\h x t- a^{\half}}{\h x  -a^{\half}t}
\ee
is a root of the mirror polynomial obtained by substituting $a\leftrightarrow q$, $t\rightarrow t^{-1}$.
Therefore, the mirror symmetry relations:
\be
\langle W_k\rangle_{g, B}^{T[SU(2)]}(q, a, t) = \langle V_k\rangle^{T[SU(2)]}_{g, A}(a, q, t^{-1})
\ee
directly follow from the statement of the mirror symmetry (at each vacuum) without the defects.

\begin{figure}[t]
\begin{center}
\includegraphics[width=10cm]{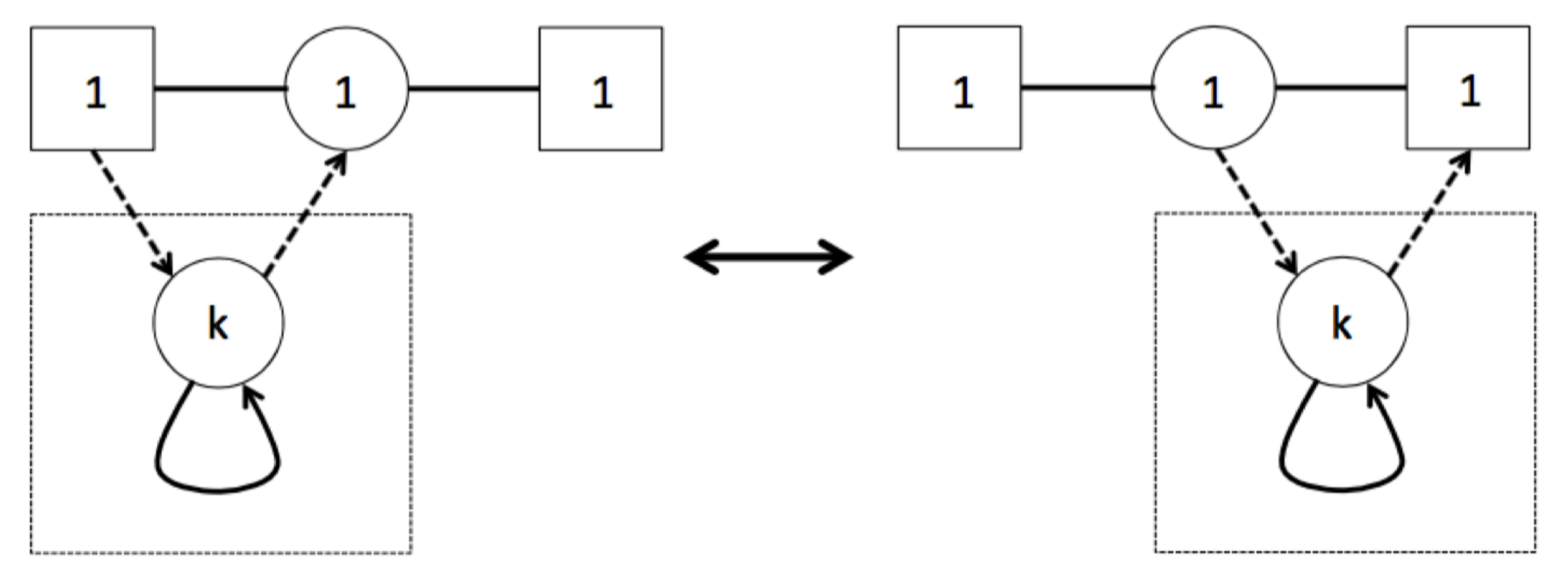}
\caption{The 1d vortex loops in 3d $\CN=4$ theories are invariant under the so-called hopping duality, as shown here for $T[SU(2)]$. This follows from the fact that the D1-brane can freely move along the D3-brane. The figure on the left corresponds to the D1-brane attached to the left $NS5$-brane, and the figure on the right corresponds to the D1-brane attached to the right NS5-brane. See Appendix \protect\ref{Appendix:loop} }
\label{hopping}
\end{center}
\end{figure}

The vortex loop defined as above is known to be invariant under the so-called hopping duality \cite{Gadde:2013ftv, Assel:2015oxa}
described in Figure \ref{hopping}. In the $A$-twisted index, this follows directly from the Bethe equation. We have
\be
q^{-\half}\frac{\h x t- a^{\half}}{\h x  -a^{\half}t}= q^{\half}\frac{\h x - a^{-\half} t}{\h x t  -a^{-\half}}
\ee
when $\h x$'s are solutions of the Bethe equation. This leads to
\be
\langle V_k^{\text{left}}\rangle_{g, A} = \langle V_k^{\text{right}} \rangle_{g, A}~,
\ee
as expected.

\subsection{Genus-zero twisted indices and Hilbert series}\label{HS}
We observed in a few examples in section \ref{examples}  that the genus-zero  $A$- and $B$-twisted indices reproduce the Coulomb branch Hilbert series and the Higgs branch Hilbert series, respectively. This is not a coincidence, but can be shown to be true for more general \emph{good} and \emph{ugly} $\CN=4$ theories, in the sense of \cite{Gaiotto:2008ak}. 
In this section, we trade the the $\CN=2^*$ deformation parameter  $t$ with
\be\label{def by}
\by =t^{-2}~.
\ee
to match with more usual conventions in the HS literature  \cite{Benvenuti:2006qr,Benvenuti:2010pq,Hanany:2011db,Cremonesi:2013lqa,Cremonesi:2014uva,Hanany:2016ezz}.
Since the Hilbert series can also be obtained from certain limits of the superconformal index \cite{Razamat:2014pta}, it follows that the twisted partition function on $S_{A}^2\times S^1$ or $S_{B}^2\times S^1$ can be obtained from the ``untwisted'' $S^2\times S^1$ partition function in the same limit. It would be worthwile to study this correspondence more thoroughly.

\subsubsection{The $B$-twisted index and the Higgs branch Hilbert series}\label{Higgs branch HS}
The equivalence of the $B$-twisted index \eqref{Btwisted index full} with the Higgs branch Hilbert series can be easily  shown whenever the contribution from infinity vanishes in the $B$-twisted index. In such cases,  only the zero flux sector contributes.
We then have:
\bea\label{HSS2B}
&Z_{S^2_B \times S^1} = (-1)^\rk \by^{\half \left[ \sum_i \dim(\FR_i)- \dim(\Fg)\right]}  \; (1-\by)^{\rk} \cr
&\qquad\qquad\times {1\ov |W_\GG|}\oint_{\rm JK} \prod_a\left[d x_a\ov 2 \pi i x_a\right] \, \prod_{\alpha\in \Fg}(1-x^\alpha)(1- x^\alpha \by) \; \CI^{\rm matter}(x)~,
\eea
with
\be
\CI^{\rm matter}(x) = \prod_i \prod_{\rho_i \in \FR_i} {1\ov (1- x^{\rho_i} y_i \by^\half)(1 - x^{-\rho_i} y_i^{-1} \by^\half)}~.
\ee
The result of the JK residue integral can be shown to be equivalent to a `unit contour'  integral $|x_a|=1$ for a large class of theories (for conveniently chosen fugacities such that all the poles from `positive' charged field components, and no `negative' pole, lie inside the unit circle).
If that is the case,  the twisted index \eqref{HSS2B}  becomes the `Molien formula'  for the Higgs branch HS \cite{Benvenuti:2006qr, Razamat:2014pta}:
\be\label{ZB to HSHiggs}
Z_{S^2_B \times S^1}(y_i, \by)  = (-1)^\rk \by^{\half \left[ \sum_i \dim(\FR_i)- \dim(\Fg)\right]}   \; {\rm HS}_{\rm Higgs}(y_i, \by)~.
\ee
 up to a power of $\by$ that could be cancelled by turning on a bare CS level 
 \be
 k_{tR}=   \dim(\Fg) -\sum_i \dim(\FR_i)
 \ee
for  the $\CN=2^\ast$ flavor symmetry $U(1)_t$. 

As an example, consider the $A_L$ quiver theory of subsection \ref{subsec: ALquiver}.
In order to show that only the $\m=0$ flux sector contributes to the $g=0$ $B$-twisted index, we need to  
prove that the residues at $x^{(s)}_a=0$ and $x^{(s)}_a=\infty$ vanish for every flux sector. Let us first examine the limit $x^{(s)}_a\rightarrow 0$. The integrand scales as:
\begin{equation}
 (x_a^{(s)})^{-2(N_s-1)-1}
(x_a^{(s)})^{M_{s}+N_{s+1}+N_{s-1}}
\end{equation}
in that limit, which converges when
\begin{equation}
M_{s}+N_{s+1}+N_{s-1}-2N_s+1 \geq 0~.
\end{equation}
This is precisely the condition for the quiver to be `good' or `ugly' in the classification of \cite{Gaiotto:2008ak}. In these cases, there
is no singularity at $x_a^{(s)}=0$, nor at $x_a^{(s)}\rightarrow \infty$ by a similar argument.
Then, one can choose $\eta = -\m$ so that all $\m\neq 0$  flux sectors contribute trivially to the JK residue. We are then left with the expression  \eqref{HSS2B} for the $A_L$ quiver. 

For the $T[SU(N)]$ theory (the case $N=(1,2,\cdots, N-1)$ and $M=(0,\cdots,0,N)$), we can show that the unit circle integral defined by $|x_a|=1$ is indeed equivalent to the JK residue integral. First of all, we choose $\eta = (1,\cdots,1)$ and fix $\by >1$. Let us start with the condition for the first node. From the JK condition, the charge set should contain one of the poles defined by
\be
1-x^{(1)}(x_i^{(2)})^{-1}\by^{1/2}=0
\ee
for $i=1,2$. If it contains another pole of the form $1-(x^{(1)})^{-1}x_j^{(2)}\by^{1/2}=0$ with $j\neq i$,
then these relations impose a equation $x_j^{(2)}\by^{1/2}-x_i^{(2)}\by^{-1/2}=0$ which is the position of the zero in the vector multiplet of the second node. Hence only the chiral multiplet which are positively charged under the first node contributes. When $|x^{(s)}_i|=1$ for $s\neq 1$, these are all the singularities inside the unit circle $|x_1|=1$. The same argument holds for the second node. We need at least one positively charged chiral fields in a form
$1-x_i^{(2)}(x_j^{(3)})^{-1}\by^{1/2}=0$ for each $i$. If there are charges in a form $1-(x_i^{(2)})^{-1}x_j^{(3)}\by^{1/2}=0$, by the same reasoning as above, the residues are zero due to the vector multiplet. This continues to the $(n-1)$-th node of the quiver, which completes the proof of the equivalence between the two prescriptions.

\subsubsection{The $A$-twisted index and the Coulomb branch Hilbert series}\label{subsec: Atwist HS}
The relation \eqref{ZB to HSHiggs} combined with mirror symmetry implies that the $Z_{S^2_B \times S^1}$ parition function is similarly related to the Coulomb branch Hilbert series first constructed in \cite{Cremonesi:2013lqa}. 
Since the genus-zero $A$-twisted index receives contributions from an infinite number of flux sectors, a direct proof of this  equivalence is expected to be rather more complicated. Here and in Appendix \ref{appF: U2}, we check that relation in some of the simplest examples. 
 We leave a more general study for future work.

For $\CN=4$ SQED with $N_f$ hypermultiplets,  the genus-zero $A$-twisted index \eqref{ZA SQED} can be evaluated by trading the residues over the fundamental chiral multiplets, which are picked by the JK residue prescription for $\eta>0$, with the residues at infinity on $\fM\cong \C^\ast$. 
For $\eta>0$, only the flux sectors $\m >0$ contribute. 
The poles of the integrand of  \eqref{ZA SQED} (for $g=0$) are located at $x= 0$, $x=\infty$, and  $x=y_i \by^\half$, using the notation \eqref{def by}. We have:
\bea
&Z_{g=0, A}^{{\rm SQED}[N_f]} \;&=&\; -{\by^\half\ov 1- \by} \sum_{\m =1}^\infty   \left((-1)^{N_f} q\right)^\m \oint_{\rm JK} {dx\ov 2 \pi i x }  \prod_{i=1}^{N_f} \left({x \by^\half - y_i \ov y_i \by^\half -x}\right)^\m \cr
&&=&  \;{\by^\half\ov 1- \by} \sum_{\m =1}^\infty   q^\m \left[ \by^{-\half N_f \m}- \by^{\half N_f \m}\right]\cr
&&=&  - {\by^\half\ov 1- \by} \, {1- \by^{N_f}\ov (1- q \by^{\half N_f})(1- q^{-1}\by^{\half N_f})}
\eea
This  reproduces \eqref{Zg0ASQED exp1} and the Coulomb branch series of \cite{Cremonesi:2013lqa} as advertised. Similar manipulations can be performed for higher-rank gauge groups, as demonstrated for $U(2)$ in Appendix \ref{appF: U2}.

\section*{Acknowledgements} We would like to thank  Davide Gaiotto, Jaume Gomis,  Zohar Komargodski, Peter Koroteev, Dario Rosa and  Brian Willett for interesting discussions and correspondence.   CC would like to especially thank  Stefano Cremonesi,  Daniel~S. Park and Noppadol Mekareeya for discussions and collaborations on related projects.
We are especially grateful to the 2015 Summer Simons Workshop in Mathematics and Physics, where this project was initiated. 
HK gratefully acknowledges support from the Simons Center for Geometry and Physics, Stony Brook University at which some of the research for this paper was performed. 
This research was supported in part by Perimeter Institute for
Theoretical Physics. Research at Perimeter Institute is supported by the Government of Canada through Industry Canada and by the Province of Ontario through the Ministry of Economic Development \& Innovation. 
The work of HK was made possible through the support of a grant from the John
Templeton Foundation. The opinions expressed in this publication are those of the
author and do not necessarily reflect the views of the John Templeton Foundation.


\appendix
\section{Conventions: geometry and quasi-topological twisting}\label{Appendix: conv and susy}
We  follow the conventions of \cite{Closset:2012ru, Closset:2013vra, Closset:2015rna} for geometry, spinors and supersymmetry multiplets. We consider a compact Euclidean space-time $\CM_3=\Sigma_g \times S^1$ with Riemannian metric:
\be
ds^2 = \beta dt^2 + 2 g_{z\bz}(z, \bz) dz d\bz~.
\ee
Here $t\sim t+2\pi$ is the coordinate on $S^1$ and $z, \bz$ are local complex coordinates on the Riemann surface $\Sigma_g$.  
We have the standard spin connection:
\be
{\omega_{\mu a}}^b = {e^b}_\nu \nabla_\mu {e_a}^\nu~,
\ee
in terms of the Levi-Civita connection $\nabla_\mu$. We generally denote by $D_\mu$ the covariant derivatives on spinors and tensors in the frame basis. The Riemann tensor is defined in the standard way.~\footnote{We follow the conventions of \cite{Closset:2012ru} except that we flip the sign of the Ricci scalar $R$. In our conventions, $R>0$ on the round $S^3$ or on the round $S^1 \times S^2$.}

We use the canonical frame $e^a= e^a_\mu dx^\mu$ with:
\be
e^0= dt~, \qquad e^1= \sqrt{2 g_{z\bz}} dz~, \qquad e^{\b1}= \sqrt{2 g_{z\bz}} dz~, 
\ee
Here $a=0, 1, \b1$ are the frame indices in complex coordinates; they are lowered using $\delta_{ab}$ with $\delta_{00}=1$ and $\delta_{1\b 1}= \half$.  We also chose the orientation such that $\epsilon^{0 1 \b 1}= -2 i$. The $\gamma$-matrices in this frame are:
\be
\left\{{(\gamma^\mu)_\alpha}^\beta\right\} = \left\{\gamma^0, \gamma^1, \gamma^{\b1}\right\}= \left\{ \mat{1 & 0 \cr 0 &-1}~,\; \mat{0& -2\cr 0&0}~,\; \mat{0& 0\cr -2&0}  \right\}~. 
\ee
Three-dimensional Dirac spinors are denoted by:
\be
\psi_\alpha=\mat{\psi_- \cr \psi_+}~.
\ee 
Dirac indices can be raised and lowered with $\epsilon^{\alpha\beta}$, $\epsilon_{\alpha\beta}$ with $\epsilon^{-+}=\epsilon_{+-}=1$. 
When reducing to two dimensions along $\d_t$, the spinor components $\psi_\mp$ become kinematically independent Weyl spinors of spin $\pm \half$, respectively. The covariant derivative on a Dirac spinor is given by:
\be
D_\mu \psi = \big(\d_\mu - {i\over 4}\omega_{\mu ab} \epsilon^{abc}\gamma_c\big)\psi~.
\ee
In section \ref{sec: susy n all}, we generally use explicit frame indices for all quantities including derivatives.

The coordinates $(t, z, \bz)$ are adapted to a choice of transverse holomorphic foliation (THF) on $\CM_3$ as explained in \cite{Closset:2012ru, Closset:2013vra}. 
Let us define $\eta^\mu$ a nowhere-vanishing vector such that
\be
\eta_\mu \eta^\mu=1~.
\ee
We can define:
\be
{\Phi^\mu}_\nu = - {\epsilon^\mu}_{\nu\rho} \eta^\rho~,
\ee
which satisfies $ {\Phi^\mu}_\nu  {\Phi^\nu}_\rho = -{\delta^\mu}_\rho + \eta^\mu \eta_\rho$.
The THF can be characterized by such an $\eta_\mu$~\footnote{We inverted the sign of $\eta_\mu$ with respect to \cite{Closset:2012ru, Closset:2013vra}---that is, $\eta_\mu= - \eta_\mu^{\rm there}$.} satisfying the integrability condition:
\be
 {\Phi^\mu}_\nu {\left(\CL_\eta \Phi\right)^\nu}_\rho=0~.
\ee
The object ${\Phi^\mu}_\nu$ reduces to a complex structure on the normal bundle of the foliation (i.e. for vectors orthogonal to $\eta_\mu$). In our case, this is just the complex structure on the Riemann surface $\Sigma_g$. We then have natural three-dimensional notions of holomorphic vectors and one-forms~\cite{Closset:2013vra}. 

\subsection{Quasi-topological twisting}
The quasi-topological twisting that we use in this paper is best understood in the context of curved-space rigid supersymmetry \cite{Festuccia:2011ws, Closset:2014uda}.
In section \ref{sec: susy n all}, we used a `twisted field' notation for all the fields. This corresponds to a field redefinition of the fermionic and bosonic fields, where the ``$A$-twisted fields'' are obtained by various contractions with the Killing spinors \eqref{KS explicit}.  On $\Sigma_g \times S^1$, we can label the fields by their $U(1)_L$  spin $L$ on $\Sigma_g$. The quasi-topological twisting is equivalent to the standard  topological $A$-twist on $\Sigma_g$, which assigns to all the fields a twisted spin:
\be
S= L+ \half R~,
\ee
with $R$ the $U(1)_R$ $\CN=2$ $R$-charge. 
We refer to the Appendix of \cite{Closset:2015rna} for a more thorough discussion in two-dimensions.
As an example, consider the $\CN=2$ vector multiplet $\CV$. In the standard notation  of \cite{Closset:2012ru}, it has components:
\be
\CV= \left(a_\mu~, \; \sigma~, \; \lambda_\alpha~, \; \t \lambda_\alpha~, \;  D\right)~.
\ee
Using the Killing spinors $\zeta$, $\t\zeta$ on $\Sigma_g\times S^1$, we defined the `twisted' gaugini:
\be
\Lambda_\mu \equiv \t\zeta\gamma_\mu\lambda~, \qquad \qquad
\t\Lambda_\mu \equiv -\zeta\gamma_\mu\t\lambda~.
\ee
They are holomorphic and anti-holomorphic one-forms with respect to the THF, as can be shown from the Killing spinor equations or by explicit computation in components. This gives \eqref{gauginiA}. The $A$-twist of the chiral multiplets discussed in \cite{Closset:2015rna} can also be given a three-dimensional uplift along the lines of \cite{Closset:2014uda}.

\section{Localization of $\CN=2$ YM-CS-matter theories}\label{app: derivation formula}
In this Appendix, we derive the main localization formula \eqref{main formula} for the twisted index of $\CN=2$ gauge theories. The main technical difficulty lies in the treatment of the fermionic zero modes, and we can  mostly follow the previous literature on the subject \cite{Benini:2013nda,Benini:2013xpa, Hori:2014tda, Benini:2015noa, Closset:2015rna}. The new ingredient is the integration of the $g$ additional  one-forms gaugini and flat connections present due to the non-trivial topology of $\Sigma_g$.

\subsection{One-loop determinant: $\h D=0$}
Consider a chiral multiplet of $U(1)$ charge $Q$ and $R$-charge $r$, coupled to a supersymmetric background $U(1)$ vector multiplet \eqref{susy eq vector} with gauge flux $\m$ on $\Sigma_g$. By supersymmetry, all the bosonic and fermionic modes cancel out, except for some unpaired `zero-modes'. The bosonic zero-modes correspond to a pair of boson and fermions $(\CA, \CB)$ (together with their charge conjugates), related by supersymmetry, which satisfy: 
\be
D_\bz \CA= 0~, \qquad D_\bz \CB=0~.
\ee
They correspond to holomorphic sections of $\CK^{r\ov 2} \otimes L^Q$ (of total  degree $d=r(g-1)+ Q\m$) on $\Sigma_g$, with $L$ the $U(1)$ line bundle.
The fermionic zero modes correspond to modes of the fermionic field $\CC$ such that:
\be
D_z \CC=0~,
\ee
corresponding to holomorphic sections of $\CK^{2-r\ov 2} \otimes L^{-Q}$.
Let $n_B$ and $n_C$ denote the number of bosonic and fermionic zero-modes, respectively. By the Riemann-Roch theorem:
\be
n_B - n_C = Q\m + (g-1) (r-1)~.
\ee
Resumming the KK tower from the $S^1$, we find the one-loop determinant \cite{Benini:2015noa}:
\be\label{Zphi app}
Z^\Phi = \left( x^{Q\ov 2} \ov 1- x^Q\right)^{Q \m +(g-1)(r-1)}~,
\ee
with $x= e^{2\pi i u}$ as defined in section \ref{sec: coulomb branch}.
This leads to the contribution \eqref{oneloop matter} in a general theory. (The $W$-boson contribution \eqref{oneloop vector} is also the same as for a chiral multiplet of $R$-charge $2$ and gauge charges given by the simple roots \cite{Benini:2015noa, Closset:2015rna}.)

\subsection{Localization for $\GG=U(1)$}
Consider a $U(1)$ YM-CS-matter theory with CS level $k$ and chiral multiplets $\Phi_i$ of charges $Q_i$ and $R$-charges $r_i$. (More generally, we could consider any $\GG$ with rank $1$.) The path integral can be localized onto the Coulomb branch by considering the localizing action:
\be
\SL_{\rm loc} ={1\ov e^2}\SL_{\rm YM}+ {1\ov g^2}\SL_{\t\Phi\Phi}~.
\ee
For a given flux $\m$, the one-loop determinant \eqref{Zphi app} can have a pole at $x^Q=1$ on the classical Coulomb branch, corresponding to additional massless modes.  The natural way to deal with this singularity is by keeping a constant mode of the auxiliary field $D$ in intermediate steps of the localization computation. We define the field $\h D$ by:
\be
D= 2 i f_{1\b 1} + i \h D~,
\ee
so that $\h D=0$ on the supersymmetric locus. 
A general supersymmetric configuration also includes flat connections along $\Sigma_g$:
\bea
& a_z dz = \sum_{I=1}^g \alpha_I \omega^I~, \qquad\quad \omega^I \in H^{1,0}(\Sigma_g, \Z)~, \cr
& a_\bz d\bz = \sum_{I=1}^g \t\alpha_I \t\omega^I~, \qquad\quad \t\omega^I \in H^{0,1}(\Sigma_g, \Z)~.
\eea
There are also fermionic zero-modes:
\be
\Lambda_0~, \quad \t\Lambda_0~, \qquad  \Lambda_1= \sum_{I=1}^g \Lambda_I \omega^I_1~,
 \qquad  \t\Lambda_{\t1}= \sum_{I=1}^g \t\Lambda_I \t\omega^I_{\b 1}~.
\ee
Here $\Lambda_0, \t\Lambda_0$ are constant and the one-form-valued gaugini satisfy  $D_{\b1 }\Lambda_1=0$, $D_1 \t\Lambda_{\t1}=0$. All these constant modes organize themselves into supersymmetry multiplets:
\be\label{V0 VI def}
\CV_0 = (\sigma~,\, a_0~,\, \lambda~,\, \t\lambda~,\, \h D)~, \qquad\qquad \CV_I = (\alpha_I~,\, \t \alpha_I~,\, \Lambda_I~,\, \t\Lambda_I)~, \quad I=1, \cdots, g~.
\ee
Consider the chiral multiplet $\Phi$ with $Q=1$,  in the background \eqref{V0 VI def}. We have:
\be\label{superdeterminant}
Z^\Phi(\sigma, a_\mu, \h D, \cdots) = \int [d\Phi] e^{-S_{\t\Phi\Phi}} = {\rm SDet}^{-1}(\CK)~, 
\ee
in terms of the kinetic Lagrangian \eqref{kin chiral}, which can be used for localization since it is $Q$-exact:
\be
\SL_{\t\Phi\Phi}= \left(\t\CA~, \, \t\CB~, \, \t\CC\right) \CK \mat{\CA\cr \CB\cr \CC} - \t\CF \CF~.
\ee

Integrating out all the massive fields in the Coulomb branch background \eqref{V0 VI def}, we obtain a complicated supersymmetric matrix model for the constant modes \eqref{V0 VI def}.
Schematically, we find:
 \be\label{Zg U1 schematic}
Z_g =\lim_{\epsilon, e^2\rightarrow 0}\; \sum_{\m\in \Z}  \int \prod_{I=1}^g d\CV_I\, \int_{\Gamma} d\h D\int_{\t\fM} {du d\t u \ov \beta}  \int d\Lambda_0 d\t\Lambda_0  
\; \CZ_\m(\CV_0, \CV_I)~,
\ee
where the limit in front is a particular scaling that we will discuss in a moment.
Here we defined the measure:~\footnote{We are being slightly careless about normalization. We fixed the overall normalization in the final formula by comparing our  result  to known results for pure $\CN=2$ Chern-Simons theory (see section \ref{sec: expl2}).}
\be
d\CV_I\equiv {1\ov  \beta {\rm vol}(\Sigma_g)} \,d \alpha_I d\t\alpha_I \,  d \Lambda_I d \t\Lambda_I~.
\ee
At this point, for future convenience, we perform a change of variable $\t u\rightarrow \t u'$ and $\t\Lambda_0 \rightarrow \t\Lambda'_0$, according to the relation
\be
\t u = \t u'/k^2,~~~ \t\Lambda_0 = \t\Lambda'_0/k^2\ ,
\ee
for a small positive number $k^2$,  leaving $u$ unchanged.
 Note that the measure in \eqref{Zg U1 schematic} is invariant under this
change of variable. The purpose of this rescaling will become clear momentarily.

Since the one-loop determinant contributions to $\CZ_\m$ potentially have singularities at points where chiral multiplets become massless, let us examine these dangerous regions of the integrand before performing the path integral, following \cite{Benini:2013nda}. Near a singular point region $u=0$ (any other singularity of the form $u=u_*$ in the bulk  can be considered similarly  by translation) the bosonic part of the chiral multiplet reads:
\be
I=\int \prod_{i=1}^Nd\t \CA^i d\CA^i ~\exp\left[-\frac{1}{g^2}\t\CA \left(u\t u'/k^2\right)\CA - \frac{e^2}{2}\left(\t\CA \CA-\xi_{FI}\right)^2\right]\ ,
\ee 
where $N$ is the number of  chiral multiplets which become massless at $u=0$.
Note that the point $\{u=0\} \in \t\fM$ is singular when we take the localization limit $e \rightarrow 0$.
This singularity can be regularized by keeping $e$ finite until we perform the $u$ integrals. Then the integral is bounded by 
\be
I \sim \frac{C}{e^{2N}}\ ,
\ee
where $C$ is a numerical factor which is independent of $e$. Given this, we divide the integral \eqref{Zg U1 schematic} into two pieces:
\be
\int_{\tfM} du d\t u' ~\CZ_m = \int_{\tfM \backslash \Delta_{\epsilon}} du d\t u' ~\CZ_m + \int_{ \Delta_{\epsilon}} du d\t u' ~\CZ_m\ ,
\ee
where $\Delta_\epsilon$ is the epsilon neighborhood of the singular region defined by $u\t u' \leq \epsilon^2$.
When $e$ is small but finite, the second factor is bounded by $C\pi \epsilon^2/e^{2N}$, which vanishes after we take the limit $\epsilon\rightarrow 0$ first.
Then we are left with the contribution from the first term, given that the condition $\epsilon \ll e^N \ll 1$ is satisfied. This is the scaling limit implied in \eqref{Zg U1 schematic}.

Now, let us first perform the integral over the scalar gaugino zero-modes $\Lambda_0, \t\Lambda'_0$. Due to the residual supersymmetry, the integrand of \eqref{Zg U1 schematic} satisfies:
\bea
&\delta \CZ_\m = \left(-2 i \beta \t\Lambda'_0 \d_{\b u'} - \h D \d_{\Lambda_0} + i \t\Lambda_I \d_{\t\alpha_I}\right)\CZ_\m=0~. \cr
\eea
We can use this relations to perform the integral over $\Lambda_0$, since:
\bea\label{susy rel integration abel}
&\d_{\Lambda_0} \d_{\t\Lambda'_0} \CZ_\m \Big|_{\Lambda_0 =\t\Lambda'_0=0} \,&=& \, {1\ov  \h D} \left(2 i \beta \d_{\b u'} + i \t\Lambda_I \d_{\t\alpha_I} \d_{\t\Lambda'_0}\right) \CZ_\m \Big|_{\Lambda_0= \t\Lambda'_0=0} ~.\cr
\eea
We have the sum of two total derivatives. The integration over the $\Sigma_g$ flat connections $\alpha_I$  is a compact domain and the integrand has no singularities as long as $\epsilon >0$, therefore the total derivatives $\d_{\t\alpha}$ in \eqref{susy rel integration abel} do not contribute to the path integral. We are left with:
 \be\label{Zg U1 deriv step1}
Z_g =\lim_{\epsilon, e^2\rightarrow 0}\; \sum_{\m\in \Z}  \int \prod_{I=1}^g d\CV_I\,  \int_{\Gamma} {d \h D\ov \h D}\int_{\t\fM\backslash \Delta_\epsilon} {du d\b u'} \; \d_{\b u'}  \CZ_\m \Big|_{\Lambda_0= \t\Lambda_0=0}~,
\ee
which reduces the integral over $\t\fM$ to an integral over the boundary  $\d\Delta_\epsilon$, by Stokes theorem.

Next, let us evaluate $\CZ_{\m}$. In addition to the classical contribution, the important contributions are the one-loop superdeterminant \eqref{superdeterminant} at ${\Lambda_0= \t\Lambda_0=0}$, for every chiral multiplets in the theory.
To compute \eqref{superdeterminant}, we first  expand any three-dimensional field in Fourier modes on $S^1$:
\be
\Phi = \sum_{n\in \Z} \Phi_n e^{i n t}~.
\ee
 It is  convenient to define the two-dimensional variables:
\be
Q\sigma_n = \frac{1}{i\beta}\left(Q u +  n\right)~, \qquad \qquad
Q\t\sigma'_n = -\frac{1}{i\beta}\left(Q \t u'/k^2 +  n\right)~.
\ee
Note that we are using the rescaled variable $\t u'$. 
Let us also denote by $\{\lambda\}$ the spectrum of the twisted Laplacian on $\Sigma_g$:
\be\label{eigenvalues}
- 4 D_1 D_{\b1} \phi = \lambda \phi~.
\ee
 We then have:
\be\label{Zphi gen}
Z^\Phi\Big|_{\Lambda_0=\t\Lambda_0=0} = Z_{\rm zero}^\Phi \; Z_{\rm massive}^\Phi~.
\ee
The first factor in \eqref{Zphi gen} is the contribution from the chiral multiplet zero-modes at $\Lambda_0=\t\Lambda_0=0$:
\be
Z_{\rm zero}^\Phi = \prod_{n\in \Z} (Q\sigma_n)^{n_C} \left( {Q\t\sigma'_n \ov  Q^2\t\sigma'_n\sigma_n+ i Q  \h D }
 \right)^{n_B}~.
\ee
At $D=0$, this gives \eqref{Zphi app} after regularizing the product over $n$:%
~\footnote{Note that the regularized product is not invariant under large gauge transformations $u\rightarrow u+1$. This is a manifestation of the so-called parity anomaly \cite{Redlich:1983dv}. In a physical theory, this lack of gauge invariance for an odd number of Dirac fermions must be compensated by an half-integer CS level.}
\be
\prod_{n\in \Z}  {1\ov Q\sigma_n}= \prod_{n\in \Z}  {i \beta \ov Qu+n}   = {x^{Q/2} \ov 1-x^Q}~.
\ee
The second factor in \eqref{Zphi gen} is the contribution from all the other modes:
\be\label{Zmassive full}
Z_{\rm massive}^\Phi=\prod_{n\in \Z} \prod_\lambda \left[{\lambda+ Q^2\t\sigma'_n \sigma_n\ov \lambda+ Q^2\t\sigma'_n \sigma_n+ i Q\h  D} \right] \left(1- 2 i {(Q\t\sigma'_n)(Q \t\Lambda_{\b1})(Q\Lambda_1)  \ov (\lambda+ Q^2\t\sigma'_n\sigma_n)(\lambda+ Q^2\t\sigma'_n \sigma_n+ iQ \h D)}\right)~.
\ee
Note the appearance of the gaugino zero-modes, with the short-hand notation:
\be
\t\Lambda_{\b1}\Lambda_1= \sum_{I=1}^g\t\Lambda_I\Lambda_I~.
\ee

We first perform the $\h D$-integrals in  \eqref{Zg U1 deriv step1}.
This is essentially the same  the discussion in the previous literatures \cite{Hori:2014tda,Benini:2015noa}.
Let $\Delta_\epsilon$ be the union of small circular neighborhoods of radius $\epsilon^2$ around the potential singularities on the classical Coulomb branch $\t \fM\cong \C^\ast$ at:
\be
H_i =\{u\; |\; Q_i u + \nu_i \in \Z\}~, \;\; \forall i~, \qquad \qquad H_\pm = \{ u\; |\;u = \mp i \infty\}~, 
\ee
corresponding to matter field and monopole operator singularities, respectively.
To each potential singularity, we associate its charge, as explained in the main text:
\be
H_i \rightarrow Q_i~, \qquad \qquad H_\pm \rightarrow Q_{\pm}~, 
\ee
where 
\be
Q_{\pm} = \pm k -\half \sum_i |Q_i| Q_i
\ee
are the monopole operator gauge charges.
In each flux sector $\m$, only some of the potential singularities are actual singularities. We have a singularity at $H_i$ if $Q_i \m + \n_i +(g-1)(r_i-1) >0$ and a singularity at $H_\pm$ if 
$Q_\pm \m + Q^F_\pm \n_F + (g-1) r_\pm\geq 0$---see equation \eqref{singularity monopole}. We denote by $ \tfM^{\m}_\text{sing}$ the union of all the singularities in a given flux sector. As alluded to in the main text, we have to {\it assume} that each singularity is {\it projective}, meaning that to each singular point we only associate either positive or negative charges. A non-projective singularity can often be rendered projective by turning on  generic fugacities. We denote by $\Delta_{\epsilon, \m}$ the circular neighborhood of the singularities in a given flux sector. Since every singularity is projective by assumption,  $\Delta_{\epsilon, \m}$  is the union of  `positive' and `negative' singularities:
\be
\Delta_{\epsilon, \m} = \Delta_{\epsilon, \m}^{(+)}\cup \Delta_{\epsilon, \m}^{(-)}~.
\ee
The integration contour of $\h D$ is taken along the real direction with a slight shift along the imaginary axis:
\be
\Gamma =\{ \h D\, |\, \h D \in \R + i \delta~, \; \delta\in \R~, \; 0<|\delta|\ll \epsilon/k\}~. 
\ee
\begin{figure}[t]
\begin{center}
\includegraphics[width=15cm]{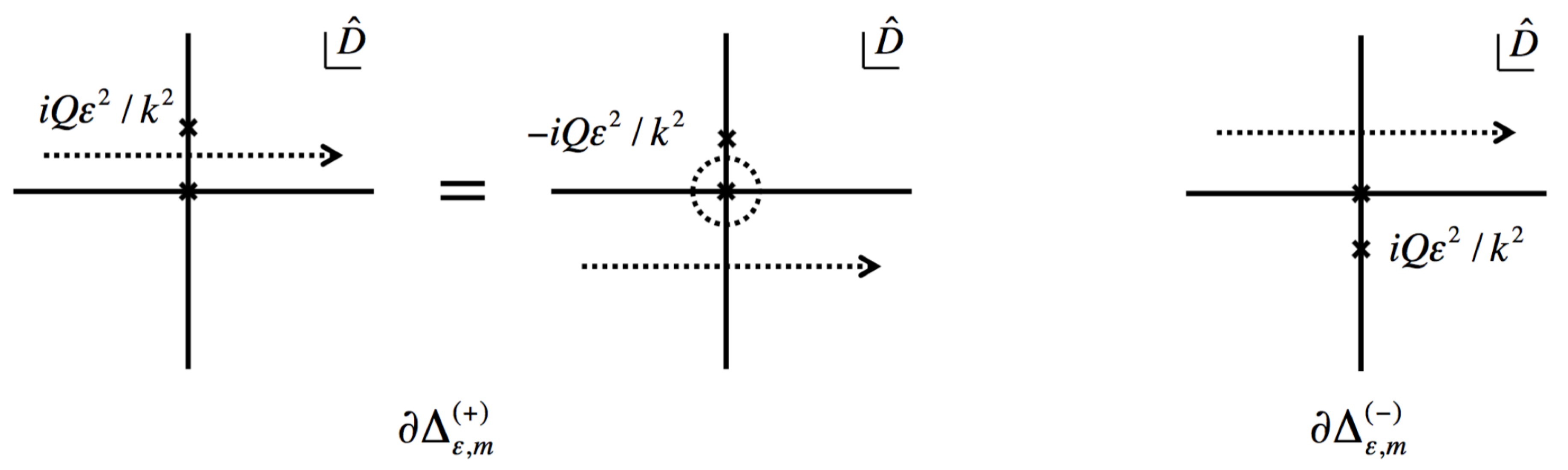}
\caption{The singularities in the $\hat D$ plane. When we choose $\delta>0$, the $\hat D$  integral from $\Delta_{\epsilon,m}^{(+)}$ are modified to that of $\hat D=0$ and a contour that passes negative imaginary axis, where the latter contribution can be deformed away to give a vanishing contribution. For 
$\Delta_{\epsilon,m}^{(-)}$, the contour can be deformed away to infinity. We set $\beta=1$ for simplicity.}
\label{contour1}
\end{center}
\end{figure}
The auxiliary parameter $\eta$ in the JK residue \eqref{main formula} is such that $\eta \delta >0$. Let us choose $\eta>0$ for definiteness. Then, for the contour $\partial\Delta_{\epsilon, m}^{(+)}$, the singularities in the $\hat D$ plane are depicted on the leftmost figure in Figure \ref{contour1}. Note that, as long as $|\delta|>0$, the integrand is bounded. 
The $\h D$ contour can be deformed to the one shown in the middle of Figure \ref{contour1}, which consists of a small contour around $\h D=0$ and of a contour in the lower-half plane. 
The latter contribution can be deformed away along the negative imaginary axis. 
Since the integrand evaluated on this latter contour is finite, the contour integral around $\partial\Delta_{\epsilon,m}^{(+)}$ gives a vanishing contribution. Hence we are left with the contour integral around $\h D=0$. For the $\partial\Delta_{\epsilon, m}^{(-)}$ contour, the $\h D$ contour is depicted in the last figure in Figure \ref{contour1}, which can be
similarly deformed away to give a vanishing answer. 
To summarize, at the singularity defined by the hyperplane $H_i$, we get
\be
\int_{\Gamma}\frac{d\h D}{\h D}\oint_{\partial\Delta^{(Q_i)}_{\epsilon}}Z_\m = \begin{cases}
\oint_{Q^2 u_i \t u'_i=\epsilon^2}  du\; \CZ_\m \Big|_{\Lambda_0=\t \Lambda_0=\h D=0}\, ~, &\qquad {\rm if}\qquad Q_i>0 \cr
0 &\qquad {\rm if}\qquad Q_i <0 
\end{cases}~,
\ee
in the case $\eta>0$. 
Similarly, in the case $\eta <0$ one can show that  $Z_i =0$ if $Q_i>0$ while we pick minus the $\h D=0$ pole if $Q_i <0$.  We will come back to the contributions of the `monopole singularities' $u= \mp i \infty$ in a moment, but for the time being we can note that they can be treated essentially like in \cite{Benini:2015noa,Closset:2015rna}.

Note that, until this point,  $\CZ_\m|_{\Lambda_0=\t \Lambda_0=0}$ still has a dependence on the $\Lambda_{1}, \t\Lambda_{\b 1}$ zero modes and on the $\Sigma_g$ flat connections, which must be integrated over.
In order to integrate out these   zero-modes, let us define
\be
f(\lambda,n) = \frac{Q\t \sigma'_n}{(\lambda+ Q^2\t\sigma'_n \sigma_n)(\lambda+ Q^2\t\sigma'_n \sigma_n+ i Q D)}~, 
\ee
and
\be
g= \prod_{\lambda, n} \left[{\lambda+ Q^2\t\sigma'_n \sigma_n\ov \lambda+ Q^2\t\sigma'_n \sigma_n+ i Q D} \right]\ . 
\ee
Then \eqref{Zmassive full} reads
\bea
Z_{\rm massive}^\Phi&= g
\exp \sum_{\lambda,n,Q}\ln\left[1-2if(\lambda,n)(Q\t\Lambda_{\b1})(Q\Lambda_1)\right]\\
&= g\exp \sum_{\lambda, n,Q}\sum_{s=1}^g \frac{-(2i)^{s}}{s} \left[f(\lambda,n) (Q\t\Lambda_{\b1})(Q\Lambda_1)\right]^s\ .
\eea
We are interested in the quantity
\be
\h \CF_s = \sum_{\lambda, n} f(\lambda_n)^s
\ee
evaluated at $\h D=0$. We can rewrite this as:
\be
\hat \CF_s(D=0) = \sum_{n\in \mathbb Z}(Q\t\sigma'_n)^s\zeta_n(2s)\ ,
\ee
where we defined
\bea\label{zeta heat kernel}
\zeta_{n}(2s) &&=& ~\sum_{\lambda}\frac{1}{(\lambda+Q^2\t\sigma'_n\sigma_n)^{2s}}\\
&&=& ~ \frac{1}{\Gamma(2s)} \int_{0}^\infty dt~t^{2s-1}\left(\sum_{\lambda}e^{-t\,\lambda}\right)e^{-t\,Q^2\t\sigma'_n\sigma_n}~.
\eea
We can now use the fact that we have introduced a rescaled variable  $\t u= \t u'/k^2$ with a positive number $k^2$.
Note that we are free to choose $k$ to our convenience in order to compute $f(\lambda, n)$, since all the contributions from the one-loop determinants, from the classical action and from the measure are independent of $\t u$ after the $\h D$ integral.
We will take $k$ arbitrarily small (which is equivalent to a large $\t \sigma_n'$ limit),
so that only the small  $t$ expansion of the heat kernel:
\be\label{heat kern}
\sum_{\lambda} e^{-t\lambda} = \frac{1}{4\pi t }\sum_{l=1}^\infty a_l t^l
\ee 
contributes to \eqref{zeta heat kernel}.
The first few coefficients $a_0, a_1, \cdots$ of \eqref{heat kern} are known to be spectral invariants \cite{berline1992heat,buser2010geometry}. In particular, we have:
\be
a_0 = \text{vol}(\Sigma)~,
\ee
which is also known as Weyl's law. Performing the $t$ integral in \eqref{zeta heat kernel}, we obtain:
\be
(Q\t\sigma'_n)^s\zeta_n(2s) = \frac{a_0 (Q\t\sigma'_n)^s}{4\pi(2s-1) (Q^2\t\sigma'_n\sigma_n)^{2s-1}} + \frac{a_1 (Q\t\sigma'_n)^s}{4\pi(2s) (Q^2\t\sigma'_n\sigma_n)^{2s}}+\cdots~.
\ee 
First, let us consider contributions from $s=1$. On the contour $\partial \Delta_\epsilon$ where $\t\sigma'_n\sigma_n=\epsilon^2$, the $l\geq 1$ terms are bounded by the
expression
\be
\frac{(\t\sigma'_n)^s}{(\t \sigma'_n\sigma_n)^{2s+l-1}} \sim \frac{1}{(\sigma_n)^s}\frac{k^{2(s+l-1)}}{\epsilon^{2(s+l-1)}} \rightarrow 0
\ee 
which vanishes if we take the limit $k\ll \epsilon$. When a contour is defined for the boundary component $u\t u'\rightarrow \infty$, the first term dominates as well.
Therefore, only the first term remains:
\bea
\lim_{k\rightarrow 0}\h\CF_1 (\h D=0)&= \lim_{k\rightarrow 0}(Q\t\sigma'_n/2e^2)\zeta_n(2)  = \sum_{n\in \mathbb{Z}} \frac{\text{vol}(\Sigma)}{4\pi Q\sigma_n} \\
&= {\beta\ov 2} {\rm vol}(\Sigma_g) \half \left(1+x^Q\ov 1-x^Q\right)\ .
\eea
Similarly,  we have
\be
\lim_{k\rightarrow 0}\h\CF_{s} (D=0) =0~, \qquad \text{ if } s>1\ .
\ee
To summarize, the dependence on $\Lambda_1$ and $\t \Lambda_{\b 1}$ can be written as\footnote{Here $a,b$ labels the gauge group indices for any higher-rank $\GG$.}
\be
\left.Z_{\rm massive}^{\Phi}\right|_{D=0} = g \exp\left[{-i\beta\rm vol(\Sigma_g)}\t\Lambda_{\bar 1}^a\Lambda_{1}^b H_{ab}(x)\right]\ ,
\ee
where 
\be\label{ner}
H_{ab}(x) = \half\sum_{Q} Q^a Q^b\left(1+x^Q\ov 1-x^Q\right)\ .
\ee
Note that it can be written in terms of the three-dimensional twisted effective superpotential $\CW_{\text{matter}}$:
\be
H_{ab} = \d_{u_a} \d_{u_b}\CW_{\text{matter}}\ ,
\ee
where 
\be
\CW_{\text{matter}} = \sum_i\sum_{\rho_i \in \CR_i}\left[\frac{1}{(2\pi i )^2}\text{Li}_2 (x^Qy_i)+\frac14 (Q_i(u)+\nu_i)^2\right]  
\ee
in general.
As an important consistency check, consider the classical Chern-Simons action \eqref{classical CS Lag} on this background (with $\Lambda_0=\t\Lambda_0=D=0$):
\be
e^{-S_{CS}} = x^{k\m}\, \exp\left( i\beta {\rm vol}(\Sigma_g) \, k^{ab}\,  \t\Lambda^a_{\b1} \Lambda^b_{1} \right)~.
\ee
The classical and one-loop terms come with the correct relative coefficients to reproduce the full twisted superpotential. 

This one-loop contribution and the contributions from the classical action are independent of $\alpha_I, \t\alpha_I$, and they have a simple dependence in the gaugini $\Lambda_I, \t\Lambda_I$. This allows us to perform the integral over these zero modes explicitly, which leads to the insertion of the Hessian determinant of the twisted superpotential:
\be
\int \prod_{I=1}^g  d\CV_I \,\CZ_\m \Big|_{\Lambda_0= \t\Lambda_0=\h D=0} =   H(u)^g\; \CZ_\m \Big|_{\Lambda_0= \t\Lambda_0=\Lambda_I= \t\Lambda_I=\h D=0}~.
\ee
Note that all the contributions are holomorphic in $u$ after the $\h D$ integral and after taking the $k\rightarrow 0$ limit. This allows us to tune $\epsilon \rightarrow 0$, while the result of the $u$-plane residue integral does not change. 
\footnote{We encountered several order of limits that we should be careful about. To summarize, the correct prescription is the following: 1) perform the $\h D$-integral;  2) take $k\rightarrow 0$;  3) take $\epsilon\rightarrow 0 $; 4) take $e\rightarrow 0$.}

The monopole singularities $H_\pm$ at $u = \mp i \infty$ can be discussed in the similar way as in \cite{Benini:2015noa,Closset:2015rna}, which we briefly summarize below. For 
this purpose, we need
to compute the dependence of $\h D$ linear part in the $\ln Z_{\rm massive}$ in the limit  $u = \mp i \infty$. It reads
\be
\left.\ln Z_{\rm massive}\right|_{\h D^a\text{-linear}} = -\sum_{\lambda}\frac{iQ^a}{(\lambda+Q^2\sigma_n\t\sigma_n)}\ ,
\ee
in large Im$(u)$. This can be evaluated from the observation
\be
\partial_{u_b}\left.\ln Z_{\rm massive}\right|_{\h D^a\text{-linear}} = \sum_{\lambda}\frac{Q^a Q^b (Q\t\sigma')/\beta}{(\lambda+Q^2\sigma_n\t\sigma_n)^2} = \frac{1}{2} H_{ab}
\ee
with $H_{ab}$ defined in \eqref{ner}. Integrating back, we find
\footnote{We added the anti-holomorphic piece to recover the fact that the expression is real. When we diffentiate the formula and integrate back, we lost the information of the phase in the argument of the $\log$.}
\be
\left.\ln Z_{\rm massive}\right|_{\h D\text{-linear}} = \frac{1}{2}\text{vol}(\Sigma_g)\left(\partial_u \CW_{\text{1-loop}}-\partial_{\t u'} \t\CW_{\text{1-loop}}\right)
\ee
where
\be
\partial_{u_a}\CW_{\text{1-loop}}
= -\frac{1}{2\pi i}\sum_i\sum_Q Q_i^a\left[\ln(1-x^Q y_i)-\pi i (\rho_i (u)+\nu_i)
\right]\ .
\ee
From here and onwards, we will set $\text{vol}(\Sigma_g)=1$. Taking the limit Im$(u)\rightarrow \mp \infty$, we get the $\h D$ dependence at infinity which is
\be
\int_{\Gamma(\eta)} \frac{d\h D}{\h D}\exp \left[-\frac{\pi \beta}{e^2}\h D^2\pm iQ_{\pm}\h D\text{Im}(u)\right]\ ,
\ee
where $Q_{\pm}$ is defined in \eqref{Q from W}. It is convenient to work with the rescaled variable $\hat D = e^2 \hat D'$. We have
\be
\int_{\Gamma(\eta)} \frac{d\h D'}{\h D'}\exp \left[-\pi \beta e^2\h D'^2\pm iQ_{\pm}e^2\h D'\text{Im}(u)\right]\ .
\ee
For the singularity at infinity, we can take $e\rightarrow 0$ before doing the $\hat D$ integral since the matter integrals are regulated with infinite mass. 
We take the limit $e\rightarrow 0$ at the same time as taking $|u| \rightarrow \infty$ in such a way that $e^2 |u|\rightarrow a$ for some finite number $a>0$. Then we have
\be
\int_{\Gamma(\eta)} \frac{d\h D'}{\h D'}\exp \left[- ia~Q_{\pm}\h D'\right]\ .
\ee
Suppose that we have a $\hat D$ integral defined at $\Gamma_+$ with positive $\delta$ as in Figure \ref{contour1}. Then the $\hat D$ contour integral can be done as follow. When $\text{Im}(u)\rightarrow -\infty$, we have
\be
\int_{\Gamma(\eta)} \frac{d\h D'}{\h D'}\exp \left[- ia~Q_{\pm}\h D'\right]= \begin{cases}
2\pi i~, &\qquad {\rm if}\qquad Q_+>0 \cr
0 &\qquad {\rm if}\qquad Q_+ <0 
\end{cases}~,\qquad 
{\rm with} \; \eta>0~
\ee
On the other hand, when $\text{Im}(u)\rightarrow \infty$, we have
\be
\int_{\Gamma(\eta)} \frac{d\h D'}{\h D'}\exp \left[- ia~Q_{\pm}\h D'\right]= \begin{cases}
2\pi i~, &\qquad {\rm if}\qquad Q_->0 \cr
0 &\qquad {\rm if}\qquad Q_- <0 
\end{cases}~,\qquad 
{\rm with} \; \eta>0~
\ee
If we choose $\eta<0$, the poles associated to $Q_\pm <0$ contribute instead.

Finally, let us consider theories with $k_{\text{eff}}=0$ (at either infinity). In this case, we can turn on an auxiliary ($Q$-exact) FI parameter $\t\xi/e^2$ which only couples to $\hat D$. Then the integral at infinity reads:
\be
\int_{\Gamma(\eta>0)} \frac{d\h D'}{\h D'}\exp \left[i \t\xi\h D '\right] = 2\pi i\Theta(-\t\xi)\ .
\ee
Since the choice of $\eta$ is arbitrary, we can set $\eta = \t\xi$ such that there is never any contribution from the singularities at infinity. 
Since the 3d theory does not suffer from wall-crossing phenomena,  the answer should not depend on the choice of auxiliary FI parameter $\t\xi = \eta$. 
The integration over $\Lambda_I, \t\Lambda_{\b I}$ and $\alpha_I$, $\t\alpha_I$ can be done in exactly same way as in the bulk singularities discussed above, resulting in a $H(u)^g$ insertion to the path integral.

\subsection{The general case}
The generalization to the higher rank $\GG$ involves technical difficulties due to the non-trivial topology of the $\tfM\backslash \tfM^{\m}_\text{sing}$. However, 
given the detailed discussion of rank one theory, the generalization to the higher rank $\GG$ follows directly from the discussions in the previous literatures \cite{Benini:2013xpa,Hori:2014tda,Benini:2015noa,Closset:2015rna}. The additional ingredient is
the insertion of the $H(u)^g$, resulting from the one-form gaugino zero modes.
The final answer can be written as a Jeffrey-Kirwan residue:
\bea
 {1 \ov |W_\GG|} \sum_{\m \in \Gamma_{\mathbf{G}^\vee}}
\sum_{u_* \in \tfM^{\m}_\text{sing}}
\underset{u=u_*}{ {\text{JK-Res}}} \left[ \mathbf{Q}(u_*),  \eta  \right]
Z_{\text{1-loop}}^{\text{vector}}(u,\m,g)Z_{\text{1-loop}}^{\Phi}(u,\m,g)~ H^g(u)\ ,
\eea
where $\tfM^{\m}_\text{sing}$ contains all the singularities from $H_i$ and $H_\pm$. This formula is discussed in details in section \ref{subsec: loc formula}.

\section{Decoupling limits for 3d $\CN=2$ SQCD in flat space}\label{appendix: Seiberg duals}
In this Appendix, we briefly review Seiberg dualities for the three-dimensional $\CN=2$ supersymmetric SQCD$[k, N_c, N_f, N_a]$ of section \ref{sec: sqcd and seiberg duals}. 
Starting from Aharony duality \cite{Aharony:1997gp} for SQCD$[0,N_c, N_f, N_f]$, we derive all the other Seiberg dualities  \cite{Giveon:2008zn, Benini:2011mf} by real mass deformations.%
~\footnote{We follow the analysis of \cite{Benini:2011mf} but choose  somewhat better conventions.   Thus the results of this Appendix for the relative flavor CS terms across dualities look a bit different from the ones of  \cite{Benini:2011mf}. (In \cite{Benini:2011mf}, the $U(1)_A$ and $U(1)_R$ symmetries were mixed with the gauge symmetry, corresponding to setting $k_{gR}= k_{gA}=0$. For that reason, the $R$- and flavor charges of the monopole operators  in that reference were not necessarily integer-quantized.) }

In three dimensions,  the two-point function of conserved currents contains an interesting conformally-invariant contact term, whose corresponding local term is a Chern-Simons functional for  background gauge fields \cite{Witten:2003ya, Benini:2011mf, Closset:2012vp}.  Whenever  the CS levels are quantized---that is, if the corresponding symmetry group is compact, these contact terms are physical up to integer shifts of the `global' CS levels \cite{Closset:2012vp}. 
While the global CS  levels of a given theory can be specified arbitrarily,  their relative values might differ across dualities. As part of the description of the duality, we need to specify the relative CS levels:
\be\label{def rel CSf}
\Delta k_{F} \equiv   k_{F}^D-    k_{F}~,
\ee
where $k_{F}$, $k_{F}^D$ are the global CS levels in the original theory and in the  dual theory, respectively.

\subsection{Aharony duality and real mass deformations}\label{app subsec: aharony}
Consider a $U(N_c)$ YM theory with vanishing CS level, with $N_f$ pair of fundamental and anti-fundamental chiral multiplets $Q_i$ ($i=1, \cdots, N_f$) and $\t Q^j$ ($j=1, \cdots, N_f$), and a vanishing superpotential. The theory has a flavor symmetry group $ SU(N_f) \times SU(N_f)  \times U(1)_A \times  U(1)_T$ and a $R$-symmetry $U(1)_R$, under which the matter fields have charges:
\be\nn
\begin{array}{c|c|ccccc}
    &  U(N_c)& SU(N_f) & SU(N_f)  & U(1)_A &  U(1)_T & U(1)_R  \\
\hline
Q_i        & \bm{N_c}& \bm{\overline{N_f}} & \bm{1}& 1   & 0   &r \\
\tilde{Q}^j   & \bm{\overline{N_c}}  &  \bm{1}& \bm{N_a}  & 1   & 0   &r 
\end{array}
\ee
Most of the classical $U(N_c)$ Coulomb branch of this theory is lifted by an instanton-generated superpotential \cite{Aharony:1997bx, deBoer:1997kr}, but the overall $U(1)$ direction remains, parameterized with the two monopole operators $T^\pm$ with charge $\pm 1$ under the topological symmetry $U(1)_T$. (The operator $T^\pm(x)$  inserts a magnetic flux $(\pm 1, 0, \cdots, 0)$ at $x\in \R^3$.) The two operators $T^\pm$ have induced $U(1)_A$ and $R$-charges given by:
\be
Q^A_\pm =   -N_f~,  \qquad\qquad r_T \equiv r_\pm = -N_f(r-1)-N_c+1~. 
\ee
Let ${M^j}_i= \t Q^j Q_i$ be the gauge-invariant `mesons', which parameterize the Higgs branch.
We consider the case $N_f \geq N_c$, which preserves both the $R$-charge and supersymmetry. For $N_f= N_c$, the IR theory can be described as a $\sigma$-model for the mesons and for two additional chiral multiplets $T^\pm$ identified with the monopole operators, interacting through the superpotential \cite{Aharony:1997bx}:
\be
W= T^+ T^- \det(M)~.
\ee
A particular instance is for $N_f=N_c=1$, which is the SQED/$XYZ$-model duality considered in section \ref{subsec:SQED XYZ}. 
For $N_f > N_c$, there is a dual description in terms of an $U(N_f-N_c)$ gauge group with $N_f$ fundamental and antifundamental chiral multiplets $q^i$, $\t q_j$ and the gauge singlets ${M^j}_i$, $T^+$ and $T^-$, with superpotential:
\be\label{W aharony}
W=\t q_j {M^j}_i q^i + T^+ t_+ + T^- t_-~,
\ee
where $t_\pm$ are the monopole operators of the dual gauge group \cite{Aharony:1997gp}. The quantum numbers of the dual matter fields are summarized in Table \ref{tab: Aharony duality charges} on page \pageref{tab: Aharony duality charges}.
Finally, all the relative flavor CS levels \eqref{def rel CSf} vanish for this duality.

Starting from this duality, we derive the Seiberg-like dualities of the other $\CN=2$ $U(N_c)$ YM-CS-matter theories with fundamental and antifundamental matter, which we dubbed SQCD$[k,N_c, N_f, N_a]$ in section \ref{sec: sqcd and seiberg duals}. If we turn on a large real mass $m_0$ for a global symmetry $U(1)_0$,  we generate the CS levels:
\bea
&\delta k_{IJ} =\half \sum_{i} \sign{\left(Q^0_i m_0\right)}   \;Q^I_i\,Q^J_i~, \cr
&\delta k_{IR} =\half \sum_i \sign{\left(Q^0_i m_0\right)} \; Q^I_i \,(r_i-1)~,
\eea
for all  abelian symmetries $U(1)_I$, $U(1)_J$ and $U(1)_R$, and similarly for any non-abelian symmetry. Here the sum runs over all chiral multiplet field components with charges $Q_i^I$ and $R$-charge $r_i$.

\subsubsection{Seiberg  duality with $k>k_c \geq 0$}
Consider SQCD$[k, N_c, N_f, N_a]$, a $U(N_c)$ theory with CS level $k >0$, $N_f$ fundamental and $N_a$ antifundamental chiral multiplets. We consider $k_c \equiv \half(N_f-N_a)\geq 0$ and $k > k_c$. 
This theory can be obtained from SQCD$[0,N_c, n, n]$ with
\be
n= k+ \half (N_f+ N_a)~,
\ee
by integrating out $k-k_c$ fundamental chiral multiplets $Q_\alpha$ with positive real mass and $k+k_c$ antifundamental chiral multiplets $\t Q^\beta$ with positive real mass, while the remaining $N_f$ fundamental chiral multiplets $Q_i$ and $N_a$ antifundamental chiral multiplets $Q_j$ remain light. The corresponding real mass $m_0 >0$ is such that:
\be
\sigma_a- m_i =0~, \qquad \sigma_a- m_\alpha = m_0~, \qquad -\sigma_a + \t m_j =0~, \qquad 
-\sigma_a + \t m_\beta =m_0~,
\ee
in the limit $m_0\rightarrow \infty$. We also need to scale the FI term as:
\be
\xi = k_c m_0~,
\ee
in order for the effective FI parameter $\xi_{\rm eff} = \xi - k_c |m_0|$ to remain finite. This means that the symmetry $U(1)_0$ contains a mixing with $U(1)_T$. The charges of the `electric' theory $U(N_c)$ with $n_f$ flavors are:
\be\nn
\begin{array}{c|c|cc|ccccc|c}
    &  U(N_c)& SU(N_f) & SU(N_a)  & U(k-k_c) &U(k+k_c) &U(1)_A &  U(1)_T & U(1)_R & U(1)_0  \\
\hline
Q_i        & \bm{N_c}& \bm{\overline{N_f}} & \bm{1}&  \bm{1}& \bm{1}& 1   & 0   &r  &0 \\
Q_\alpha       & \bm{N_c}& \bm{1} & \bm{1}&  \bm{\overline{k-k_c}} &  \bm{1} & 1   & 0   &r  & 1\\
\tilde{Q}^j   & \bm{\overline{N_c}}  &  \bm{1}& \bm{N_a}  &  \bm{1}& \bm{1}& 1   & 0   &r &0 \\
\tilde{Q}^\beta   & \bm{\overline{N_c}}  &  \bm{1}& \bm{1}  &  \bm{1}& \bm{k+k_c}& 1   & 0   &r &1 
\end{array}
\ee
Here the $U(1)_0$ charge is indicated in the last  column. Sending $m_0 \rightarrow \infty$, we integrate out $Q_\alpha$ and $\t Q^\beta$ and obtain the CS levels:
\be
k_{gg} = k~, \qquad k_{g A}= - k_c~, \qquad k_{gR} = - k_c (r-1)~,
\ee
for the gauge CS levels. We also generate the following flavor CS levels:
\be\label{dual 2 CS f}
k_{AA} = N_c k~, \qquad k_{AR} = N_c k (r-1)~.
\ee
We also generate a level $k_{RR}$, which we will ignore throughout because such terms  do not play any role on $\Sigma_g \times S^1$. All other flavor CS levels vanish. 

\begin{table}[t]
\centering
\be\nn
\begin{array}{c|c|cc|ccccc|c}
    &  U(n-N_c)& SU(N_f) & SU(N_a)  & U(k-k_c) &U(k+k_c) &U(1)_A &  U(1)_T & U(1)_R & U(1)_0  \\
\hline
q^i        & \bm{\overline{n-N_c}}& \bm{N_f} & \bm{1}&  \bm{1}& \bm{1}& -1   & 0   &1-r  &0 \\
q^\alpha       &\bm{\overline{n-N_c}} & \bm{1} & \bm{1}&  \bm{k-k_c} &  \bm{1} & -1   & 0   &1-r  & -1\\
\tilde{q}_j   & \bm{n-N_c}  &  \bm{1}& \bm{\overline{ N_a}}  &  \bm{1}& \bm{1}& -1   & 0   &1-r &0 \\
\tilde{q}_\beta   & \bm{n-N_c}  &  \bm{1}& \bm{1}  &  \bm{1}& \bm{\overline{k+k_c}}& -1   & 0   &1-r &-1 \\
{M^j}_i &  \bm{1}    &  \bm{\overline{N_f}}   &   \bm{N_a}   &  \bm{1}     &  \bm{1}   & 2 & 0 & 2r & 0 \\
{M^\beta}_i    &  \bm{1}   &   \bm{\overline{N_f}}   &  \bm{1}   &  \bm{1}  &  \bm{k+k_c}    & 2 & 0 & 2r & 1 \\
{M^j}_\alpha    &  \bm{1}   &   \bm{1}   &  \bm{N_a}   &  \bm{\overline{k-k_c}}   &  \bm{1}   & 2 & 0 & 2r & 1 \\
{M^\beta}_\alpha    &  \bm{1}   &   \bm{1}   &  \bm{1}   &\bm{\overline{k-k_c}}    &  \bm{k+k_c}    & 2 & 0 & 2r & 2 \\
T^+   &  \bm{1}   &   \bm{1}   &  \bm{1}   &  \bm{1}   &  \bm{1}   & -n & 1 &r_T & -k + k_c  \\
T^-   &  \bm{1}   &   \bm{1}   &  \bm{1}   &  \bm{1}   &  \bm{1}   & -n & -1 &r_T & -k - k_c  
\end{array}
\ee
\caption{Charges of the matter fields in the $U(n-N_c)$ Aharony dual theory used to derive the Seiberg dual of SQCD with $k> k_c \geq 0$. Here $r_T= -n (r-1) -N_c +1$.}
\label{tab: Aharony duality charges for RG flow 1}
\end{table}
We can follow the same RG flow in the Aharony dual $U(n-N_c)$ theory. The dual matter fields are summarized in Table \ref{tab: Aharony duality charges for RG flow 1}. Integrating out all the fields with $Q_0 \neq 0$, we generate the gauge CS levels:
\be
k^D_{gg} = -k~, \qquad k^D_{g A}=  k_c~, \qquad k^D_{gR} =  k_c r~,
\ee
and the flavor CS levels:
\bea\label{dual 2 CS fD}
& k^D_{SU(N_f)}=  \half(k+k_c)~, \qquad   && k^D_{SU(N_a)}=  \half(k-k_c)~,\cr
& k^D_{AA}= k N_c+ \half(N_f+N_a) n - 2 N_f N_a~, \qquad && k^D_{TT}= -1~,\cr
&k^D_{AR} = {N_f+ N_a\ov 2}(n-N_c) - N_f N_a + (r-1)  k^D_{AA}~,\quad &&
\eea
and all other mixed CS levels vanishing. From \eqref{dual 2 CS f} and \eqref{dual 2 CS fD}, we find the relative global CS levels \eqref{rel CS 02}-\eqref{rel CS 2}.

\subsubsection{Seiberg  duality with $k_c > k >  0$}
Consider SQCD$[k, N_c, N_f, N_a]$ with CS level $k >0$ and $k_c >k$. This theory can be obtained from SQCD$[0, N_c, N_f, N_f]$ by integrating out $k_c+ k$ antifundamental multiplets $\t Q^\beta$ with positive real mass and $k_c-k$ antifundamental multiplets $\t Q^\gamma$ with negative real mass. The relevant real mass $m_0$ is such that:
\be
\sigma_a- m_i =0~, \qquad -\sigma_a+ \t m_j = 0~, \qquad -\sigma_a + \t m_\beta = m_0~, \qquad 
-\sigma_a + \t m_\gamma =-m_0~,
\ee
in the limit $m_0\rightarrow \infty$. We also need to scale the FI term as:
\be
\xi = k_c m_0~.
\ee
The charges of the fields in the `electric' theory with $N_f$ flavors are:
\be\nn
\begin{array}{c|c|cc|ccccc|c}
    &  U(N_c)& SU(N_f) & SU(N_a)  & U(k_c+k) &U(k_c-k) &U(1)_A &  U(1)_T & U(1)_R & U(1)_0  \\
\hline
Q_i        & \bm{N_c}& \bm{\overline{N_f}} & \bm{1}&  \bm{1}& \bm{1}& 1   & 0   &r  &0 \\
\tilde{Q}^j   & \bm{\overline{N_c}}  &  \bm{1}& \bm{N_a}  &  \bm{1}& \bm{1}& 1   & 0   &r &0 \\
\tilde{Q}^\gamma   & \bm{\overline{N_c}}  &  \bm{1}& \bm{1}  &  \bm{k_c+k}& \bm{1}& 1   & 0   &r &1 \\
\tilde{Q}^\gamma   & \bm{\overline{N_c}}  &  \bm{1}& \bm{1}  &  \bm{1}& \bm{k_c-k}& 1   & 0   &r &-1 
\end{array}
\ee
Integrating out the massive fields generates the gauge CS levels:
\be
k_{gg} = k~, \qquad k_{g A}= - k~, \qquad k_{gR} = - k  (r-1)~,
\ee
and the global CS levels:
\be
k_{AA}= k N_c~, \qquad k_{RA}= k N_c (r-1)~.
\ee
\begin{table}[t]
\centering
\be\nn
\begin{array}{c|c|cc|ccccc|c}
    &  U(n-N_c)& SU(N_f) & SU(N_a)  & U(k_c+k_c) &U(k_c-k_c) &U(1)_A &  U(1)_T & U(1)_R & U(1)_0  \\
\hline
q^i        & \bm{\overline{n-N_c}}& \bm{N_f} & \bm{1}&  \bm{1}& \bm{1}& -1   & 0   &1-r  &0 \\
\tilde{q}_j   & \bm{n-N_c}  &  \bm{1}& \bm{\overline{ N_a}}  &  \bm{1}& \bm{1}& -1   & 0   &1-r &0 \\
\tilde{q}_\beta   & \bm{n-N_c}  &  \bm{1}& \bm{1}  &  \bm{\overline{k_c+k}}& \bm{1}& -1   & 0   &1-r &-1 \\
\tilde{q}_\gamma   & \bm{n-N_c}  &  \bm{1}& \bm{1}  &  \bm{1}& \bm{\overline{k_c-k}}& -1   & 0   &1-r &1 \\
{M^j}_i &  \bm{1}    &  \bm{\overline{N_f}}   &   \bm{N_a}   &  \bm{1}     &  \bm{1}   & 2 & 0 & 2r & 0 \\
{M^\beta}_i    &  \bm{1}   &   \bm{\overline{N_f}}   &  \bm{1}  & \bm{k_c+k}  &  \bm{1}     & 2 & 0 & 2r & 1 \\
{M^\gamma}_i    &  \bm{1}   &   \bm{1}   &  \bm{N_a}     &  \bm{1} &  \bm{k_c-k}   & 2 & 0 & 2r & -1 \\
T^+   &  \bm{1}   &   \bm{1}   &  \bm{1}   &  \bm{1}   &  \bm{1}   & -N_f & 1 &r_T & -k + k_c  \\
T^-   &  \bm{1}   &   \bm{1}   &  \bm{1}   &  \bm{1}   &  \bm{1}   & -N_f & -1 &r_T & -k - k_c  
\end{array}
\ee
\caption{Charges of the matter fields in the $U(N_f-N_c)$ Aharony dual theory used to derive the Seiberg dual of SQCD with $k_c \geq k > 0$. Here $r_T= -N_f(r-1) -N_c +1$.}
\label{tab: Aharony duality charges for RG flow 2}
\end{table}
The charges of the fields in the dual field theory in the UV are given in Table \ref{tab: Aharony duality charges for RG flow 2}. Integrating out the massive fields, we obtain a $U(N_f-N_c)$ theory at CS level $-k$ with the mixed CS levels \eqref{kmixed dual 3}. We easily verify that the relative CS levels are given by \eqref{rel CS 3}.

\subsubsection{Seiberg  duality with $k_c = k >  0$}
The limiting case $k=k_c$ is obtained by the same reasoning as in the previous subsection. The only difference is that the singlet $T^+$ in the Aharony dual remains massless---see Table \ref{tab: Aharony duality charges for RG flow 2}. 

In this case, the singlet $T^+$ is dual to the `half' Coulomb branch that survives in the $U(N_c)_{k_c}$ theory. The $U(N_f-N_c)$ dual theory also contain a superpotential 
\be\label{W aharony 2}
W=\t q_j {M^j}_i q^i + T^+ t_+~,
\ee
coupling $T^+$ to a monopole of the dual gauge group. In the particular case  $N_c= N_f$, the gauge theory is  dual to a free theory of $N_f N_a +1$ chiral multiplets  ${M^{j}}_i$ and $T^+$. 
The case with $N_c=N_f=1$ and $N_a=0$ was considered in section \ref{subsec: U1half}.

\section{Proving the equality of Seiberg-dual indices}\label{appendix: proof equality indices}
In this Appendix, we briefly explain how to prove the equality of the twisted indices between the Seiberg dual theories considered in section \ref{subsec: bethe and duals}. 
Consider SQCD$[k,N_c, N_f, N_a]$ with $k\geq 0$ and $k_c \geq 0$, which is governed by the characteristic polynomial of degree $n$:
\be\label{Px full app}
P(x) = \prod_{i=1}^{N_f} (x-y_i ) - q\, y_A^{Q_+^A} x^{k+ k_c} \prod_{j=1}^{N_a} (x-\t y_j)~,
\ee
Let us denote by $\{\h x_\alpha\}_{\alpha=1}^n$ the $n$ distinct roots of $P(x)$. 
Given the quantities $\CU, \CH$ and $\CU_D, \CH_D$ defined in \eqref{CU sqcd}-\eqref{CH sqcd} and \eqref{CU sqcd dual}-\eqref{CH sqcd dual}, respectively, we can show that:
\be\label{U H U H dual app}
\CU(\h x)= \cu \; \CU_D(\h x_D)~,\qquad \qquad \CH(\h x)=\ch\; \CH_D(\h x_D)~, 
\ee
where $\h x\equiv \{\h x_a\}_{a=1}^{N_c} \subset \{\h x_\alpha\}$ is a choice of  $N_c$ distinct roots of $P(x)$, and $\h x_D\equiv \{\h x_{\b a}\}_{\b a=1}^{n-N_c}$ its complement.

\paragraph{Identities satisfied by $P(x)$.}
 From the factorization:
\be\label{poly}
 P(x) = C(q) \prod_{\alpha=1}^n (x-\h x_\alpha)~,\qquad C(q) = \begin{cases}
1 -q\, y_A^{-N_f} &\qquad {\rm if}\qquad k=k_c\geq 0 \cr
-q\, y_A^{-N_f} &\qquad {\rm if}\qquad k>k_c\geq 0 \cr
1 &\qquad {\rm if}\qquad k_c>k \geq 0 
\end{cases}~,
\ee
we obtain a useful identity for the product of all the roots:
 \be
  \prod_{\alpha=1}^n  \h x_\alpha = {(-1)^n\ov C(q)} P(0) = {(-1)^{n+N_f} \ov C(q)}  \h p_0~,
\ee
where we defined:
\be
 \h p_0\equiv 
  \begin{cases}
 y_A^{-N_f}-q &\; {\rm if}\; k=k_c=0 \cr
y_A^{-N_f} &\; {\rm if}\; k+k_c > 0 
\end{cases}~.
\ee
Note that we used \eqref{kgA explicit} in the above equations. 
Similarly, we find:
\bea
& \prod_{\alpha=1}^n (y_i- \h x_\alpha) = {1\ov C(q)} P(y_i)= { (-1)\ov C(q)}  q\,  y_A^{-N_f} y_i^{k+k_c} \, \prod_{j=1}^{N_a}(y_i-\t y_j)~, \cr
 & \prod_{\alpha=1}^n (\h x_\alpha -\t y_j) = {(-1)^n\ov C(q)} P(\t y_j)={(-1)^{n+N_f} \ov C(q)}  \prod_{i=1}^{N_f}(y_i-\t y_j)~.
 \eea
We also need the following lemma. Consider partitioning the set of roots $\{\h x_\alpha\}_{a=1}^n$ into a subset $\h x\equiv \{\h x_a\}_{a=1}^{N_c}$ and its complement $\h x_D\equiv \{\h x_{\b a}\}_{\b a=1}^{n-N_c}$. It is easy to show that:
\be
{\prod_a \d_x P(\h x_a) \ov \prod_{a \neq b} (\h x_a- \h x_b)} = (-1)^{N_c(n-N_c)} C(q)^{2 N_c-n} {\prod_{\b a} \d_x P(\h x_{\b a}) \ov \prod_{\b a \neq \b b} (\h x_{\b a}-\h x_{\b b})}~,
\ee
for any polynomial $P(x)$.

\vskip0.5cm
\paragraph{Explicit form of $\cu$ and $\ch$.}
By direct computation, we can show that:
\be\label{u h explicit}
\cu =(-1)^{s_\cu}\;\cu_M  \; Z_{CS}^{SU(N_f)} Z_{CS}^{SU(N_a)} \; \h \cu~, 
\qquad\qquad \ch =(-1)^{s_\ch} \; \ch_M \; \h \ch~.
\ee
Here $ \cu_M$ and  $\ch_M$ are  the contributions of the mesons ${M^j}_i$ defined in \eqref{def uM hM}. We also introduced the quantities
\be
Z_{CS}^{SU(N_f)}= \left(\prod_{i=1}^{N_f} y_i^{s_i}\right)^{k+k_c -\half(n-N_a)}~, \qquad\qquad Z_{CS}^{SU(N_a)} = \left(\prod_{j=1}^{N_a} y_i^{\t s_j}\right)^{\half(n-N_f)}~,
\ee
with the $SU(N_f)\times SU(N_a)$ fluxes defined by $s_i = \n_i +\n_A$ and $\t s_j= \n_j- \n_A$. These are the contributions from the $SU(N_f)\times SU(N_a)$ flavor Chern-Simons terms at level:
\be
k_{SU(N_f)} = k+k_c-\half(n-N_a)~, \qquad\qquad k_{SU(N_a)} = \half(n-N_f)~.
\ee
The signs in \eqref{u h explicit} are given by:
\bea
&(-1)^{s_\cu} = (-1)^{(n-N_c)(Nf-N_a)} (-1)^{(n+N_f) \n_T + N_f^2 \n_A}~, \cr
&(-1)^{s_\ch}=(-1)^{(n-N_c)(N_f-N_a) + N_f^2 r}~.
\eea
The remaining factors in \eqref{u h explicit} read:
\be\label{full cu app}
 \h {\cu} =  {{\h p}_0}^{\;\;  \n_T - Q_-^A \n_A} \; C(q)^{- \n_T + N_f \n_A}\; q^{-N_f\n_A} \; y_A^{\left[\half n (N_f+N_a) - N_a N_f -N_f Q_+^A \right]\n_A}~,
\ee
and 
\bea\label{full ch app}
& \h {\ch} =    {{\h p}_0}^{\, -(r_- -1)} C(q)^{r N_f+ N_c- n} q^{- r N_f + n- N_c} y_A^{\left[ (N_f+N_c-n)(k-Q_+^A) + N_f k_c \right]}\cr
&\qquad \times y_A^{(r-1) \left[\half n (N_f+N_a) - N_a N_f -N_f Q_+^A \right]}~,
\eea
with $Q_\pm^A$ and $r_-$ given by \eqref{def QA rpm} and \eqref{kgA explicit}.
One can evaluate these terms in the four cases $k=k_c=0$, $k>k_c \geq 0$, $k_c>k\geq 0$ and $k=k_c >0$, to complete the proof the equality of the twisted indices across the corresponding Seiberg dualities.

\section{Vortex-Wilson loop duality in $\CN=4$ theories} \label{Appendix:loop}
In this section, we briefly review some of the  results of \cite{Assel:2015oxa}, where the
duality mapping between  half-BPS Wilson loops and  vortex loops under  $3d$ $\CN=4$ mirror symmetry was studied. For $\CN=4$ quiver theories engineered in type IIB string theory,
it was shown that the  vortex loop mirror to  a Wilson loop in a given representation $\CR$ of $\GG$ can be described by a  1d supersymmetric quantum mechanics, which can be read off from the brane configuration. On general ground, such 1d GLSMs coupled to the three-dimensional theory provide a useful UV descriptions of vortex loop operators.

\begin{figure}[t]
\begin{center}
\includegraphics[width=13cm]{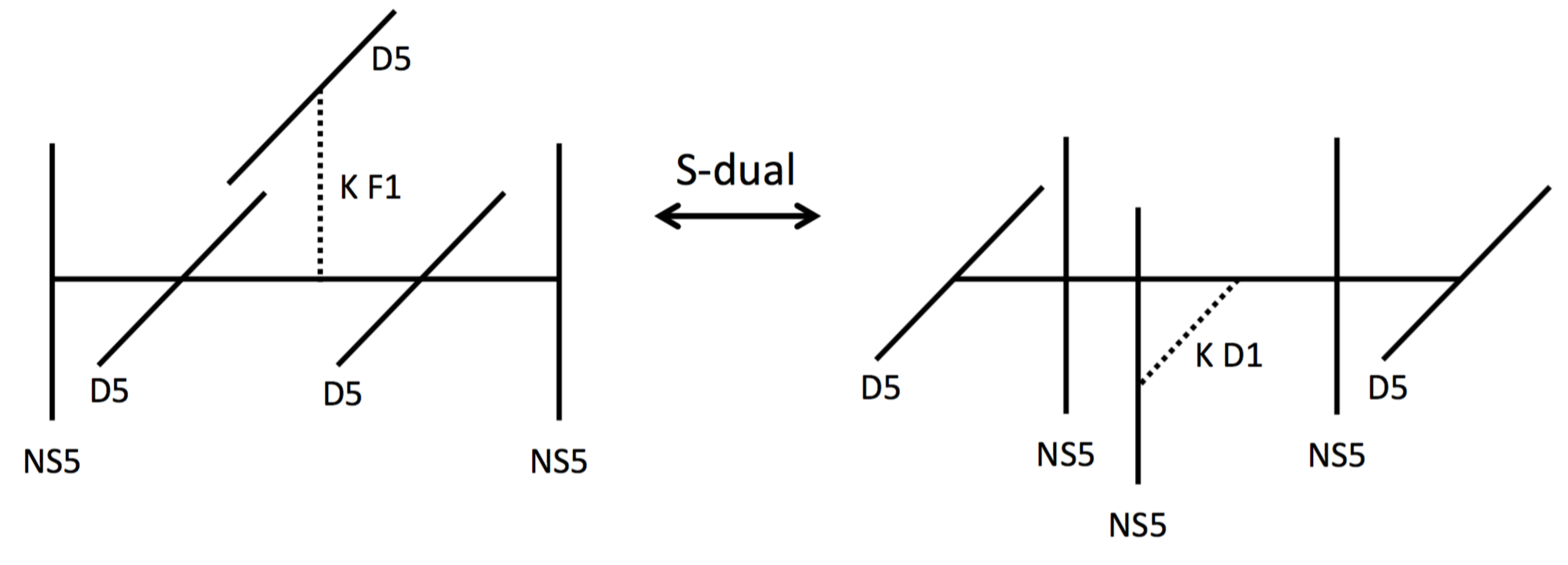}
\caption{Brane construction of the charge $k$ Wilson loop for $T[SU(2)]$, and its S-dual
configuration. The horizontal segment represents a stretched D3-brane, which is invariant under S-duality.}
\label{brane1}
\end{center}
\end{figure}

\begin{figure}[t]
\begin{center}
\includegraphics[width=13cm]{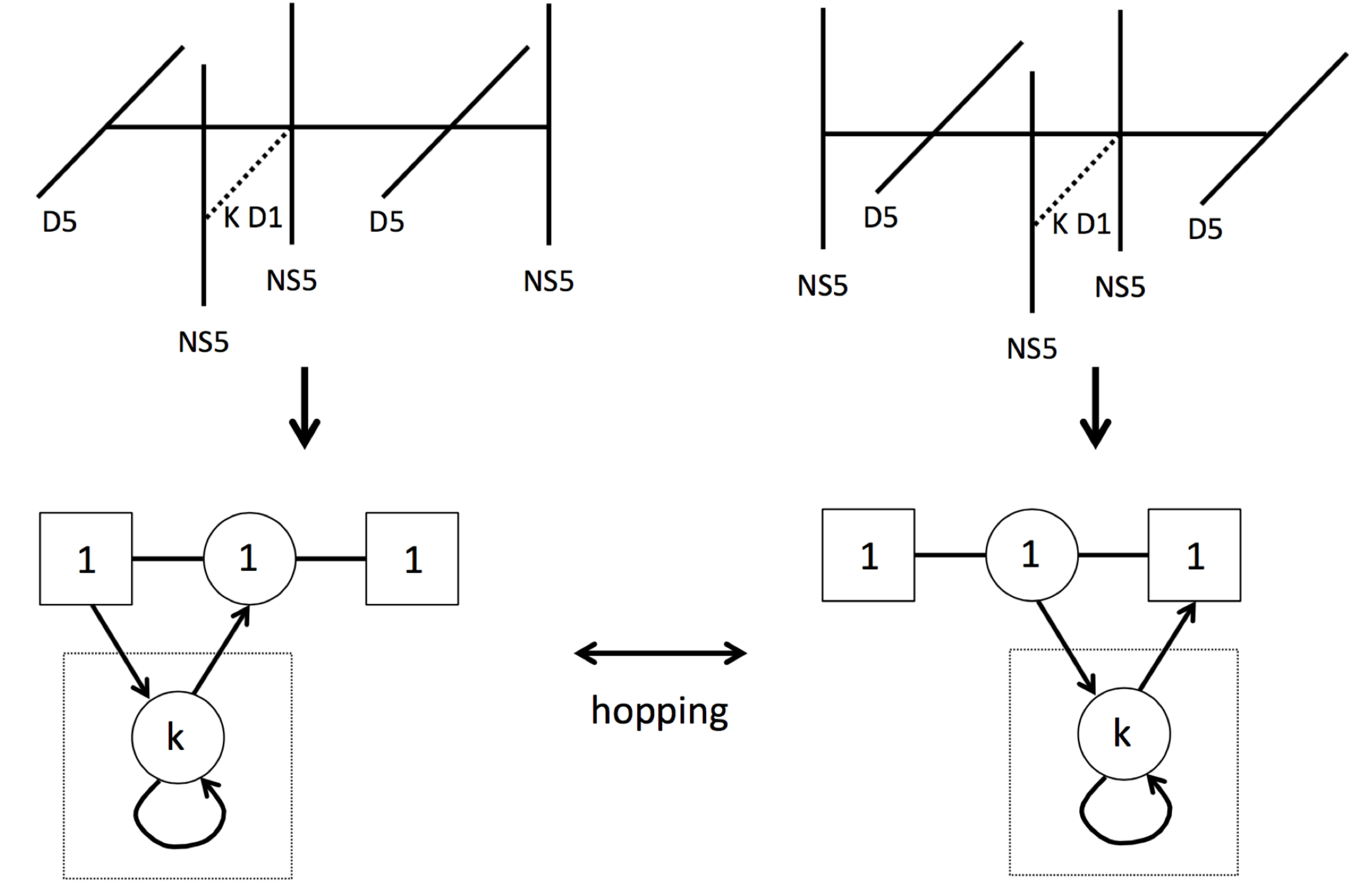}
\caption{Hanany-Witten brane move \protect\cite{Hanany:1996ie} of the S-dual configuration for $T[SU(2)]$ theory. The field contents in the dotted box are coupled 1d theory.
 If the $k$ D1-branes are attached to the left (right) NS5-brane, the 1d quiver is coupled to the (anti-)fundamental and to the gauge node of $3d$ theory.}
\label{brane2}
\end{center}
\end{figure}

For example, the charge $k$ Wilson loop in $T[SU(2)]$ has a brane construction in terms of $k$ fundamental strings, shown  on the left in Figure \ref{brane1}.
In the S-dual brane configuration, the $k$ D1-branes can be moved
along the D3-brane, so that they end up on top of the left NS5-brane or if the right NS5-brane.
The field content  of the 1d worldvolume theory on the D1-brane can be read off 
in either case as a quiver shown in Figure \ref{brane2}. The two quiver descriptions are two distinct but IR-equivalent realizations of the vortex loop, which is known as hopping duality \cite{Assel:2015oxa}.

\begin{figure}[t]
\begin{center}
\includegraphics[width=4cm]{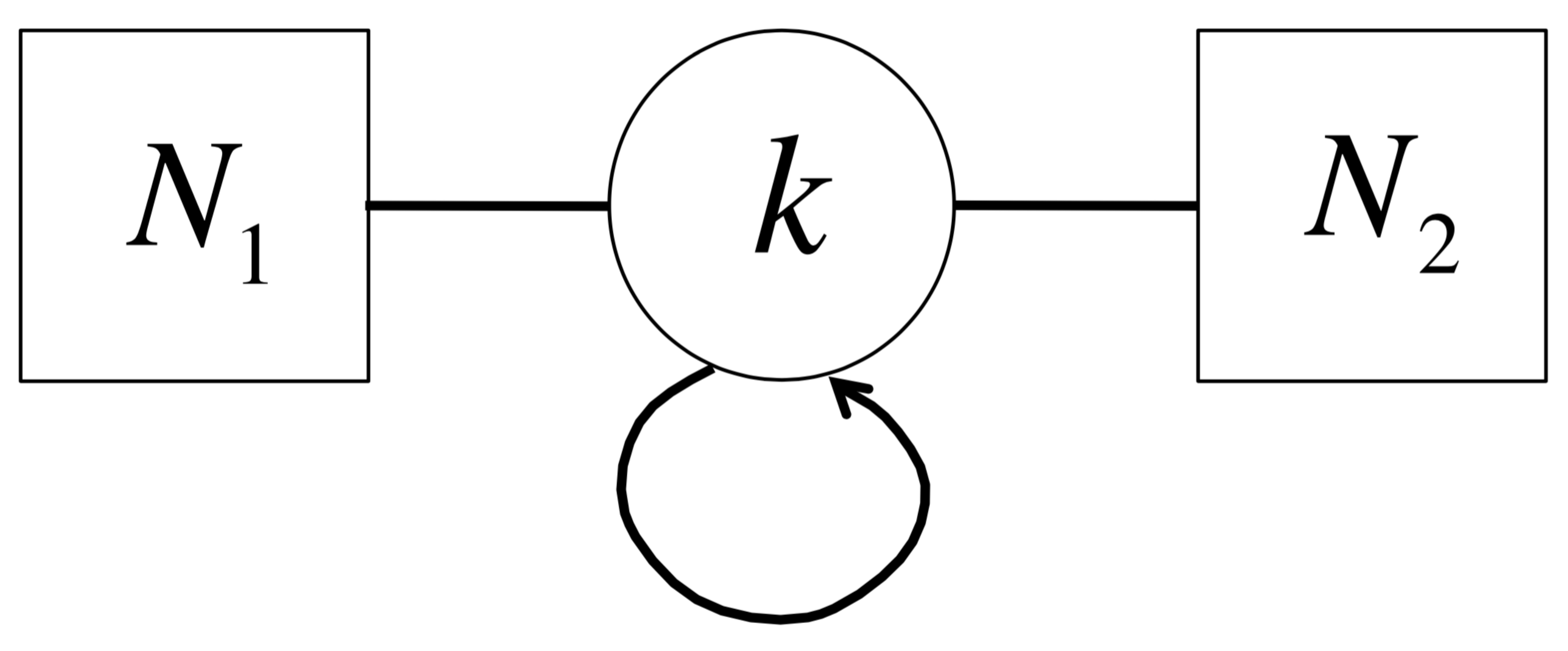}
\caption{Quiver corresponding to the vortex loop which is dual to the Wilson loop in $k$-symmetric representation. The 1d flavor symmetry $N_{1,2}$ couple to the gauge group and flavor group of the bulk 3d theory.}
\label{brane3}
\end{center}
\end{figure}

One can construct the dual vortex loops for more general non-abelian theories using a similar argument. These results have been also confirmed via the $S^3$ partition function \cite{Assel:2015oxa}. 
Let us consider a $U(N_1)$ gauge theory coupled to $N_2+ N_3$ fundamental hypermultiplets, which we split into two groups $N_2$, $N_3$ (splitting the stacks of D5-branes in two, in the analog of Figure~  \ref{brane1}). 
For simplicity, we consider a Wilson loop in the $k$-symmetric representation of $U(N_1)$, corresponding to $k$ stretched F-strings.
The 1d theory which is dual to that Wilson loop  can be obtained from the quiver in Figure~\ref{brane3}.

When considering vortex loops in the twisted theory  on $\Sigma_g$ (as compared to vortex loops in flat space-time), we have to be careful about 
the $R$-charge assignment. The cubic superpotential among 3d fundamental $(Q)$,
1d fundamental $(q)$ and 1d anti-fundamental $(\t q)$ requires that the sum of $U(1)_H$ charges to be 1. Finally, the 1d the adjoint multiplet $(A)$ is not charged under the R-symmetries. Hence the R-charge assignment reads:
\be
\begin{array}{c|ccc} & U(1)_H & U(1)_C & U(1)_{H-C}\\
\hline
Q & \frac{1}{2}&0&\frac12\\
q & 0& 0 & 0\\
\t q & \frac12 &0&\frac12\\
A & 0 &0 &0
\end{array}
\ee
Therefore, the QM index of the 1d theory reads:
\bea \label{general1d}
&V_{k(\pm)} (x,a,t)=\cr
&\quad \frac{1}{k!} \frac{q^{\pm k/2}}{(t-t^{-1})^k}\int_{\text{JK}(\xi_{1d})} \prod_{i=1}^k\frac{du_i}{u_i}
\prod_{i\neq j}^k \frac{u_i-u_j}{u_i t^{-1} -u_j t}
\prod_{i\neq j}^k\frac{u_i t^{R_{\text{adj}}-1}-u_j t^{-R_{\text{adj}}+1}}{u_i t^{R_{\text{adj}}}-u_j t^{-R_{\text{adj}}}}\cr
&\quad \qquad\qquad \quad \times \prod_{i=1}^k\prod_{a=1}^{N_1}\left(\frac{-u_i t^{-1}+ x_a t}{ u_i -x_a }\right)\prod_{i=1}^k\prod_{p=1}^{N_2}\left(\frac{-y_p t^{-1/2} +u_i t^{1/2} }{y_p t^{1/2} -u_i t^{-1/2} }\right)\ .
\eea
Here $R_{\rm adj}$ is a regulator, to be sent to zero at the end of the computation.
Note that the D1-branes in between two left (or right) NS5 branes induce an additional flavor Wilson line factor in \eqref{general1d}. When the defect is attached to the left (right) NS5-brane, we have the flavor Wilson loop factor $q_L^{|\CR|}=q^{-k/2}$ or $q_R^{|\CR|}=q^{k/2}$, respectively \cite{Assel:2015oxa}.
 Let us focus on the D1-branes attached to the left NS5-brane.
The set of  poles selected by the JK residue and with  non-vanishing residue are given as follows.
The contributing poles are classified by the integer set $\{k_a\geq 0,~a=1,\cdots, N_1\}$, which satisfies $k= \sum_{a=1}^{N_1}k_a$. For each set,  the positions of the poles are at
\be
\{u_{i=1,\cdots, k}\} = \{ x_a,x_a t^{2R}, x_at^{4R},\cdots, x_at^{(k_a-1)R},~\text{for }a=1,\cdots, N_1 \}\ .
\ee
Different mappings of $u_i$'s to  elements of the RHS give the same residue due to the Weyl symmetry, which cancels the factor $\frac{1}{k!}$ of (\ref{general1d}). Evaluating the residue and taking the limit $R_{\text{adj}}\rightarrow 0$, we end up with 
\be\label{hjr}
V_{k(-)}(x) = q^{-k/2}\sum_{\substack{k=\sum_{i=1}^{N_1}k_i\\k_i\geq 0}}
\prod_{a=1}^{N_1}
\left(\prod_{b\neq a}^{N_1}\frac{x_a t^{-1}- x_b t}{x_a-x_b}
\prod_{p=1}^{N_2}-\frac{x_a t - a_p^{1/2} }{x_a -a_p^{1/2}t}\right)^{k_a}\ .
\ee
Suppose that there exists another 3d node with rank $N_3$ which is connected to the $N_1$ node
by a 3d bifundamental. Then, applying the three-dimensional Bethe equation for the $U(N_1)$ theory to each term of (\ref{hjr}), we obtain  an alternative expression:
\be
V_{k(-)}(x) = q^{k/2} \sum_{\substack{k=\sum_{i=1}^{N_1}k_i\\k_i\geq 0}} 
\prod_{a=1}^{N_1}\left(\prod_{b\neq a}^{N_1}\frac{-x_a t^{-1}+ x_b t}{-x_a+x_b}
\prod_{k=1}^{N_3}-\frac{x_a -b_k^{1/2}t}{x_a t - b_k^{1/2} }\right)^{k_a} = V_{k(+)}(x) \ .
\ee
This is simply the expression for $V_{k(+)}(x)$, the vortex loop which is attached
to the right NS5-brane in the brane construction. The existence of two distinct UV descriptions of
an IR vortex loop is known as ``hopping duality'' \cite{Gadde:2013ftv, Assel:2015oxa}.

\section{Coulomb branch Hilbert series for an $\CN=4$ $U(2)$ theory}\label{appF: U2}
In this appendix, we show that  the $A$-twisted index for an $\CN=4$ $U(2)$ gauge  theory with $n$ fundamental hypermultiplets reproduces
the monopole formula \cite{Cremonesi:2013lqa}  of the Coulomb branch Hilbert series.

Consider the expression \eqref{ZA full AL} with $\GG=U(2)$. In order to perform the integral at each flux sector, we pick the $\eta=(1,1)$.
In this case, the sum over the flux sectors for the twisted index can be
decomposed into the following expression
\begin{equation}\label{decompose}
I_{U(2)}=\frac12\frac{\by}{(1-\by)^2}\left[\sum_{m_1=1}^{\infty} I_{(m_1,m_1)}+ 2\sum_{m_1>m_2>0}I_{(m_1,m_2)}\right]\ .
\end{equation}
Let us first consider the second term. It can be written as the residue integral at fundamental fields:
\begin{equation}
I_{(m_1,m_2)}=\sum_{q=1}^n\underset{x_2=y_q \by^{1/2}}{\text{res}}\left[\sum_{p=1}^n\underset{x_1=y_p \by^{1/2}}{\text{res}} Z_{\text{1-loop}} (x_1,x_2)\right]\ .
\end{equation}
Note that the charge sets involving $x_1=x_2\by^{-1}$ do not contribute even  though they pass the JK condition. The hyperplane equation $x_1=x_2\by^{-1}$ evaluated at $x_2 = a_q \by^{1/2}$ imposes
the condition $x_1=a_q\by^{-1/2}$, where we have a zero of order $m_1$. Since the order of the pole is
$(1+m_1-m_2) + m_2$, this singularity always has a vanishing residue. 
Since the only poles on the $x_1$ plane are $x_1=0,~\infty,~ y_p \by^{1/2},~ x_1=x_2\by^{-1}$, the residue integral on the $x_1$ plane can be converted into:
\begin{equation}\label{ahe}
I_{(m_1,m_2)}=\sum_{q=1}^n\underset{x_2=y_q \by^{1/2}}{\text{res}}\left[-\underset{x_1=0,\infty}{\text{res}} 
Z_{\text{1-loop}} (x_1,x_2)\right]\ .
\end{equation}
Then we can write (\ref{ahe}) as
\begin{eqnarray}\nonumber
I_{(m_1,m_2)}&=&\sum_{q=1}^n\underset{x_2=y_q \by^{1/2}}{\text{res}}\left[-
\underset{x_1=0,\infty}{\text{res}} 
 Z_{\text{1-loop}} (x_1,x_2)\right]\\
 &=& \underset{x_2=0,\infty}{\text{res}} \left[\underset{x_1=0,\infty}{\text{res}} 
 Z_{\text{1-loop}} (x_1,x_2)\right]
\end{eqnarray}
The last equation follows from the fact that after taking residues at $x_1=0,\infty$, the only remaining
poles on the $x_2$ plane are $x_2=y_q \by^{1/2}$ and $x=0,\infty$.\footnote{
Note that the order in which we take the residues matters for the last expression. We choose this order 
according to the magnitude of $m_1, m_2$.
} Evaluating this expression gives
\begin{eqnarray*}
2\sum_{m_1>m_2>0} I_{(m_1,m_2)} = 2\sum_{m_1>m_2>0} & q^{m_1+m_2}
\left(\by^{{n\ov 2}(m_1+m_2)-(m_1-m_2)}+\by^{-{n\ov 2}(m_1+m_2)+(m_1-m_2)}\right.\\
& \left.-\by^{{n\ov 2}(m_1-m_2)-(m_1-m_2)}-\by^{-{n\ov 2}(m_1-m_2)+(m_1-m_2)}\right)~.
\end{eqnarray*}
Rearranging each infinite sums, we can show  the following identities:
\bea\nn
&2 \sum_{m_1>m_2>0} q^{m_1+m_2}\by^{n(m_1+m_2)/2-(m_1-m_2)} &=& \sum_{\substack{m_1>0, m_2>0\\ m_1\neq m_2}} q^{m_1+m_2} \by^{n|m_1|/2+n|m_2|/2-|m_1-m_2|}\ ,\cr
&2 \sum_{m_1>m_2>0} q^{m_1+m_2}\by^{n(-m_1-m_2)/2+(m_1-m_2)}& =&  \sum_{\substack{m_1\leq 0, m_2\leq 0\\m_1\neq m_2}} q^{m_1+m_2} \by^{n|m_1|/2+n|m_2|/2-|m_1-m_2|} \\
&&&+ 2\sum_{m_1=-\infty}^0 q^{2m_1} \by^{n|m_1|}~,
\eea
and
\bea\nn
&-2\sum_{m_1>m_2>0}q^{m_1+m_2} \left(\by^{n(-m_1+m_2)/2 +(m_1-m_2)}
+\by^{n(m_1-m_2)/2 + (m_2-m_1)}\right) \\
&= \sum_{m_1>0, m_2\leq 0} q^{m_1+m_2} \by^{n|m_1|/2+n|m_2|/2-|m_1-m_2|} 
+  \sum_{m_1\leq 0, m_2> 0} q^{m_1+m_2} \by^{n|m_1|/2+n|m_2|/2-|m_1-m_2|} \\
&\quad + 2\sum_{m_1=1}^\infty q^{2m_1}
\eea
Using these, we have
\bea\label{m1m2}
2\sum_{m_1>m_2>0} I_{(m_1,m_2)} =& \sum_{\substack{(m_1,m_2)\in \mathbb{Z}\times \mathbb{Z}\\ m_1\neq m_2 }}q^{m_1+m_2} \by^{n|m_1|/2+n|m_2|/2-|m_1-m_2|} \\
&+2 \sum_{m_1=-\infty}^0 q^{2m_1} \by^{n|m_1|}+ 2\sum_{m_1=1}^\infty q^{2m_1}~.
\eea

Next let us evaluate the first term of (\ref{decompose}), for the case when the $U(1)^2$ gauge symmetry enhances to $U(2)$.
The residue formula reads:
\begin{equation}
I_{(m_1,m_1)}= \sum_{q=1}^n\underset{x_2=y_q \by^{1/2}}{\text{res}}\left[\sum_{p=1}^n\underset{x_1=y_p \by^{1/2}}{\text{res}} Z_{\text{1-loop}} (x_1,x_2)\right]\ ,
\end{equation}
which can be converted into
\begin{eqnarray}\nonumber
I_{(m_1,m_1)}
 &=&\sum_{q=1}^n\underset{x_2=y_q \by^{1/2}}{\text{res}}\left[
-\left(\underset{x_1=0,\infty}{\text{res}} +\underset{x_1=x_2\by}{\text{res}}\right)
 Z_{\text{1-loop}} (x_1,x_2)\right] \\
 &=& \underset{x_2=0,\infty}{\text{res}} \left(\underset{x_1=0,\infty}{\text{res}} +\underset{x_1=x_2\by}{\text{res}}\right)
 Z_{\text{1-loop}} (x_1,x_2)\ .
\end{eqnarray}
Then we can evaluate the residue integral explicitly, which yields
\bea \label{m1m1}
\sum_{m_1=1}^{\infty} I_{(m_1,m_1)} =& \sum_{m_1=1}^\infty q^{2m_1}(\by^{nm_1/2}-\by^{-nm_1/2})^2 \\
&
+\sum_{m_1=1}^{\infty}q^{2m_1} (\by^{nm_1}-\by^{-nm_1})
\oint_{x_1=x_2 \by} \frac{dx_1}{x_1} \prod_\alpha \frac{x^\alpha-1}{x^\alpha \by^{1/2}-\by^{-1/2}}~.
\eea
Using the formal identity:
\be
\sum_{m_1=-\infty}^\infty q^{2m_1} \by^{nm_1}=0\ , 
\ee
which can be checked by analytic continuation,
we can show that the sum of \eqref{m1m2} and \eqref{m1m1} can be written in the following form:
\bea\nn
&I_{U(2)}= \frac12 \left(\frac{\by^{1/2}}{1-\by}\right)^2\left[ \sum_{m_1=1}^{\infty} I_{(m_1,m_1)}+ 2\sum_{m_1>m_2>0}I_{(m_1,m_2)} \right]\cr
&\qquad = \frac12\left(\frac{\by^{1/2}}{1-\by}\right)^2 \cr
&\times \sum_{(m_1,m_2)\in \mathbb{Z}\times \mathbb{Z}}
 q^{m_1+m_2}\by^{{n\ov 2}(|m_1|+|m_2|)-|m_1-m_2|}
  \left(\oint_{|x_1|=1}
\frac{dx_1}{x_1} \prod_\alpha \frac{x^\alpha-1}{x^\alpha \by^{1/2}-\by^{-1/2}}\right)^{\delta_{m_1,m_2}}
\eea
which reproduces the monopole formula of the $\CN=4$ $U(2)$ gauge theory with $n$  fundamental hypermultiplets, up to prefactor which can be defined away by turning on a background CS level for $U(1)_t$. Note that the integral in the last factor is  a unit circle contour integral, which includes the residue at $x_1=x_2\by$ and $x_1=0$.
This factor can be evaluated as
\begin{equation}
\frac{1}{2}\left(\frac{\by^{1/2}}{1-\by}\right)^2
\oint_{|x_1|=1}
\frac{dx_1}{x_1} \prod_\alpha \frac{x^\alpha-1}{x^\alpha \by^{1/2}-\by^{-1/2}} = \by^2\cdot \frac{1}{(1-\by)(1-\by^2)}
\end{equation}
where the second factor in the RHS corresponds to the Casimir invariant for the $U(2)$ gauge group.

\bibliographystyle{utphys}
\bibliography{bib2d}{}

\providecommand{\href}[2]{#2}\begingroup\raggedright\begin{thebibliography}{10}

\bibitem{Witten:1982df}
E.~Witten, ``{Constraints on Supersymmetry Breaking},''
\href{http://dx.doi.org/10.1016/0550-3213(82)90071-2}{{\em Nucl. Phys.}
  {\bfseries B202} (1982) 253}.

\bibitem{Witten:1999ds}
E.~Witten, ``{Supersymmetric index of three-dimensional gauge theory},''
\href{http://arxiv.org/abs/hep-th/9903005}{{\ttfamily arXiv:hep-th/9903005
  [hep-th]}}.

\bibitem{Kinney:2005ej}
J.~Kinney, J.~M. Maldacena, S.~Minwalla, and S.~Raju, ``{An Index for 4
  dimensional super conformal theories},''
  \href{http://dx.doi.org/10.1007/s00220-007-0258-7}{{\em Commun. Math. Phys.}
  {\bfseries 275} (2007) 209--254},
\href{http://arxiv.org/abs/hep-th/0510251}{{\ttfamily arXiv:hep-th/0510251
  [hep-th]}}.

\bibitem{Romelsberger:2005eg}
C.~Romelsberger, ``{Counting chiral primaries in N = 1, d=4 superconformal
  field theories},''
  \href{http://dx.doi.org/10.1016/j.nuclphysb.2006.03.037}{{\em Nucl. Phys.}
  {\bfseries B747} (2006) 329--353},
\href{http://arxiv.org/abs/hep-th/0510060}{{\ttfamily arXiv:hep-th/0510060
  [hep-th]}}.

\bibitem{Gadde:2013ftv}
A.~Gadde and S.~Gukov, ``{2d Index and Surface operators},''
  \href{http://dx.doi.org/10.1007/JHEP03(2014)080}{{\em JHEP} {\bfseries 03}
  (2014) 080},
\href{http://arxiv.org/abs/1305.0266}{{\ttfamily arXiv:1305.0266 [hep-th]}}.

\bibitem{Benini:2013xpa}
F.~Benini, R.~Eager, K.~Hori, and Y.~Tachikawa, ``{Elliptic Genera of 2d
  ${\mathcal{N}}$ = 2 Gauge Theories},''
  \href{http://dx.doi.org/10.1007/s00220-014-2210-y}{{\em Commun. Math. Phys.}
  {\bfseries 333} no.~3, (2015) 1241--1286},
\href{http://arxiv.org/abs/1308.4896}{{\ttfamily arXiv:1308.4896 [hep-th]}}.

\bibitem{Hori:2014tda}
K.~Hori, H.~Kim, and P.~Yi, ``{Witten Index and Wall Crossing},''
  \href{http://dx.doi.org/10.1007/JHEP01(2015)124}{{\em JHEP} {\bfseries 1501}
  (2015) 124},
\href{http://arxiv.org/abs/1407.2567}{{\ttfamily arXiv:1407.2567 [hep-th]}}.

\bibitem{Benini:2015noa}
F.~Benini and A.~Zaffaroni, ``{A topologically twisted index for
  three-dimensional supersymmetric theories},''
  \href{http://dx.doi.org/10.1007/JHEP07(2015)127}{{\em JHEP} {\bfseries 07}
  (2015) 127},
\href{http://arxiv.org/abs/1504.03698}{{\ttfamily arXiv:1504.03698 [hep-th]}}.

\bibitem{Nekrasov:2014xaa}
N.~A. Nekrasov and S.~L. Shatashvili, ``{Bethe/Gauge correspondence on curved
  spaces},'' \href{http://dx.doi.org/10.1007/JHEP01(2015)100}{{\em JHEP}
  {\bfseries 01} (2015) 100},
\href{http://arxiv.org/abs/1405.6046}{{\ttfamily arXiv:1405.6046 [hep-th]}}.

\bibitem{Nekrasov:2009uh}
N.~A. Nekrasov and S.~L. Shatashvili, ``{Supersymmetric vacua and Bethe
  ansatz},'' \href{http://dx.doi.org/10.1016/j.nuclphysbps.2009.07.047}{{\em
  Nucl. Phys. Proc. Suppl.} {\bfseries 192-193} (2009) 91--112},
\href{http://arxiv.org/abs/0901.4744}{{\ttfamily arXiv:0901.4744 [hep-th]}}.

\bibitem{Nekrasov:2009ui}
N.~A. Nekrasov and S.~L. Shatashvili, ``{Quantum integrability and
  supersymmetric vacua},'' \href{http://dx.doi.org/10.1143/PTPS.177.105}{{\em
  Prog. Theor. Phys. Suppl.} {\bfseries 177} (2009) 105--119},
\href{http://arxiv.org/abs/0901.4748}{{\ttfamily arXiv:0901.4748 [hep-th]}}.

\bibitem{Closset:2015rna}
C.~Closset, S.~Cremonesi, and D.~S. Park, ``{The equivariant A-twist and gauged
  linear sigma models on the two-sphere},''
  \href{http://dx.doi.org/10.1007/JHEP06(2015)076}{{\em JHEP} {\bfseries 06}
  (2015) 076},
\href{http://arxiv.org/abs/1504.06308}{{\ttfamily arXiv:1504.06308 [hep-th]}}.

\bibitem{Doroud:2012xw}
N.~Doroud, J.~Gomis, B.~Le~Floch, and S.~Lee, ``{Exact Results in D=2
  Supersymmetric Gauge Theories},''
  \href{http://dx.doi.org/10.1007/JHEP05(2013)093}{{\em JHEP} {\bfseries 05}
  (2013) 093},
\href{http://arxiv.org/abs/1206.2606}{{\ttfamily arXiv:1206.2606 [hep-th]}}.

\bibitem{Benini:2012ui}
F.~Benini and S.~Cremonesi, ``{Partition Functions of ${\mathcal{N}=(2,2)}$
  Gauge Theories on S$^{2}$ and Vortices},''
  \href{http://dx.doi.org/10.1007/s00220-014-2112-z}{{\em Commun. Math. Phys.}
  {\bfseries 334} no.~3, (2015) 1483--1527},
\href{http://arxiv.org/abs/1206.2356}{{\ttfamily arXiv:1206.2356 [hep-th]}}.

\bibitem{Benini:2013nda}
F.~Benini, R.~Eager, K.~Hori, and Y.~Tachikawa, ``{Elliptic genera of
  two-dimensional N=2 gauge theories with rank-one gauge groups},''
  \href{http://dx.doi.org/10.1007/s11005-013-0673-y}{{\em Lett. Math. Phys.}
  {\bfseries 104} (2014) 465--493},
\href{http://arxiv.org/abs/1305.0533}{{\ttfamily arXiv:1305.0533 [hep-th]}}.

\bibitem{Closset:2015ohf}
C.~Closset, W.~Gu, B.~Jia, and E.~Sharpe, ``{Localization of twisted $
  \mathcal{N}=\left(0,\;2\right) $ gauged linear sigma models in two
  dimensions},'' \href{http://dx.doi.org/10.1007/JHEP03(2016)070}{{\em JHEP}
  {\bfseries 03} (2016) 070},
\href{http://arxiv.org/abs/1512.08058}{{\ttfamily arXiv:1512.08058 [hep-th]}}.

\bibitem{Benini:2015eyy}
F.~Benini, K.~Hristov, and A.~Zaffaroni, ``{Black hole microstates in AdS$_4$
  from supersymmetric localization},''
\href{http://arxiv.org/abs/1511.04085}{{\ttfamily arXiv:1511.04085 [hep-th]}}.

\bibitem{Hosseini:2016tor}
S.~M. Hosseini and A.~Zaffaroni, ``{Large $N$ matrix models for 3d ${\cal N}=2$
  theories: twisted index, free energy and black holes},''
\href{http://arxiv.org/abs/1604.03122}{{\ttfamily arXiv:1604.03122 [hep-th]}}.

\bibitem{Hosseini:2016ume}
S.~M. Hosseini and N.~Mekareeya, ``{Large $N$ topologically twisted index:
  necklace quivers, dualities, and Sasaki-Einstein spaces},''
\href{http://arxiv.org/abs/1604.03397}{{\ttfamily arXiv:1604.03397 [hep-th]}}.

\bibitem{Gukov:2015sna}
S.~Gukov and D.~Pei, ``{Equivariant Verlinde formula from fivebranes and
  vortices},''
\href{http://arxiv.org/abs/1501.01310}{{\ttfamily arXiv:1501.01310 [hep-th]}}.

\bibitem{Gukov:2016gkn}
S.~Gukov, P.~Putrov, and C.~Vafa, ``{Fivebranes and 3-manifold homology},''
\href{http://arxiv.org/abs/1602.05302}{{\ttfamily arXiv:1602.05302 [hep-th]}}.

\bibitem{Intriligator:2013lca}
K.~Intriligator and N.~Seiberg, ``{Aspects of 3d N=2 Chern-Simons-Matter
  Theories},'' \href{http://dx.doi.org/10.1007/JHEP07(2013)079}{{\em JHEP}
  {\bfseries 07} (2013) 079},
\href{http://arxiv.org/abs/1305.1633}{{\ttfamily arXiv:1305.1633 [hep-th]}}.

\bibitem{Ohta:1999iv}
K.~Ohta, ``{Supersymmetric index and s rule for type IIB branes},''
  \href{http://dx.doi.org/10.1088/1126-6708/1999/10/006}{{\em JHEP} {\bfseries
  10} (1999) 006},
\href{http://arxiv.org/abs/hep-th/9908120}{{\ttfamily arXiv:hep-th/9908120
  [hep-th]}}.

\bibitem{Aharony:1997gp}
O.~Aharony, ``{IR duality in d = 3 N=2 supersymmetric USp(2N(c)) and U(N(c))
  gauge theories},''
  \href{http://dx.doi.org/10.1016/S0370-2693(97)00530-3}{{\em Phys. Lett.}
  {\bfseries B404} (1997) 71--76},
\href{http://arxiv.org/abs/hep-th/9703215}{{\ttfamily arXiv:hep-th/9703215
  [hep-th]}}.

\bibitem{Seiberg:1994pq}
N.~Seiberg, ``{Electric - magnetic duality in supersymmetric nonAbelian gauge
  theories},'' \href{http://dx.doi.org/10.1016/0550-3213(94)00023-8}{{\em Nucl.
  Phys.} {\bfseries B435} (1995) 129--146},
\href{http://arxiv.org/abs/hep-th/9411149}{{\ttfamily arXiv:hep-th/9411149
  [hep-th]}}.

\bibitem{Giveon:2008zn}
A.~Giveon and D.~Kutasov, ``{Seiberg Duality in Chern-Simons Theory},''
  \href{http://dx.doi.org/10.1016/j.nuclphysb.2008.09.045}{{\em Nucl. Phys.}
  {\bfseries B812} (2009) 1--11},
\href{http://arxiv.org/abs/0808.0360}{{\ttfamily arXiv:0808.0360 [hep-th]}}.

\bibitem{Cremonesi:2010ae}
S.~Cremonesi, ``{Type IIB construction of flavoured ABJ(M) and fractional M2
  branes},'' \href{http://dx.doi.org/10.1007/JHEP01(2011)076}{{\em JHEP}
  {\bfseries 01} (2011) 076},
\href{http://arxiv.org/abs/1007.4562}{{\ttfamily arXiv:1007.4562 [hep-th]}}.

\bibitem{Benini:2011mf}
F.~Benini, C.~Closset, and S.~Cremonesi, ``{Comments on 3d Seiberg-like
  dualities},'' \href{http://dx.doi.org/10.1007/JHEP10(2011)075}{{\em JHEP}
  {\bfseries 10} (2011) 075},
\href{http://arxiv.org/abs/1108.5373}{{\ttfamily arXiv:1108.5373 [hep-th]}}.

\bibitem{Witten:1993xi}
E.~Witten, ``{The Verlinde algebra and the cohomology of the Grassmannian},''
\href{http://arxiv.org/abs/hep-th/9312104}{{\ttfamily arXiv:hep-th/9312104
  [hep-th]}}.

\bibitem{Kapustin:2013hpk}
A.~Kapustin and B.~Willett, ``{Wilson loops in supersymmetric
  Chern-Simons-matter theories and duality},''
\href{http://arxiv.org/abs/1302.2164}{{\ttfamily arXiv:1302.2164 [hep-th]}}.

\bibitem{Rozansky:1996bq}
L.~Rozansky and E.~Witten, ``{HyperKahler geometry and invariants of three
  manifolds},'' \href{http://dx.doi.org/10.1007/s000290050016}{{\em Selecta
  Math.} {\bfseries 3} (1997) 401--458},
\href{http://arxiv.org/abs/hep-th/9612216}{{\ttfamily arXiv:hep-th/9612216
  [hep-th]}}.

\bibitem{Intriligator:1996ex}
K.~A. Intriligator and N.~Seiberg, ``{Mirror symmetry in three-dimensional
  gauge theories},'' \href{http://dx.doi.org/10.1016/0370-2693(96)01088-X}{{\em
  Phys. Lett.} {\bfseries B387} (1996) 513--519},
\href{http://arxiv.org/abs/hep-th/9607207}{{\ttfamily arXiv:hep-th/9607207
  [hep-th]}}.

\bibitem{Assel:2015oxa}
B.~Assel and J.~Gomis, ``{Mirror Symmetry And Loop Operators},''
  \href{http://dx.doi.org/10.1007/JHEP11(2015)055}{{\em JHEP} {\bfseries 11}
  (2015) 055},
\href{http://arxiv.org/abs/1506.01718}{{\ttfamily arXiv:1506.01718 [hep-th]}}.

\bibitem{Noppadol:2016}
A.~Zaffaroni and N.~Mekareeya, ``{{The relation between the $S^2 \times S^1$
  twisted index and the Hilbert series}}.''. {unpublished}.

\bibitem{Benvenuti:2006qr}
S.~Benvenuti, B.~Feng, A.~Hanany, and Y.-H. He, ``{Counting BPS Operators in
  Gauge Theories: Quivers, Syzygies and Plethystics},''
  \href{http://dx.doi.org/10.1088/1126-6708/2007/11/050}{{\em JHEP} {\bfseries
  11} (2007) 050},
\href{http://arxiv.org/abs/hep-th/0608050}{{\ttfamily arXiv:hep-th/0608050
  [hep-th]}}.

\bibitem{Benvenuti:2010pq}
S.~Benvenuti, A.~Hanany, and N.~Mekareeya, ``{The Hilbert Series of the One
  Instanton Moduli Space},''
  \href{http://dx.doi.org/10.1007/JHEP06(2010)100}{{\em JHEP} {\bfseries 06}
  (2010) 100},
\href{http://arxiv.org/abs/1005.3026}{{\ttfamily arXiv:1005.3026 [hep-th]}}.

\bibitem{Hanany:2011db}
A.~Hanany and N.~Mekareeya, ``{Complete Intersection Moduli Spaces in N=4 Gauge
  Theories in Three Dimensions},''
  \href{http://dx.doi.org/10.1007/JHEP01(2012)079}{{\em JHEP} {\bfseries 01}
  (2012) 079},
\href{http://arxiv.org/abs/1110.6203}{{\ttfamily arXiv:1110.6203 [hep-th]}}.

\bibitem{Cremonesi:2013lqa}
S.~Cremonesi, A.~Hanany, and A.~Zaffaroni, ``{Monopole operators and Hilbert
  series of Coulomb branches of $3d$ $\mathcal{N} = 4$ gauge theories},''
  \href{http://dx.doi.org/10.1007/JHEP01(2014)005}{{\em JHEP} {\bfseries 01}
  (2014) 005},
\href{http://arxiv.org/abs/1309.2657}{{\ttfamily arXiv:1309.2657 [hep-th]}}.

\bibitem{Cremonesi:2014uva}
S.~Cremonesi, A.~Hanany, N.~Mekareeya, and A.~Zaffaroni, ``{T$_{\rho}^{\sigma}$
  (G) theories and their Hilbert series},''
  \href{http://dx.doi.org/10.1007/JHEP01(2015)150}{{\em JHEP} {\bfseries 01}
  (2015) 150},
\href{http://arxiv.org/abs/1410.1548}{{\ttfamily arXiv:1410.1548 [hep-th]}}.

\bibitem{Hanany:2016ezz}
A.~Hanany and M.~Sperling, ``{Coulomb branches for rank 2 gauge groups in 3d
  N=4 gauge theories},''
\href{http://arxiv.org/abs/1605.00010}{{\ttfamily arXiv:1605.00010 [hep-th]}}.

\bibitem{Benini:2016hjo}
F.~Benini and A.~Zaffaroni, ``{Supersymmetric partition functions on Riemann
  surfaces},''
\href{http://arxiv.org/abs/1605.06120}{{\ttfamily arXiv:1605.06120 [hep-th]}}.

\bibitem{Closset:2012ru}
C.~Closset, T.~T. Dumitrescu, G.~Festuccia, and Z.~Komargodski,
  ``{Supersymmetric Field Theories on Three-Manifolds},''
  \href{http://dx.doi.org/10.1007/JHEP05(2013)017}{{\em JHEP} {\bfseries 1305}
  (2013) 017},
\href{http://arxiv.org/abs/1212.3388}{{\ttfamily arXiv:1212.3388 [hep-th]}}.

\bibitem{Closset:2013vra}
C.~Closset, T.~T. Dumitrescu, G.~Festuccia, and Z.~Komargodski, ``{The Geometry
  of Supersymmetric Partition Functions},''
  \href{http://dx.doi.org/10.1007/JHEP01(2014)124}{{\em JHEP} {\bfseries 01}
  (2014) 124},
\href{http://arxiv.org/abs/1309.5876}{{\ttfamily arXiv:1309.5876 [hep-th]}}.

\bibitem{Closset:2012vp}
C.~Closset, T.~T. Dumitrescu, G.~Festuccia, Z.~Komargodski, and N.~Seiberg,
  ``{Comments on Chern-Simons Contact Terms in Three Dimensions},''
  \href{http://dx.doi.org/10.1007/JHEP09(2012)091}{{\em JHEP} {\bfseries 09}
  (2012) 091},
\href{http://arxiv.org/abs/1206.5218}{{\ttfamily arXiv:1206.5218 [hep-th]}}.

\bibitem{Aganagic:2001uw}
M.~Aganagic, K.~Hori, A.~Karch, and D.~Tong, ``{Mirror symmetry in
  (2+1)-dimensions and (1+1)-dimensions},''
  \href{http://dx.doi.org/10.1088/1126-6708/2001/07/022}{{\em JHEP} {\bfseries
  07} (2001) 022},
\href{http://arxiv.org/abs/hep-th/0105075}{{\ttfamily arXiv:hep-th/0105075
  [hep-th]}}.

\bibitem{Dimofte:2011jd}
T.~Dimofte and S.~Gukov, ``{Chern-Simons Theory and S-duality},''
  \href{http://dx.doi.org/10.1007/JHEP05(2013)109}{{\em JHEP} {\bfseries 05}
  (2013) 109},
\href{http://arxiv.org/abs/1106.4550}{{\ttfamily arXiv:1106.4550 [hep-th]}}.

\bibitem{Kapustin:2009kz}
A.~Kapustin, B.~Willett, and I.~Yaakov, ``{Exact Results for Wilson Loops in
  Superconformal Chern-Simons Theories with Matter},''
  \href{http://dx.doi.org/10.1007/JHEP03(2010)089}{{\em JHEP} {\bfseries 03}
  (2010) 089},
\href{http://arxiv.org/abs/0909.4559}{{\ttfamily arXiv:0909.4559 [hep-th]}}.

\bibitem{Blau:1994rk}
M.~Blau and G.~Thompson, ``{On diagonalization in map(M,G)},''
  \href{http://dx.doi.org/10.1007/BF02104681}{{\em Commun.Math.Phys.}
  {\bfseries 171} (1995) 639--660},
\href{http://arxiv.org/abs/hep-th/9402097}{{\ttfamily arXiv:hep-th/9402097
  [hep-th]}}.

\bibitem{Blau:1995rs}
M.~Blau and G.~Thompson, ``{Localization and diagonalization: A review of
  functional integral techniques for low dimensional gauge theories and
  topological field theories},'' \href{http://dx.doi.org/10.1063/1.531038}{{\em
  J.Math.Phys.} {\bfseries 36} (1995) 2192--2236},
\href{http://arxiv.org/abs/hep-th/9501075}{{\ttfamily arXiv:hep-th/9501075
  [hep-th]}}.

\bibitem{Englert:1976ng}
F.~Englert and P.~Windey, ``{Quantization Condition for 't Hooft Monopoles in
  Compact Simple Lie Groups},''
\href{http://dx.doi.org/10.1103/PhysRevD.14.2728}{{\em Phys.Rev.} {\bfseries
  D14} (1976) 2728}.

\bibitem{Kapustin:2005py}
A.~Kapustin, ``{Wilson-'t Hooft operators in four-dimensional gauge theories
  and S-duality},'' \href{http://dx.doi.org/10.1103/PhysRevD.74.025005}{{\em
  Phys. Rev.} {\bfseries D74} (2006) 025005},
\href{http://arxiv.org/abs/hep-th/0501015}{{\ttfamily arXiv:hep-th/0501015
  [hep-th]}}.

\bibitem{Closset:2012vg}
C.~Closset, T.~T. Dumitrescu, G.~Festuccia, Z.~Komargodski, and N.~Seiberg,
  ``{Contact Terms, Unitarity, and F-Maximization in Three-Dimensional
  Superconformal Theories},''
  \href{http://dx.doi.org/10.1007/JHEP10(2012)053}{{\em JHEP} {\bfseries 10}
  (2012) 053},
\href{http://arxiv.org/abs/1205.4142}{{\ttfamily arXiv:1205.4142 [hep-th]}}.

\bibitem{Aharony:1997bx}
O.~Aharony, A.~Hanany, K.~A. Intriligator, N.~Seiberg, and M.~J. Strassler,
  ``{Aspects of N=2 supersymmetric gauge theories in three-dimensions},''
  \href{http://dx.doi.org/10.1016/S0550-3213(97)00323-4}{{\em Nucl. Phys.}
  {\bfseries B499} (1997) 67--99},
\href{http://arxiv.org/abs/hep-th/9703110}{{\ttfamily arXiv:hep-th/9703110
  [hep-th]}}.

\bibitem{deBoer:1997kr}
J.~de~Boer, K.~Hori, and Y.~Oz, ``{Dynamics of N=2 supersymmetric gauge
  theories in three-dimensions},''
  \href{http://dx.doi.org/10.1016/S0550-3213(97)00328-3}{{\em Nucl. Phys.}
  {\bfseries B500} (1997) 163--191},
\href{http://arxiv.org/abs/hep-th/9703100}{{\ttfamily arXiv:hep-th/9703100
  [hep-th]}}.

\bibitem{Borokhov:2002ib}
V.~Borokhov, A.~Kapustin, and X.-k. Wu, ``{Topological disorder operators in
  three-dimensional conformal field theory},''
  \href{http://dx.doi.org/10.1088/1126-6708/2002/11/049}{{\em JHEP} {\bfseries
  11} (2002) 049},
\href{http://arxiv.org/abs/hep-th/0206054}{{\ttfamily arXiv:hep-th/0206054
  [hep-th]}}.

\bibitem{Borokhov:2002cg}
V.~Borokhov, A.~Kapustin, and X.-k. Wu, ``{Monopole operators and mirror
  symmetry in three-dimensions},''
  \href{http://dx.doi.org/10.1088/1126-6708/2002/12/044}{{\em JHEP} {\bfseries
  12} (2002) 044},
\href{http://arxiv.org/abs/hep-th/0207074}{{\ttfamily arXiv:hep-th/0207074
  [hep-th]}}.

\bibitem{Gaiotto:2009tk}
D.~Gaiotto and D.~L. Jafferis, ``{Notes on adding D6 branes wrapping RP**3 in
  AdS(4) x CP**3},'' \href{http://dx.doi.org/10.1007/JHEP11(2012)015}{{\em
  JHEP} {\bfseries 11} (2012) 015},
\href{http://arxiv.org/abs/0903.2175}{{\ttfamily arXiv:0903.2175 [hep-th]}}.

\bibitem{Jafferis:2009th}
D.~L. Jafferis, ``{Quantum corrections to $\mathcal{N} = 2$ Chern-Simons
  theories with flavor and their AdS$_{4}$ duals},''
  \href{http://dx.doi.org/10.1007/JHEP08(2013)046}{{\em JHEP} {\bfseries 08}
  (2013) 046},
\href{http://arxiv.org/abs/0911.4324}{{\ttfamily arXiv:0911.4324 [hep-th]}}.

\bibitem{Benini:2009qs}
F.~Benini, C.~Closset, and S.~Cremonesi, ``{Chiral flavors and M2-branes at
  toric CY4 singularities},''
  \href{http://dx.doi.org/10.1007/JHEP02(2010)036}{{\em JHEP} {\bfseries 02}
  (2010) 036},
\href{http://arxiv.org/abs/0911.4127}{{\ttfamily arXiv:0911.4127 [hep-th]}}.

\bibitem{Benini:2011cma}
F.~Benini, C.~Closset, and S.~Cremonesi, ``{Quantum moduli space of
  Chern-Simons quivers, wrapped D6-branes and AdS4/CFT3},''
  \href{http://dx.doi.org/10.1007/JHEP09(2011)005}{{\em JHEP} {\bfseries 09}
  (2011) 005},
\href{http://arxiv.org/abs/1105.2299}{{\ttfamily arXiv:1105.2299 [hep-th]}}.

\bibitem{Aharony:2013hda}
O.~Aharony, N.~Seiberg, and Y.~Tachikawa, ``{Reading between the lines of
  four-dimensional gauge theories},''
  \href{http://dx.doi.org/10.1007/JHEP08(2013)115}{{\em JHEP} {\bfseries 08}
  (2013) 115},
\href{http://arxiv.org/abs/1305.0318}{{\ttfamily arXiv:1305.0318 [hep-th]}}.

\bibitem{Verlinde:1988sn}
E.~P. Verlinde, ``{Fusion Rules and Modular Transformations in 2D Conformal
  Field Theory},''
\href{http://dx.doi.org/10.1016/0550-3213(88)90603-7}{{\em Nucl. Phys.}
  {\bfseries B300} (1988) 360}.

\bibitem{Witten:1988hf}
E.~Witten, ``{Quantum Field Theory and the Jones Polynomial},''
\href{http://dx.doi.org/10.1007/BF01217730}{{\em Commun. Math. Phys.}
  {\bfseries 121} (1989) 351--399}.

\bibitem{Blau:1993tv}
M.~Blau and G.~Thompson, ``{Derivation of the Verlinde formula from
  Chern-Simons theory and the G/G model},''
  \href{http://dx.doi.org/10.1016/0550-3213(93)90538-Z}{{\em Nucl. Phys.}
  {\bfseries B408} (1993) 345--390},
\href{http://arxiv.org/abs/hep-th/9305010}{{\ttfamily arXiv:hep-th/9305010
  [hep-th]}}.

\bibitem{Ohta:2012ev}
K.~Ohta and Y.~Yoshida, ``{Non-Abelian Localization for Supersymmetric
  Yang-Mills-Chern-Simons Theories on Seifert Manifold},''
  \href{http://dx.doi.org/10.1103/PhysRevD.86.105018}{{\em Phys. Rev.}
  {\bfseries D86} (2012) 105018},
\href{http://arxiv.org/abs/1205.0046}{{\ttfamily arXiv:1205.0046 [hep-th]}}.

\bibitem{Hanany:1997vm}
A.~Hanany and K.~Hori, ``{Branes and N=2 theories in two-dimensions},''
  \href{http://dx.doi.org/10.1016/S0550-3213(97)00754-2}{{\em Nucl. Phys.}
  {\bfseries B513} (1998) 119--174},
\href{http://arxiv.org/abs/hep-th/9707192}{{\ttfamily arXiv:hep-th/9707192
  [hep-th]}}.

\bibitem{JK1995}
L.~C. {Jeffrey} and F.~C. {Kirwan}, ``{Localization for nonabelian group
  actions},'' {\em Topology} {\bfseries 34} (1995) 291--327.

\bibitem{1999math......3178B}
M.~{Brion} and M.~{Vergne}, ``{Arrangements of hyperplanes I: Rational
  functions and Jeffrey-Kirwan residue},'' {\em ArXiv Mathematics e-prints}
  (Mar., 1999) , \href{http://arxiv.org/abs/math/9903178}{{\ttfamily
  math/9903178}}.

\bibitem{2004InMat.158..453S}
A.~{Szenes} and M.~{Vergne}, ``{Toric reduction and a conjecture of Batyrev and
  Materov},'' \href{http://dx.doi.org/10.1007/s00222-004-0375-2}{{\em
  Inventiones Mathematicae} {\bfseries 158} (June, 2004) 453--495},
  \href{http://arxiv.org/abs/math/0306311}{{\ttfamily math/0306311}}.

\bibitem{Gaiotto:2013bwa}
D.~Gaiotto and P.~Koroteev, ``{On Three Dimensional Quiver Gauge Theories and
  Integrability},'' \href{http://dx.doi.org/10.1007/JHEP05(2013)126}{{\em JHEP}
  {\bfseries 05} (2013) 126},
\href{http://arxiv.org/abs/1304.0779}{{\ttfamily arXiv:1304.0779 [hep-th]}}.

\bibitem{Imbimbo:2014pla}
C.~Imbimbo and D.~Rosa, ``{Topological anomalies for Seifert 3-manifolds},''
  \href{http://dx.doi.org/10.1007/JHEP07(2015)068}{{\em JHEP} {\bfseries 07}
  (2015) 068},
\href{http://arxiv.org/abs/1411.6635}{{\ttfamily arXiv:1411.6635 [hep-th]}}.

\bibitem{Dimofte:2011ju}
T.~Dimofte, D.~Gaiotto, and S.~Gukov, ``{Gauge Theories Labelled by
  Three-Manifolds},'' \href{http://dx.doi.org/10.1007/s00220-013-1863-2}{{\em
  Commun. Math. Phys.} {\bfseries 325} (2014) 367--419},
\href{http://arxiv.org/abs/1108.4389}{{\ttfamily arXiv:1108.4389 [hep-th]}}.

\bibitem{Zagier}
D.~Zagier, ``{Elementary aspects of the Verlinde formula and the
  Harder-Narasimhan-Atiyah-Bott formula}.''
\newblock
  \url{http://people.mpim-bonn.mpg.de/zagier/files/mpim/94-5/fulltext.pdf}.

\bibitem{Willett:2011gp}
B.~Willett and I.~Yaakov, ``{N=2 Dualities and Z Extremization in Three
  Dimensions},''
\href{http://arxiv.org/abs/1104.0487}{{\ttfamily arXiv:1104.0487 [hep-th]}}.

\bibitem{Gaiotto:2013sma}
D.~Gaiotto, S.~Gukov, and N.~Seiberg, ``{Surface Defects and Resolvents},''
  \href{http://dx.doi.org/10.1007/JHEP09(2013)070}{{\em JHEP} {\bfseries 1309}
  (2013) 070},
\href{http://arxiv.org/abs/1307.2578}{{\ttfamily arXiv:1307.2578 [hep-th]}}.

\bibitem{Benini:2014mia}
F.~Benini, D.~S. Park, and P.~Zhao, ``{Cluster algebras from dualities of 2d
  N=(2,2) quiver gauge theories},''
\href{http://arxiv.org/abs/1406.2699}{{\ttfamily arXiv:1406.2699 [hep-th]}}.

\bibitem{Kapustin:2010ag}
A.~Kapustin and K.~Vyas, ``{A-Models in Three and Four Dimensions},''
\href{http://arxiv.org/abs/1002.4241}{{\ttfamily arXiv:1002.4241 [hep-th]}}.

\bibitem{Brooks:1994nn}
R.~Brooks and S.~J. Gates, Jr., ``{Extended supersymmetry and superBF gauge
  theories},'' \href{http://dx.doi.org/10.1016/0550-3213(94)90600-9}{{\em Nucl.
  Phys.} {\bfseries B432} (1994) 205--224},
\href{http://arxiv.org/abs/hep-th/9407147}{{\ttfamily arXiv:hep-th/9407147
  [hep-th]}}.

\bibitem{Kapustin:1999ha}
A.~Kapustin and M.~J. Strassler, ``{On mirror symmetry in three-dimensional
  Abelian gauge theories},''
  \href{http://dx.doi.org/10.1088/1126-6708/1999/04/021}{{\em JHEP} {\bfseries
  04} (1999) 021},
\href{http://arxiv.org/abs/hep-th/9902033}{{\ttfamily arXiv:hep-th/9902033
  [hep-th]}}.

\bibitem{Gadde:2015wta}
A.~Gadde, S.~S. Razamat, and B.~Willett, ``{On the reduction of 4d $
  \mathcal{N}=1 $ theories on $ {\mathbb{S}}^2 $},''
  \href{http://dx.doi.org/10.1007/JHEP11(2015)163}{{\em JHEP} {\bfseries 11}
  (2015) 163},
\href{http://arxiv.org/abs/1506.08795}{{\ttfamily arXiv:1506.08795 [hep-th]}}.

\bibitem{Kapustin:2010xq}
A.~Kapustin, B.~Willett, and I.~Yaakov, ``{Nonperturbative Tests of
  Three-Dimensional Dualities},''
  \href{http://dx.doi.org/10.1007/JHEP10(2010)013}{{\em JHEP} {\bfseries 10}
  (2010) 013},
\href{http://arxiv.org/abs/1003.5694}{{\ttfamily arXiv:1003.5694 [hep-th]}}.

\bibitem{Gaiotto:2008ak}
D.~Gaiotto and E.~Witten, ``{S-Duality of Boundary Conditions In N=4 Super
  Yang-Mills Theory},''
  \href{http://dx.doi.org/10.4310/ATMP.2009.v13.n3.a5}{{\em Adv. Theor. Math.
  Phys.} {\bfseries 13} no.~3, (2009) 721--896},
\href{http://arxiv.org/abs/0807.3720}{{\ttfamily arXiv:0807.3720 [hep-th]}}.

\bibitem{Hanany:1996ie}
A.~Hanany and E.~Witten, ``{Type IIB superstrings, BPS monopoles, and
  three-dimensional gauge dynamics},''
  \href{http://dx.doi.org/10.1016/S0550-3213(97)00157-0,
  10.1016/S0550-3213(97)80030-2}{{\em Nucl. Phys.} {\bfseries B492} (1997)
  152--190},
\href{http://arxiv.org/abs/hep-th/9611230}{{\ttfamily arXiv:hep-th/9611230
  [hep-th]}}.

\bibitem{Razamat:2014pta}
S.~S. Razamat and B.~Willett, ``{Down the rabbit hole with theories of class $
  \mathcal{S} $},'' \href{http://dx.doi.org/10.1007/JHEP10(2014)099}{{\em JHEP}
  {\bfseries 10} (2014) 99},
\href{http://arxiv.org/abs/1403.6107}{{\ttfamily arXiv:1403.6107 [hep-th]}}.

\bibitem{Festuccia:2011ws}
G.~Festuccia and N.~Seiberg, ``{Rigid Supersymmetric Theories in Curved
  Superspace},'' \href{http://dx.doi.org/10.1007/JHEP06(2011)114}{{\em JHEP}
  {\bfseries 1106} (2011) 114},
\href{http://arxiv.org/abs/1105.0689}{{\ttfamily arXiv:1105.0689 [hep-th]}}.

\bibitem{Closset:2014uda}
C.~Closset, T.~T. Dumitrescu, G.~Festuccia, and Z.~Komargodski, ``{From Rigid
  Supersymmetry to Twisted Holomorphic Theories},''
  \href{http://dx.doi.org/10.1103/PhysRevD.90.085006}{{\em Phys. Rev.}
  {\bfseries D90} no.~8, (2014) 085006},
\href{http://arxiv.org/abs/1407.2598}{{\ttfamily arXiv:1407.2598 [hep-th]}}.

\bibitem{Redlich:1983dv}
A.~N. Redlich, ``{Parity Violation and Gauge Noninvariance of the Effective
  Gauge Field Action in Three-Dimensions},''
\href{http://dx.doi.org/10.1103/PhysRevD.29.2366}{{\em Phys. Rev.} {\bfseries
  D29} (1984) 2366--2374}.

\bibitem{berline1992heat}
N.~Berline, E.~Getzler, and M.~Vergne, {\em Heat kernels and Dirac operators}.
\newblock Springer Science \& Business Media, 1992.

\bibitem{buser2010geometry}
P.~Buser, {\em Geometry and spectra of compact Riemann surfaces}.
\newblock Springer Science \& Business Media, 2010.

\bibitem{Witten:2003ya}
E.~Witten, ``{SL(2,Z) action on three-dimensional conformal field theories with
  Abelian symmetry},''
\href{http://arxiv.org/abs/hep-th/0307041}{{\ttfamily arXiv:hep-th/0307041
  [hep-th]}}.

\end{thebibliography}\endgroup

\end{document}